\newif\ifllncs

\ifllncs
    \documentclass{llncs}
    \newcommand{\authnote}[3]{\textcolor{#3}{[{\footnotesize {#2} {\bf -- #1}}]}}
    \newcommand{\prabhanjan}[1]{}
    \newcommand{\pnote}[1]{}
    \newcommand{\john}[1]{}
    \newcommand{\yaoting}[1]{}
    \newcommand{\aditya}[1]{}
\else
    \documentclass{article}
    \usepackage{fullpage}
    
    \newcommand{\authnote}[3]{\textcolor{#3}{[{\footnotesize {#2} {\bf - #1}}]}}
    \newcommand{\pnote}[1]{{\color{blue} Prabhanjan: #1}}
    \newcommand{\prabhanjan}[1]{{\color{blue} [Prabhanjan: #1]}}
    \newcommand{\john}[1]{\authnote{John}{#1}{purple}}
    \newcommand{\yaoting}[1]{\authnote{Yao-Ting}{#1}{teal}}
    \newcommand{\aditya}[1]{\authnote{Aditya}{#1}{brown}}
\fi
\usepackage{diagbox}
\usepackage{graphicx} 
\usepackage{amsmath,amsfonts,amssymb}
\usepackage{ninecolors}

\ifllncs
\else
    \usepackage{amsthm}
\fi

\usepackage{graphicx} 
\usepackage{hyperref}
\usepackage[nameinlink]{cleveref}
\usepackage[T1]{fontenc}
\usepackage{physics}
\usepackage{todonotes}
\usepackage{mdframed} 
\usepackage{lipsum}
\usepackage{multicol}
\usepackage{bbm}
\usepackage{longfbox}
\usepackage{enumitem}
\usepackage{float} 
\usepackage{nicefrac} 
\usepackage{dsfont}
\usepackage{longtable} 
\usepackage{circuitikz} 
\usepackage{caption}
\usepackage{subcaption}
\usepackage{mathtools} 
\usepackage{quantikz}
\usetikzlibrary{backgrounds}

\newcommand{\regA}{{\color{gray} \mathbf{A}}}
\newcommand{\regB}{{\color{gray} \mathbf{B}}}
\newcommand{\regC}{{\color{gray} \mathbf{C}}}

\newcommand{\fwd}{\mathsf{fwd}}
\newcommand{\inv}{\mathsf{inv}}
\newcommand{\Dom}{\operatorname{Dom}}
\newcommand{\rcolor}[1]{{\color{red4} #1}}
\newcommand{\bcolor}[1]{{\color{blue4} #1}}

\allowdisplaybreaks

\hypersetup{colorlinks={true},linkcolor={blue},citecolor=magenta}

\usepackage{comment}

\usepackage[style=alphabetic,minalphanames=3,maxalphanames=4,maxnames=99,backref=true]{biblatex}
\newcommand{\myparagraph}[1]{\vspace{.5em} \noindent \textbf{#1.}\,}

\ifllncs

\spnewtheorem{myclaim}[theorem]{Claim}{\bfseries}{\itshape}
\Crefname{myclaim}{Claim}{Claims}

\renewenvironment{proof}[1][Proof]
{\par\noindent\textit{#1. }}
{\hfill$\square$\par}

\else
\newtheorem{theorem}{Theorem}[section]

\newtheorem{definition}[theorem]{Definition}
\newtheorem{lemma}[theorem]{Lemma}

\newtheorem{myclaim}[theorem]{Claim}

\newtheorem{corollary}[theorem]{Corollary}

\Crefname{fact}{Fact}{Facts}
\Crefname{myclaim}{Claim}{Claims}
\fi

\newmdenv[
  linewidth=1pt,
  roundcorner=5pt,
  linecolor=black,
  innerleftmargin=10pt,
  innerrightmargin=10pt,
  innertopmargin=5pt,
  innerbottommargin=5pt
]{protocolbox}

\newcommand{\st}{\ \text{s.t.}\ }

\newcommand{\N}{\mathbb{N}}

\newcommand{\G}{\mathbb{G}}
\newcommand{\F}{\mathbb{F}}

\renewcommand{\bra}[1]{\langle#1\rvert}
\renewcommand{\braket}[2]{\langle #1 \mid #2 \rangle}
\renewcommand{\ket}[1]{\lvert#1\rangle}
\newcommand{\set}[1]{\{ #1 \}}
\newcommand{\bit}{\{0,1\}}

\newcommand{\cB}{{\mathcal B}}

\newcommand{\cD}{{\mathcal D}}

\newcommand{\cG}{{\mathcal G}}

\newcommand{\cI}{{\mathcal I}}

\newcommand{\cK}{{\mathcal K}}
\newcommand{\cL}{{\mathcal L}}

\newcommand{\cP}{{\mathcal P}}

\newcommand{\cR}{{\mathcal R}}

\newcommand{\cU}{{\mathcal U}}
\newcommand{\cV}{{\mathcal V}}

\newcommand{\frF}{\mathfrak{F}}
\newcommand{\frG}{\mathfrak{G}}

\newcommand{\frl}{\mathfrak{l}}
\newcommand{\frm}{\mathfrak{m}}

\newcommand{\frp}{\mathfrak{p}}
\newcommand{\frq}{\mathfrak{q}}
\newcommand{\frr}{\mathfrak{r}}

\newcommand{\secp}{{\lambda}}
\newcommand{\poly}{\mathsf{poly}}

\newcommand{\Exp}{\operatorname*{\mathbb{E}}}
\newcommand{\Ex}{\Exp}

\newcommand{\negl}{\mathsf{negl}}

\newcommand{\TD}{\mathsf{TD}}

\usepackage[normalem]{ulem}

\newcommand{\E}{\mathop{\mathbb{E}}}

\newcommand{\Haar}{\mathcal{H}}

\renewcommand{\ketbra}[2]{\ket{#1}\bra{#2}}
\renewcommand{\braket}[2]{\langle #1 | #2 \rangle}

\newcommand{\good}{\mathsf{Good}}

\renewcommand{\bra}[1]{\langle#1\rvert}
\renewcommand{\braket}[2]{\langle #1 | #2 \rangle}
\renewcommand{\ket}[1]{\lvert#1\rangle}
\renewcommand{\ketbra}[2]{\ket{#1}\!\bra{#2}}
\newcommand{\reg}[1]{\mathsf{#1}}

\newcommand{\avg}{\mathop{\mathbb{E}}}

\newcommand{\haarstates}{\Haar}
\newcommand{\haarunitaries}{\mu}
\newcommand{\Symgp}{\mathsf{Sym}}

\newcommand{\Adversary}{{\cal A}}

\newcommand{\pr}{\mathsf{PR}}

\newcommand{\dist}{\mathrm{dist}}

\newcommand{\oracle}{\mathcal{O}}

\newcommand{\indic}{\mathds{1}}

\newcommand{\hyb}{\mathbf{H}}

\renewcommand{\op}{\mathrm{op}}

\newcommand{\comp}{\mathsf{comp}}
\newcommand{\Ocomp}{\oracle_{\comp}}
\newcommand{\glued}{\mathsf{glued}}
\newcommand{\gluedfwd}{\mathsf{glued\text{-}fwd}}
\newcommand{\gluedinv}{\mathsf{glued\text{-}inv}}

\newcommand{\sG}{\sf start}
\newcommand{\eG}{\sf end}

\newcommand{\vect}[1]{\overrightarrow{#1}}
\newcommand{\leng}{\mathsf{len}}
\newcommand{\coun}{\mathsf{count}}
\newcommand{\midd}{\mathsf{mid}}

\title{Gluing Random Unitaries with Inverses\\
and Applications to Strong Pseudorandom Unitaries\thanks{A preliminary version, merging this paper and~\cite{ABGL25arXivPartI}, appears in the proceedings of the \textit{45th Annual International Cryptology Conference} (\textsc{CRYPTO~2025}) under the title ``Pseudorandom Unitaries in the Haar Random Oracle Model''~\cite{CRYPTO:ABGL25}. This is Part~II of the full version.}}

\ifllncs
    \author{}
    \institute{}
\else
    \author{Prabhanjan Ananth\thanks{\texttt{prabhanjan@cs.ucsb.edu}}\\ \small{UCSB} \and John Bostanci\thanks{\texttt{johnb@cs.columbia.edu}}\\ \small{Columbia} \and Aditya Gulati\thanks{\texttt{adityagulati@ucsb.edu}}\\ \small{UCSB} \and Yao-Ting Lin\thanks{\texttt{yao-ting\_lin@ucsb.edu}}\\ \small{UCSB}}
    \date{}
\fi

\begin{document}

\maketitle

\begin{abstract}
    \noindent Gluing theorem for random unitaries [Schuster, Haferkamp, Huang, QIP 2025] have found numerous applications, including designing low depth random unitaries [Schuster, Haferkamp, Huang, QIP 2025], random unitaries in ${\sf QAC0}$ [Foxman, Parham, Vasconcelos, Yuen'25] and generically shortening the key length of pseudorandom unitaries [Ananth, Bostanci, Gulati, Lin EUROCRYPT'25].  We present an alternate method of combining Haar random unitaries from the gluing lemma from [Schuster, Haferkamp, Huang, QIP 2025] that is secure against adversaries with inverse query access to the joined unitary. As a consequence, we show for the first time that strong pseudorandom unitaries can generically have their length extended, and can be constructed using only $O(n^{1/c})$ bits of randomness, for any constant $c$, if any family of strong pseudorandom unitaries exists.


\end{abstract}

\ifllncs
\else
\newpage 
    \tableofcontents
\newpage 
\fi

\section{Introduction}
Random unitaries are fundamental objects that find applications across diverse areas of quantum information science, including quantum algorithm benchmarking~\cite{knill2008randomized}, quantum machine learning~\cite{huang2023learning}, quantum cryptography~\cite{JLS18,GJMZ23,AGKL,bostanci2024efficient}, quantum chaos~\cite{gu2024simulating,liu2018spectral} and quantum gravity~\cite{cotler2017black}. Their utility stems from their ability to model generic quantum processes and serve as building blocks for various quantum protocols.
Random unitaries are inherently complex objects—they require exponentially sized descriptions in general. To circumvent this complexity, researchers have developed the concepts of 
$t$-designs~\cite{ambainis2007quantum} and pseudorandom unitaries (PRUs)~\cite{JLS18}, which can efficiently approximate the statistical properties of truly random unitaries for many applications. 
\par Understanding the resources needed to implement random unitaries, $t$-designs and pseudorandom unitaries has been an important problem. Recently, a remarkable work by Schuster, Haferkamp and Huang~\cite{schuster2024random} presented a construction of random unitaries in extremely low depth. Specifically, they showed that pseudorandom unitaries can be constructed in logarithmic depth. The core contribution of their work is the gluing theorem which informally states the following: suppose we have two random unitaries $U_1,U_2$ such that $U_{1}$ acts on registers $\regA,\regB$ and unitaries $U_2$ acts on registers $\regB,\regC$ then $U_1U_2$ approximately computes a random unitary on registers $\regA,\regB$ and $\regC$ as long as $\regB$ is sufficiently large enough.  The gluing theorem has been proven to be quite useful in many applications:
\begin{itemize}
    \item In the same work, Schuster et al.~\cite{schuster2024random} applied the gluing theorem recursively to construct random unitaries in logarithmic depth. 
    \item Foxman, Parham, Vasconcelos, Yuen~\cite{foxman2025random} used the gluing theorem to demonstrate that pseudorandom unitaries can be approximately implemented in ${\sf QAC0}$.
    \item Ananth, Bostanci, Gulati and, Lin~\cite{ABGL24} used the gluing theorem to show that any pseudorandom unitary can be converted into another pseudorandom unitary with the key length to be much smaller than the output length. 
\end{itemize}
\par The disadvantage of the above gluing theorem is that the closeness to the joining random unitary does not hold if additionally oracle access to the inverse of the glued unitary is provided. In many applications, giving both forward and inverse access is important. As noted in \cite{foxman2025random}, to determine lightcones, entanglement entropy and displacement amplitudes, access to the inverse is required. Having a gluing theorem that holds {\em even with inverse access} could have powerful applications; we call such a gluing theorem, a {\em strong} gluing theorem. As an example, \cite{foxman2025random} showed that the non-existence of strong gluing theorem (with certain properties) would imply that ${\sf PARITY} \notin {\sf QAC0}$, settling a major open problem in quantum complexity theory. 
 
\subsection{Our Results}
We present for the first time a strong gluing theorem for random unitaries. 

\begin{theorem}[Strong gluing of random unitaries]
\label{thm:strong:gluing}
    Let $U^1$, $U^2$, and $U^3$ be three Haar random unitaries on $n$ qubits, and $\reg{A}, \reg{C}$ be registers of length $n - \lambda$ qubits, and $\reg{B}$ be a register of $\lambda$ qubits, for $\lambda = \Omega(\log^{1+\epsilon}(n))$.  Then no polynomial-query adversary can distinguish between $U^1_{\reg{AB}}V^2_{\reg{BC}}W^3_{\reg{AB}}$ and a Haar random unitary on $\reg{ABC}$ {\em even given inverse access} except with probability $\negl(n)$.
\end{theorem}

We note that our strong gluing theorem is incomparable to the gluing lemma of \cite{schuster2024random}.  The strong gluing theorem uses a different construction, and applies to Haar random unitaries with inverse access, but does \emph{not} get the same depth savings that the gluing lemma achieves.  This is perhaps to be expected, as in the stronger query model with inverse access any two-layer construction is impossible. Hence, we end-up with the following three-layer construction:

\begin{center}
\scalebox{1}{%
\begin{quantikz}
\lstick{} & \gate[2]{U^1} & \qw & \gate[2]{U^3} & \qw \\
\lstick{} &               & \gate[2]{U^2} &  & \qw             \\
\lstick{} & \qw           &               & \qw & \qw          
\end{quantikz}%
\hspace{0.1cm},
\begin{quantikz}
\lstick{} & \gate[2]{U^{3,\dagger}} & \qw & \gate[2]{U^{1,\dagger}} & \qw \\
\lstick{} &               & \gate[2]{U^{2,\dagger}} &  & \qw             \\
\lstick{} & \qw           &               & \qw & \qw          
\end{quantikz}%
\hspace{0.5cm} $\approx$ 
\begin{quantikz}
\lstick{} & \gate[3]{{\bf O}} & \qw \\
\lstick{} &  & \qw           \\
\lstick{} &  & \qw           
\end{quantikz}%
\hspace{0.1cm},
\begin{quantikz}
\lstick{} & \gate[3]{{\bf O}^{\dagger}} & \qw \\
\lstick{} &  & \qw           \\
\lstick{} &  & \qw           
\end{quantikz}%
}
\end{center}

Combining~\Cref{thm:strong:gluing} with the construction of strong PRUs in the quantum Haar random oracle model~\cite{CRYPTO:ABGL25}, we show how to shrink keys of strong PRUs for free: given a single sample of a PRU, denoted by $U$, we can sample $O(\log^{1+\epsilon}(n))$ additional bits of randomness to get sample access to two additional instances of a strong PRU, $V$, and $W$.  Then we can join those instances to form a new strong PRU family that acts on (roughly) double the qubits.  Recursively applying this strategy to the new, larger PRU, we can stretch to any arbitrary polynomial output length, giving us the following corollary.
\begin{corollary}[Key-stretched strong PRUs]
    If there exists a family of strong PRUs in the plain model, then for every constant $c$, there exists a family of strong PRUs acting on $n$ qubits with keys of length $O(n^{1/c})$.
\end{corollary}

Interestingly, our strong gluing theorem implies that the existence of strong PRUs (in plain model) implies the existence of strong PRUs with linear depth (in plain model).  In particular, given any strong PRU family that has depth $O(n^{d})$ for some constant $d$, we can construct a strong PRU family with depth almost linear (i.e. $O(n^{1+1/c})$ for any constant $c$). 

\begin{corollary}
    If there exists a family of strong PRUs in the plain model, then for every constant $c$, there exists a family of strong PRUs acting on $n$ qubits with depth $O(n^{1 + 1/c})$.  
\end{corollary}

Beyond these results, we develop a number of mathematical tools and results useful for analyzing Haar random unitaries and modeling states using the path-recording isometries from~\cite{MH24}.

\section{Technical Overview}

We structure the overview of our proof in the following steps: 
First, we will interpret Path Recording as a purification to queries to Haar unitaries and give a way to generalise it. Then we give a construction of gluing Haar unitaries and give a purification for this similar to Path Recording. Then we study the structure of this purification of glued Haar unitaries. Finally, using insights into the structure of the purification of glued Haar unitaries, we will define an operator that maps this purification to the purification of a single larger Haar unitary. We then by a query-by-query analysis show that the glued Haar unitaries is indistinguishable from a larger Haar unitary.

\subsection{Interpreting and Generalizing Ma-Huang's Path Recording Framework.}
\label{sec:tech:PR}

Before we recall the isometries described by~\cite{MH24}, we first set up some notation. A relation $R$ is defined as a \emph{multiset} $R = \set{(x_1,y_1),\ldots,(x_t,y_t)}$ of ordered pairs $(x_i,y_i) \in [N] \times [N]$, for some $N \in \mathbb{N}$. For any relation $R = \set{(x_1,y_1),\ldots,(x_t,y_t)}$, we say that $R$ is \emph{$\cD$-distinct} if the first coordinates of all elements are distinct, and \emph{injective} or \emph{$\cI$-distinct} if the second coordinates are distinct. For a relation $R$, we use $\Dom(R)$ to denote the \emph{set} $\Dom(R) := \{x: x \in [N], \exists y \st (x,y) \in R\}$ and $\Im(R)$ to denote the \emph{set} $\Im(R) := \{y: y \in [N], \exists x \st (x,y) \in R\}$. For any relation $R = \set{(x_1,y_1),\ldots,(x_t,y_t)}$, we use $R^{-1}$ to denote the relation $R^{-1} := \{(y_1,x_1),\ldots,(y_t,x_t)\}$ obtained by swapping the coordinates of all elements in $R$. \footnote{For an $\cI$-distinct or $\cD$-distinct relation $L = \{(x_1,y_1), \dots, (x_t, y_t) \}$, the corresponding \emph{relation state} $\ket{L}$ is defined to be 
\[
\ket{L} := \frac{1}{\sqrt{t!}} \sum_{\pi \in \Symgp_t} \ket{x_{\pi^{-1}(1)}} \ket{y_{\pi^{-1}(1)}} \dots \ket{x_{\pi^{-1}(t)}} \ket{y_{\pi^{-1}(t)}}.
\] In~\cite{MH24}, relation states are defined for arbitrary relations, whereas we will not require them in this work.}

\noindent To understand the path recording framework in~\cite{MH24}, we start by thinking about what querying a Haar unitary looks like. We start by thinking about only forward queries to the Haar unitary. Since a Haar unitary is highly scrambling, a single query to a Haar unitary on any state returns a maximally mixed state. \cite{MH24} notices that given half of a maximally entangled state, it looks like a maximally mixed state. 

\noindent Even on multiple queries to a Haar unitary, it acts almost like returning maximally mixed states except that if the query is made on the same state, the output should pass the swap test and while on orthogonal states, the result should be orthogonal (pass swap test with only half probability). Hence, even while returning a maximally mixed state, we want to associate these to the input. \cite{MH24} proposes the following: On any input, create a maximally entangled pair, return one half of this pair in the query register and save the other half in the purification register labeled by the input. Formally, define the following operator: for any injective relations $R$,
\[
\pr:\ket{x}_{\reg{A}}\ket{R}_{\reg{R}} \mapsto \frac{1}{\sqrt{N - |R|}} \sum_{y \notin \Im(R)} \ket{y}_{\reg{A}} \ket{R\cup\set{(x,y)}}_{\reg{R}}.
\]
\cite{MH24} shows in essence that querying $\pr$ simulated querying a Haar unitary. Notice that this seems to follow our intuition, as we can see that $\ket{y}$ returned in the query register is almost maximally entanged with a $\ket{y}$ in the purification register. 

\noindent To extend this to both forward and inverse queries to  the Haar unitary, we can think of the following intuition: The forward and backward queries to a Haar unitary look almost like independent Haar unitaries except if an inverse query is made on the output of a forward query, we should invert the forward query. In purification sense, we can do this as follows: Instantiate almost independent purification for forward and inverse queries. When an inverse query is made, check if the input is the output of a forward query (which looks like being maximally entagled with the purification), if it is, invert the query (which looks like returning the label). Else apply the independent Haar unitary corresponding to the inverse query. To formalise this, we do the following: 

\noindent We define the following two operators (which are also partial isometries) such that for any relations $L, R$,
\par 
\[
V_L: \ket{x}_{\reg{A}}\ket{L}_{\reg{S}}\ket{R}_{\reg{T}} \mapsto \frac{1}{\sqrt{N - |\Im(L\cup R^{-1})|}} \sum_{y \notin \Im(L\cup R^{-1})} \ket{y}_{\reg{A}} \ket{L \cup \{(x,y)\}}_{\reg{S}} \ket{R}_{\reg{T}},
\]
\[
V_R: \ket{x}_{\reg{A}} \ket{L}_{\reg{S}}\ket{R}_{\reg{T}} \mapsto \frac{1}{\sqrt{N - |\Dom(L\cup R^{-1})|}} \sum_{y \notin \Dom(L\cup R^{-1})} \ket{y}_{\reg{A}} \ket{L}_{\reg{S}} \ket{R \cup \{(x,y)\}}_{\reg{T}}.
\]

\noindent We define the following projector:
\[\Pi^{L} = V_LV^{\dagger}_L\]
\[\Pi^{R} = V_RV^{\dagger}_R\]
\noindent Using $V_L$ and $V_R$, they define the following partial isometry:
\[
V = \Pi^{L}\cdot V_L \cdot (I - \Pi^{R}) + (I - \Pi^{L}) \cdot V_R^{\dagger}\cdot\Pi^{R}.
\]
They then showed that oracle access to a Haar random unitary $U$ and its inverse $U^\dagger$ can be simulated by $V$ and $V^\dagger$, respectively. In more detail, consider any oracle algorithm $\Adversary$ described by a sequence of unitaries $\left( A_1,B_1,\ldots,A_t,B_t \right)$ such that $\Adversary$ alternatively makes $t$ forward queries and $t$ inverse queries. Namely, the final state of $\Adversary$ with oracle access to (fixed) $U,U^{\dagger}$ is denoted by
\[
\ket{\Adversary_t^{U, U^{\dagger}}}_{\reg{AB}} := \prod_{i=1}^{t} \left( U^{\dagger} B_i U A_i \right) \ket{0}_{\reg{A}} \ket{0}_{\reg{B}},
\]
where $\reg{A}$ is the adversary's query register, $\reg{B}$ is the adversary's auxiliary register, and each $A_i$ and $B_i$ acts on $\reg{AB}$. They then consider the final joint state of $\Adversary$ and the purification after interacting with $V,V^{\dagger}$:
\[
\ket{\Adversary_t^{V,V^{\dagger}}}_{\reg{ABST}} 
:= \prod_{i=1}^{t} \left( V^{\dagger} B_i V A_i \right) \ket{0}_{\reg{A}}\ket{0}_{\reg{B}} \ket{\varnothing}_{\reg{S}} \ket{\varnothing}_{\reg{T}}.
\]
\cite{MH24} showed that $\rho_{{\sf Haar}}$ is $O(t^2/N^{1/8})$-close in trace distance to $\rho_{{\sf MH}}$, where 
\[
\rho_{{\sf Haar}} := \E_{U \sim \haarunitaries_n} \left[\ketbra{\Adversary_t^{U,U^{\dagger}}}{\Adversary_t^{U,U^{\dagger}}}_{\reg{AB}}\right]
\quad \text{and} \quad
\rho_{{\sf MH}} := \Tr_{\reg{ST}} \left( \ketbra{\Adversary_t^{V,V^{\dagger}}}{\Adversary_t^{V,V^{\dagger}}}_{\reg{ABST}}\right),
\]
and $\haarunitaries_n$ denotes the Haar measure over $n$-qubit unitaries and $N = 2^n$. We discuss the above in more detail in~\Cref{sec:path:def}. 

\noindent We give a generalisation of Path recording framework with the following intuition. The main idea is that in most applications of Path Recording, we don't want the maximal entanglement over all strings, and want some conditions on what strings are part of the maximal entanglement. We generally want this condition to depend on the current state, the input and some auxiliary information. We notice that as long as this condition doesn't eliminate too many strings, we find the resulting operator still simulates a Haar unitary well. We formalise how to do this in~\Cref{sec:path:general}.

\subsection{The Strong Gluing Theorem and Its Purification}
\label{sec:tech:glued}

In our main result, we show that for three Haar random unitaries, $U^1$, $U^2$, and $U^3$, applying them in a shifted brickwork pattern, overlapping on some register $\reg{B}$, yields an ensemble that is indistinguishable from a larger Haar random unitary to any adversary, with inverse access, making $\poly(|\reg{B}|)$ queries. That is, let $|\reg{A}|,|\reg{C}| = n$ and $|\reg{B}|=\secp$, then
\begin{equation*}
    \avg_{U^{1}, U^{2}, U^{3}\sim\haarunitaries_{n+\secp}} \left[\mathcal{A}^{U^{3}_{\reg{AB}} U^{2}_{\reg{BC}} U^{1}_{\reg{AB}}, (U^{3}_{\reg{AB}} U^{2}_{\reg{BC}} U^{1}_{\reg{AB}})^{\dagger}}\right] \approx \avg_{O\sim\haarunitaries_{2n+\secp}} \left[\mathcal{A}^{O_{\reg{ABC}}, O^{\dagger}_{\reg{ABC}}}\right]\,.
\end{equation*}

\noindent We start by writing the two oracles the adversary has access to, i.e. $U^{3}_{\reg{AB}} U^{2}_{\reg{BC}} U^{1}_{\reg{AB}}$ and $U^{1,\dagger}_{\reg{AB}} U^{2,\dagger}_{\reg{BC}} U^{3,\dagger}_{\reg{AB}}$. 
\\

\begin{quantikz}[column sep=1cm] 
\lstick{$\reg{A}$} & \gate[wires=2]{U^1} & \qw                  & \gate[wires=2]{U^3} & \qw \\
\lstick{$\reg{B}$} &                      & \gate[wires=2]{U^2} &                      & \qw \\
\lstick{$\reg{C}$} & \qw                  &                      & \qw                  & \qw
\end{quantikz}
\hspace{1cm} 
\begin{quantikz}[column sep=1cm] 
\lstick{$\reg{A}$} & \gate[wires=2]{U^{3,\dagger}} & \qw                  & \gate[wires=2]{U^{1,\dagger}} & \qw \\
\lstick{$\reg{B}$} &                      & \gate[wires=2]{U^{2,\dagger}} &                      & \qw \\
\lstick{$\reg{C}$} & \qw                  &                      & \qw                  & \qw
\end{quantikz}

\noindent The associated trivial purification using Path recording looks like the following:

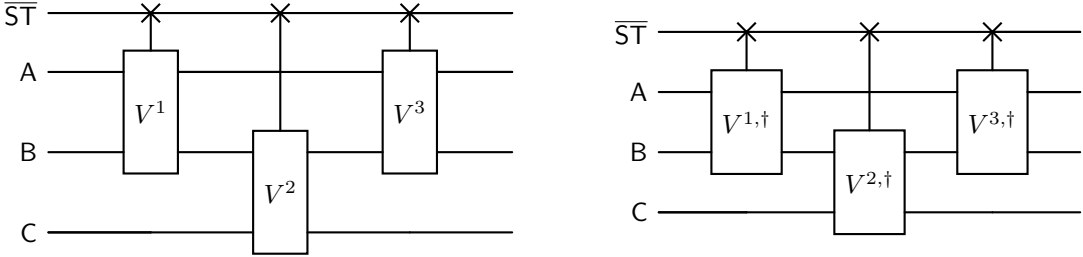
\begin{quantikz}[transparent, column sep=1cm]
\lstick{$\reg{\overline{ST}}$} & \swap{1} & \swap{2} & \swap{1} & \qw\\ 
\lstick{$\reg{A}$} & \gate[wires=2]{V^{1}} & \qw & \gate[wires=2]{V^{3}} & \qw\\
\lstick{$\reg{B}$} & & \gate[wires=2]{V^{2}} & & \\
\lstick{$\reg{C}$} & \qw & & & \\
\end{quantikz}
\hspace{1cm} 
\begin{quantikz}[transparent, column sep=0.7cm,row sep={0.8cm,between origins}]
\lstick{$\reg{\overline{ST}}$} & \swap{1} & \swap{2} & \swap{1} & \qw\\ 
\lstick{$\reg{A}$} & \gate[wires=2]{V^{1,\dagger}} & \qw & \gate[wires=2]{V^{3,\dagger}} & \qw\\
\lstick{$\reg{B}$} & & \gate[wires=2]{V^{2,\dagger}} & & \\
\lstick{$\reg{C}$} & \qw & & & \\
\end{quantikz}

\noindent Where $\reg{\overline{ST}}$ denotes the concatenation of the databases associated with the three Path Recording Framework (i.e. $\reg{\overline{ST}} = \reg{S_1T_1S_2T_2S_3T_3}$). 

\noindent Notice that the output of $V^{1}$ in the above is partially fed into $V^{2}$. Hence, if $V^{1}$ outputs one half of a maximally entangled state while saving the other half in $L_1$, then when this is fed into $V^2$, $V^2$ checks if the query is maximally entangled with anything in $R_2$, then because of monogamy of entanglement, since the query register is maximally entangled with something in $L_1$, it cannot be maximally entangled with something in $R_2$. Formally, this means $$\|V_R^{2,\dagger}V_L^{1}\|_{\op} = \negl(\secp).$$

\noindent Similarly, $$\|V_R^{3,\dagger}V_L^{2}\|_{\op} = \negl(\secp).$$ Analyzing these, we get that the construction in effect does one of four operations. To see what these are, we start by stating the purified isometry. 

\noindent We start by defining some projectors that correspond to checking entanglements (similar to $\Pi^{R}$ in case of Path Recording):
\begin{align*}
    \Pi^{R,1} &= V_R^{1} V_R^{1,\dagger}\\
    \Pi^{R,12} &= V_R^{1} V_R^{2} V_R^{2,\dagger} V_R^{1,\dagger}\\
    \Pi^{R,123} &= V_R^{1} V_R^{2} V_R^{3} V_R^{3,\dagger} V_R^{2,\dagger} V_R^{1,\dagger}
\end{align*}

\noindent Similarly, we define similar projectors in the opposite direction (similar to $\Pi^{L}$ in case of Path Recording):
\begin{align*}
    \Pi^{L,3} &= V_L^{3} V_L^{3,\dagger}\\
    \Pi^{L,32} &= V_L^{3} V_L^{2} V_L^{2,\dagger} V_L^{3,\dagger}\\
    \Pi^{L,321} &= V_L^{3} V_L^{2} V_L^{1} V_L^{1,\dagger} V_L^{2,\dagger} V_L^{3,\dagger}
\end{align*} 

\noindent Then with these in mind, we define the glued purification as follows: 
\begin{align*}
    V^{\glued} = &\left(\Pi^{L,321}\right)\cdot V_L^{3}\cdot V_L^{2}\cdot V_L^{1} \cdot\left(I-\Pi^{R,1}\right) \\
    &+ \left(\Pi^{L,32} -\Pi^{L,321}\right)\cdot V_L^{3}\cdot V_L^{2}\cdot V_R^{1,\dagger}\cdot\left(\Pi^{R,1}-\Pi^{R,12}\right) \\
    &+ \left(\Pi^{L,3} -\Pi^{L,32}\right)\cdot V_L^{3}\cdot V_R^{2,\dagger}\cdot V_R^{1,\dagger}\cdot\left(\Pi^{R,12}-\Pi^{R,123}\right) \\
    &+ \left(I-\Pi^{L,3}\right)\cdot V_R^{3,\dagger}\cdot V_R^{2,\dagger}\cdot V_R^{1,\dagger}\cdot\left(\Pi^{R,123}\right) \\
\end{align*}

\noindent Operationally, the $V^{\glued}$ works as follows:
\begin{itemize}
    \item Check if the query register is in the output of $V_R^{1}\cdot V_R^{2}\cdot V_R^{3}$, if it is, invert these queries.
    \item Else, check if the query register is in the output of $V_R^{1}\cdot V_R^{2}$, if it is, invert these queries and apply $V_L^{3}$.
    \item Else, check if the query register is in the output of $V_R^{1}$, if it is, invert this query and apply $V_L^{3}\cdot V_L^{2}$.
    \item Else, apply $V_L^{3}\cdot V_L^{2}\cdot V_L^{1}$.
\end{itemize}

\noindent We show that querying the glued Haar unitary can be purified to querying $V^{\glued}$ and $V^{\glued,\dagger}$. Notice that $V^{\glued}$ looks very similar to path recording, in the way that it checks how the query register is entangled to the purification, and depending on how it is entangled, does some operation (which is deleting some entangled pairs or adds some entangled pairs). We discuss this purification in more detail and provide proofs in~\Cref{sec:glued:PR}.

\subsection{Analyzing Purification of Querying $V^{\glued}$}
\label{sec:tech:struc}

Next, we try to study the structure of purification of querying $V^{\glued}$. We know that $V^{\glued}$ and $V^{\glued,\dagger}$ have different actions depending on how the query register is entangled with the purification. Notice that $V^{\glued,\dagger}$ just acts as adding three entangled pairs unless the query the first part of the query is entangled with the $L_3$ database. If it is entanged, we have a different action. To see how these actions behave, we look at the following example: Let we just look at what happens if we query $V^{\glued}$ followed by $V^{\glued,\dagger}$. 

{\bf{Example 1:}}
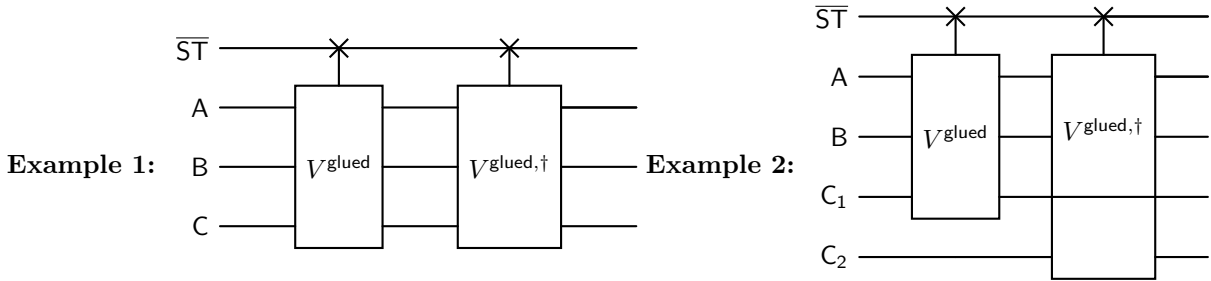
\begin{quantikz}[transparent, column sep=1cm]
\lstick{$\reg{\overline{ST}}$} & \swap{1} & \swap{1} & \qw\\ 
\lstick{$\reg{A}$} & \gate[wires=3]{V^{\glued}} & \gate[wires=3]{V^{\glued,\dagger}} & \qw\\
\lstick{$\reg{B}$} & & & \\
\lstick{$\reg{C}$} & & & \\
\end{quantikz}
{\bf{Example 2:}}
\begin{quantikz}[transparent, column sep=0.7cm,row sep={0.8cm,between origins}]
\lstick{$\reg{\overline{ST}}$} & \swap{1} & \swap{1} & \qw\\ 
\lstick{$\reg{A}$} & \gate[wires=3]{V^{\glued}} & \gate[wires=4,label
style={yshift=0.5cm}]{V^{\glued,\dagger}} & \qw\\
\lstick{$\reg{B}$} & & & \\
\lstick{$\reg{C_1}$} & & \linethrough & \\
\lstick{$\reg{C_2}$} & & & \\
\end{quantikz}

\noindent Looking at the operational definition of $V^{\glued}$ and $V^{\glued,\dagger}$, we can simplify the above as:

{\bf{Example 1:}}
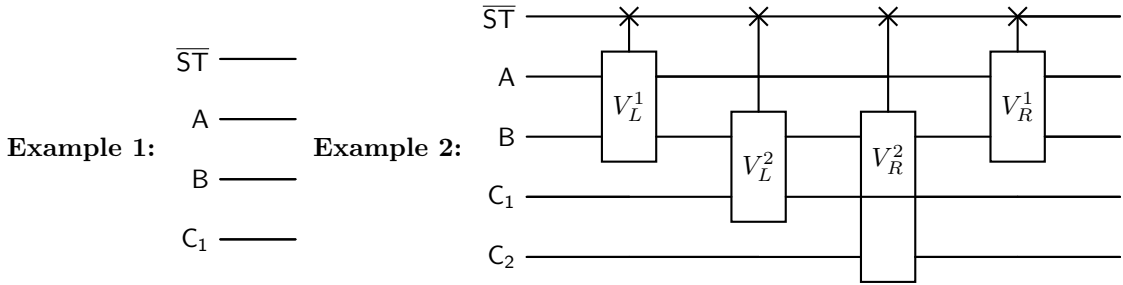
\begin{quantikz}[transparent, column sep=1cm,row sep={0.8cm,between origins}]
\lstick{$\reg{\overline{ST}}$} &
  \qw \\
\lstick{$\reg{A}$} &
  \qw \\
\lstick{$\reg{B}$} & \qw \\
\lstick{$\reg{C_1}$} & \qw 
\end{quantikz}
{\bf{Example 2:}}
\begin{quantikz}[transparent, column sep=1cm,row sep={0.8cm,between origins}]
\lstick{$\reg{\overline{ST}}$} &
  \swap{1} &
  \swap{2} & \swap{2} &
  \swap{1} & \qw \\
\lstick{$\reg{A}$} &
  \gate[wires=2]{V_{L}^{1}} &
  \qw & \qw &
  \gate[wires=2]{V_{R}^{1}} & \qw \\
\lstick{$\reg{B}$} & & \gate[wires=2]{V_{L}^{2}} & \gate[wires=3,label
style={yshift=0.5cm}]{V_{R}^{2}} & & \qw \\
\lstick{$\reg{C_1}$} & \qw & & \linethrough & \qw & \qw \\
\lstick{$\reg{C_2}$}& \qw & \qw & & \qw & \qw
\end{quantikz}

\noindent Notice next, if we query $V^{\glued}$ on Example 2, unless the $V^{\glued}$ matches up on $\reg{AB}$, $V^{\glued}$ just adds 3 new entangled pairs. The more interesting case is when $V^{\glued}$ matches up on $\reg{AB}$. This can again be seen as two cases, either the $C$ register also matches up or it doesn't. Let's analyse these examples:

{\bf{Example 2a:}} 
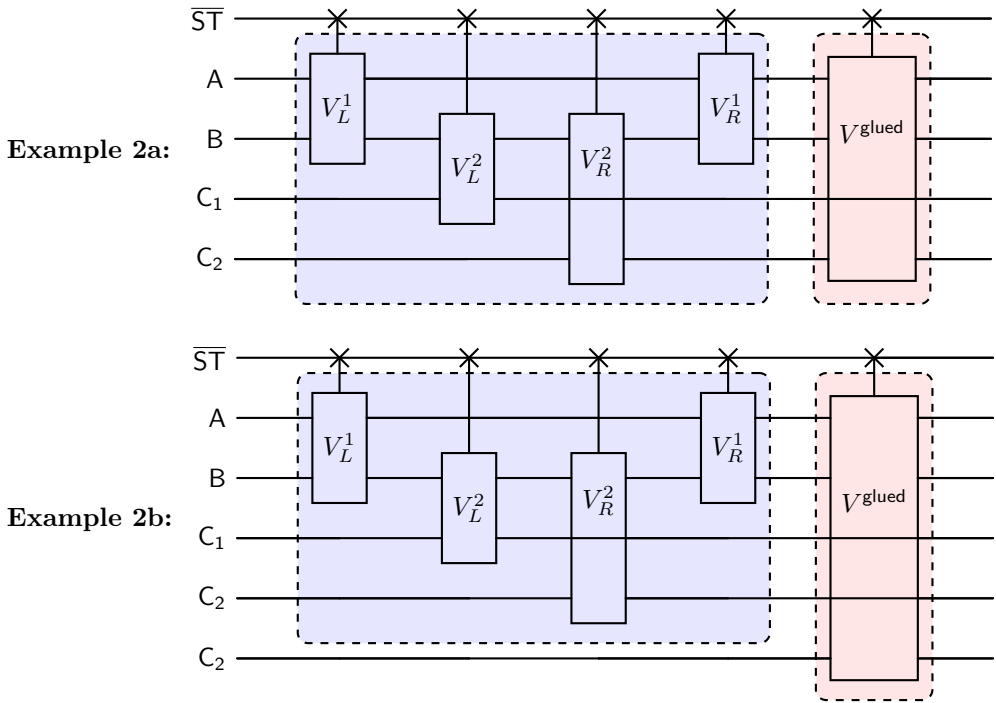
\begin{quantikz}[transparent, column sep=1cm,row sep={0.8cm,between origins}]
\lstick{$\reg{\overline{ST}}$} &
  \swap{1} &
  \swap{2} & \swap{2} &
  \swap{1} & \swap{1} & \qw\\
\lstick{$\reg{A}$} &
  \gate[wires=2]{V_L^1}\gategroup[wires=4,steps=4,
    style={dashed,rounded corners,inner xsep=2pt,fill=blue!10},
    background,
    label style={label position=below,yshift=-0.4cm}]{} &
  \qw & \qw &
  \gate[wires=2]{V_{R}^{1}} & \gate[wires=4,label
style={yshift=0.5cm}]{V^{\glued}}\gategroup[wires=4,steps=1,
    style={dashed,rounded corners,inner xsep=2pt,fill=red!10},
    background,
    label style={label position=below,yshift=-0.4cm}]{} & \qw\\
\lstick{$\reg{B}$} & & \gate[wires=2]{V_L^2} & \gate[wires=3,label
style={yshift=0.5cm}]{V_R^{2}} & & & \qw \\
\lstick{$\reg{C_1}$} & \qw & & \linethrough & \qw & \linethrough & \qw \\
\lstick{$\reg{C_2}$}& \qw & \qw & & \qw & \qw & \qw 
\end{quantikz}

{\bf{Example 2b:}} \begin{quantikz}[transparent, column sep=1cm,row sep={0.8cm,between origins}]
\lstick{$\reg{\overline{ST}}$} &
  \swap{1} &
  \swap{2} & \swap{2} &
  \swap{1} & \swap{1} & \qw\\
\lstick{$\reg{A}$} &
  \gate[wires=2]{V_L^1}\gategroup[wires=4,steps=4,
    style={dashed,rounded corners,inner xsep=2pt,fill=blue!10},
    background,
    label style={label position=below,yshift=-0.4cm}]{} &
  \qw & \qw &
  \gate[wires=2]{V_{R}^{1}} & \gate[wires=5,label
style={yshift=0.5cm}]{V^{\glued}}\gategroup[wires=5,steps=1,
    style={dashed,rounded corners,inner xsep=2pt,fill=red!10},
    background,
    label style={label position=below,yshift=-0.4cm}]{} & \qw\\
\lstick{$\reg{B}$} & & \gate[wires=2]{V_L^2} & \gate[wires=3,label
style={yshift=0.5cm}]{V_R^{2}} & & & \qw \\
\lstick{$\reg{C_1}$} & \qw & & \linethrough & \qw & \linethrough & \qw \\
\lstick{$\reg{C_2}$}& \qw & \qw & & \qw & \linethrough & \qw \\
\lstick{$\reg{C_2}$}& \qw & \qw & & \qw & \qw & \qw 
\end{quantikz}

\noindent Again, expanding by definition of $V^{\glued}$, we simplify as follows:

{\bf{Example 2a:}} 
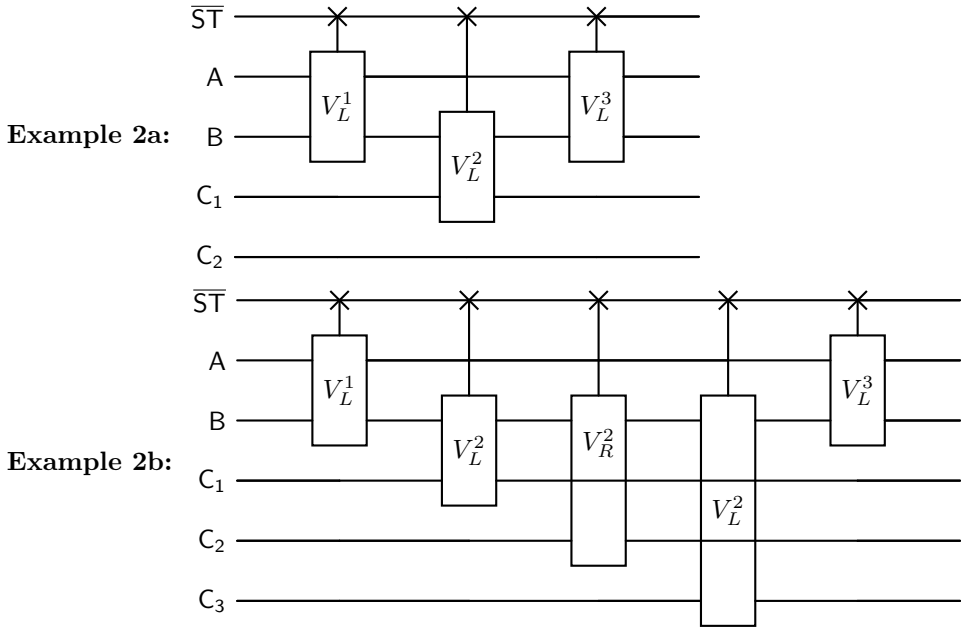
\begin{quantikz}[transparent, column sep=1cm,row sep={0.8cm,between origins}]
\lstick{$\reg{\overline{ST}}$} &
  \swap{1} &
  \swap{2} & \swap{1} & \qw\\
\lstick{$\reg{A}$} &
  \gate[wires=2]{V_L^1} &
  \qw & \gate[wires=2]{V_L^{3}} & \qw\\
\lstick{$\reg{B}$} & & \gate[wires=2]{V_L^2} & & \qw \\
\lstick{$\reg{C_1}$} & \qw & & \qw & \qw \\
\lstick{$\reg{C_2}$}& \qw & \qw & \qw & \qw 
\end{quantikz}

{\bf{Example 2b:}} \begin{quantikz}[transparent, column sep=1cm,row sep={0.8cm,between origins}]
\lstick{$\reg{\overline{ST}}$} &
  \swap{1} &
  \swap{2} & \swap{2} & \swap{2} & \swap{1} & \qw\\
\lstick{$\reg{A}$} &
  \gate[wires=2]{V_L^1} &
  \qw & \qw & \qw & \gate[wires=2]{V_L^{3}} & \qw\\
\lstick{$\reg{B}$} & & \gate[wires=2]{V_L^2} & \gate[wires=3,label
style={yshift=0.5cm}]{V_R^{2}} & \gate[wires=4]{V_L^{2}} & & \qw \\
\lstick{$\reg{C_1}$} & \qw & & \linethrough & \linethrough &  & \qw \\
\lstick{$\reg{C_2}$}& \qw & \qw & & \linethrough & & \qw  \\
\lstick{$\reg{C_3}$}& \qw & \qw & & \qw &  & \qw 
\end{quantikz}

\noindent Notice that Example 2a is just equivalent to a single query to the first oracle. Exmple 2b has multiple unitaries chained together. Notice in all of the above two properties maintained. First, the register $\reg{A}$ which has only two gates applied to it (the first, e.g. $V_L^{1}$ and the last, e.g. $V_L^{3}$). Second, the register $\reg{B}$ has all gates applied to it. Whenever a wire is the output of $V^{i}_L$ or $V^{i}_R$, it is a maximally entangled state, half of which is stored in the purification. Whenever a wire is the input of $V^{i}_L$ or $V^{i}_R$, it is saved in the purification. Hence, all wire on register $\reg{B}$ creates a maximally entagled pair between two databases in the purification register. Similarly, the wire on $\reg{A}$ creates a maximally entagled pair between two databases in the purification register.

\noindent Notice that we could extend the above example further to get multiple chained isometries. Pictorially, this looks as follows:

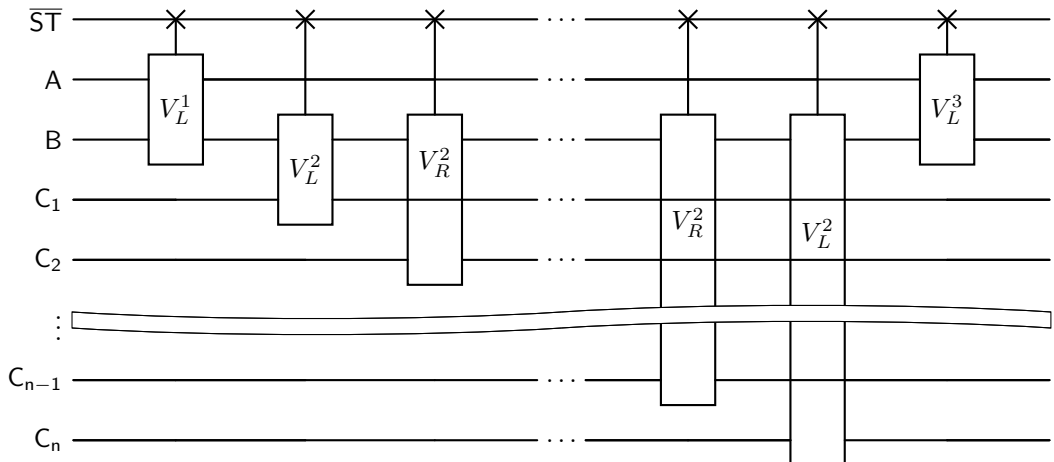
\begin{quantikz}[transparent, column sep=1cm,row sep={0.8cm,between origins}]
\lstick{$\reg{\overline{ST}}$} & \swap{1} & \swap{2} & \swap{2} & \ \ldots \ & \swap{2} & \swap{2} & \swap{1} &\\ 
\lstick{$\reg{A}$} &
  \gate[wires=2]{V_L^{1}} &
  \qw & \qw &
\ \ldots\ & \qw & \qw & \gate[wires=2]{V_L^{3}} & \qw\\
\lstick{$\reg{B}$} & & \gate[wires=2]{V_L^{2}} & \gate[wires=3,label
style={yshift=0.5cm}]{V_R^{2}} &\ \ldots\ & \gate[wires=5,label
style={yshift=0.5cm}]{V_R^{2}} & \gate[wires=6,label
style={yshift=0.75cm}]{V_L^{2}} & & \qw \\
\lstick{$\reg{C_1}$} & \qw & & \linethrough &\ \ldots\ & \linethrough & \linethrough &  & \qw \\
\lstick{$\reg{C_2}$}& \qw & \qw & &\ \ldots\ & \linethrough & \linethrough & & \qw  \\
\lstick{$\vdots$} & \wave[draw=black, fill=white, opacity=1] &&&&&&&\\ 
\lstick{$\reg{C_{n-1}}$}& \qw & \qw & &\ \ldots\ & \qw & \linethrough  & \qw & \qw \\ 
\lstick{$\reg{C_n}$}& \qw & \qw & &\ \ldots\ & \qw & \qw &  & \qw 
\end{quantikz}

\noindent Thinking of the above as a "chain" of unitaries. Then we want to imagine any adversary's circuit as some "chains" strung together. We give an example below:

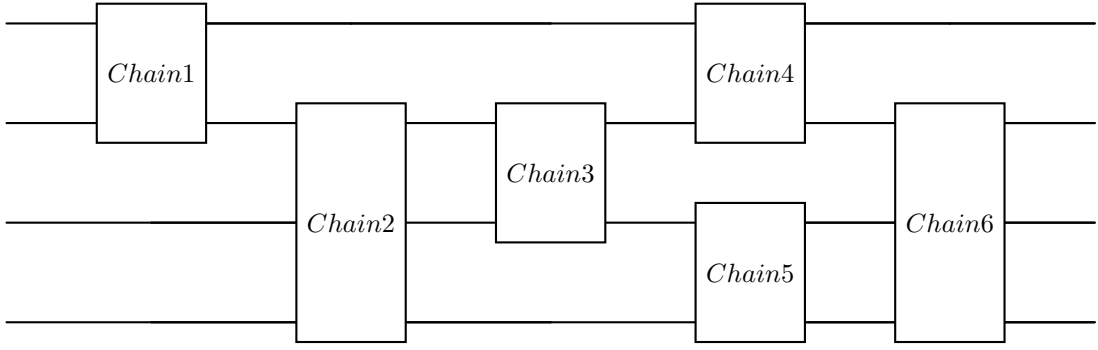
\begin{quantikz}[row sep=0.8cm, column sep=1.2cm]
& \gate[wires=2]{Chain1} & \qw & \qw & \gate[wires=2]{Chain4} & \qw & \qw \\
&  & \gate[wires=3]{Chain2} & \gate[wires=2]{Chain3} & \qw &  \gate[wires=3]{Chain6} & \qw  \\
&  & \qw & & \gate[wires=2]{Chain5} & \qw & \qw \\
&  & \qw & \qw  & & \qw & \qw
\end{quantikz}

\noindent We want to formalise the intuition above, any adversary querying the oracles can be broken as multiple chains. To formalise the above intuition, start labeling the wires as follows:

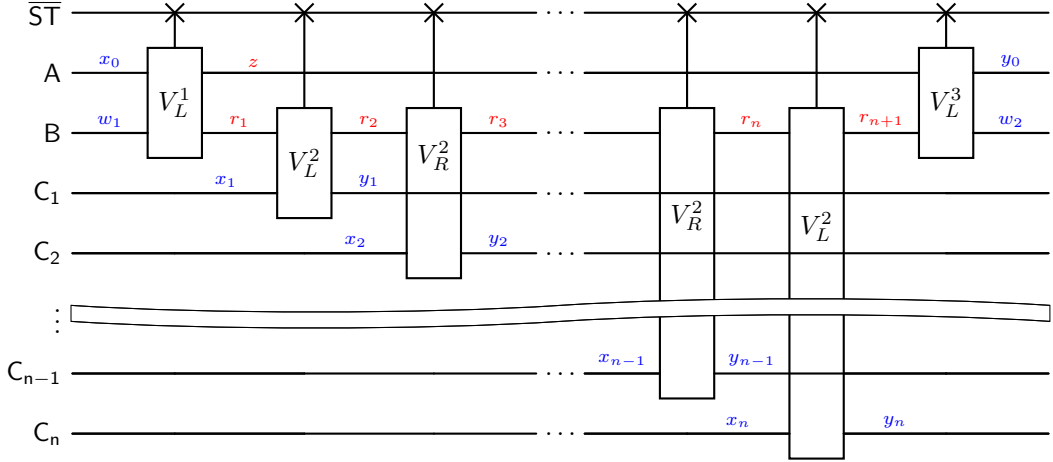
\begin{quantikz}[transparent, column sep=1cm,row sep={0.8cm,between origins}]
\lstick{$\reg{\overline{ST}}$} & \swap{1} & \swap{2} & \swap{2} & \ \ldots \ & \swap{2} & \swap{2} & \swap{1} &\\ 
\lstick{$\reg{A}$} \arrow[r,draw=none,"\textcolor{blue}{x_0}"]&
  \gate[wires=2]{V_L^{1}}\arrow[r,draw=none,"\textcolor{red}{z}"] &
  \qw & \qw &
\ \ldots\ & \qw & \qw & \gate[wires=2]{V_L^{3}}\arrow[r,draw=none,"\textcolor{blue}{y_0}"] & \qw\\
\lstick{$\reg{B}$}\arrow[r,draw=none,"\textcolor{blue}{w_1}"] & \qw\arrow[r,draw=none,"\textcolor{red}{r_1}"] & \gate[wires=2]{V_L^{2}}\arrow[r,draw=none,"\textcolor{red}{r_2}"] & \gate[wires=3,label
style={yshift=0.5cm}]{V_R^{2}}\arrow[r,draw=none,"\textcolor{red}{r_3}"] &\ \ldots\ & \gate[wires=5,label
style={yshift=0.5cm}]{V_R^{2}}\arrow[r,draw=none,"\textcolor{red}{r_{n}}"] & \gate[wires=6,label
style={yshift=0.75cm}]{V_L^{2}}\arrow[r,draw=none,"\textcolor{red}{r_{n+1}}"] & \qw\arrow[r,draw=none,"\textcolor{blue}{w_2}"] & \qw \\
\lstick{$\reg{C_1}$} & \qw\arrow[r,draw=none,"\textcolor{blue}{x_1}"]  & \qw\arrow[r,draw=none,"\textcolor{blue}{y_1}"] & \linethrough &\ \ldots\ & \linethrough & \linethrough &  & \qw \\
\lstick{$\reg{C_2}$}& \qw & \qw\arrow[r,draw=none,"\textcolor{blue}{x_2}"] & \qw\arrow[r,draw=none,"\textcolor{blue}{y_2}"] &\ \ldots\ & \linethrough & \linethrough & & \qw  \\
\lstick{$\vdots$} & \wave[draw=black, fill=white, opacity=1] &&&&&&&\\ 
\lstick{$\reg{C_{n-1}}$}& \qw & \qw & &\ \ldots\ \arrow[r,draw=none,"\textcolor{blue}{x_{n-1}}"]& \qw\arrow[r,draw=none,"\textcolor{blue}{y_{n-1}}"] & \linethrough  & \qw & \qw \\ 
\lstick{$\reg{C_n}$}& \qw & \qw & &\ \ldots\ & \qw\arrow[r,draw=none,"\textcolor{blue}{x_n}"] & \qw\arrow[r,draw=none,"\textcolor{blue}{y_n}"] & \qw & \qw 
\end{quantikz}

\noindent Recalling the properties from before, we have the all isometries are applied to $\reg{B}$, only the first and last isometry are applied to $\reg{A}$ and the labels in $\rcolor{red}$ are the maximally entangled pairs that only exist in the purification. Then on the above labels, the database register looks as follows: 
\begin{align*}
    &\ket{\set{(\bcolor{x_0}||\bcolor{w_1},\rcolor{z}||\rcolor{r_1})}}_{\reg{S}_1}\ket{\set{}}_{\reg{T}_1}\\
    &\otimes\ket{\set{(\rcolor{r_1}||\bcolor{x_1},\rcolor{r_2}||\bcolor{y_1}),\ldots,(\rcolor{r_{n}}||\bcolor{x_n},\rcolor{r_{n+1}}||\bcolor{y_n})}}_{\reg{S}_2}\\
    &\otimes\ket{\set{(\rcolor{r_2}||\bcolor{x_2},\rcolor{r_3}||\bcolor{y_2}),\ldots,(\rcolor{r_{n-1}}||\bcolor{x_{n-1}},\rcolor{r_{n}}||\bcolor{y_{n-1}})}}_{\reg{T}_2}\\
    &\otimes\ket{\set{(\rcolor{z}||\rcolor{r_{n+1}},\bcolor{y_0}||\bcolor{w_2})}}_{\reg{S}_3}\ket{\set{}}_{\reg{T}_3}
\end{align*}

\noindent A better way to think about this database is modeling it as a graph. To do this, we do the following:
\begin{itemize}
    \item {\textbf{Defining Vertices}:} For each tuple in the database, we add a vertex in the graph labelled by the touple. 
    \item {\textbf{Adding Edges from $L_1$ to $L_2$}:} For any vertices $v_1$ coming from $L_1$, say the label of this vertex is $(\bcolor{x}||\bcolor{w},\rcolor{z}||\rcolor{r})$, and any vertex $v_2$ coming from $L_2$, say the label of this vertex is $(\rcolor{r'}||\bcolor{x'},\rcolor{\tilde{r}'}||\bcolor{y'})$. We add an edge from $v_1$ to $v_2$  if the vertices are "corelated", i.e. $\rcolor{r} = \rcolor{r'}$.
    \item {\textbf{Adding Edges from $L_2$ to $R_2$}:} For any vertices $v_1$ coming from $L_2$, say the label of this vertex is $(\rcolor{r}||\bcolor{x},\rcolor{\tilde{r}}||\bcolor{y})$, and any vertex $v_2$ coming from $R_2$, say the label of this vertex is $(\rcolor{r'}||\bcolor{x'},\rcolor{\tilde{r}'}||\bcolor{y'})$. We add an edge from $v_1$ to $v_2$  if the vertices are "corelated", i.e. $\rcolor{\tilde{r}} = \rcolor{r'}$.
    \item {\textbf{Adding Edges from $R_2$ to $L_2$}:} For any vertices $v_1$ coming from $R_2$, say the label of this vertex is $(\rcolor{r}||\bcolor{x},\rcolor{\tilde{r}}||\bcolor{y})$, and any vertex $v_2$ coming from $L_2$, say the label of this vertex is $(\rcolor{r'}||\bcolor{x'},\rcolor{\tilde{r}'}||\bcolor{y'})$. We add an edge from $v_1$ to $v_2$  if the vertices are "corelated", i.e. $\rcolor{\tilde{r}} = \rcolor{r'}$. 
    \item {\textbf{Adding Edges from $L_2$ to $L_3$}:} For any vertices $v_1$ coming from $L_2$, say the label of this vertex is $(\rcolor{r}||\bcolor{x},\rcolor{\tilde{r}}||\bcolor{y})$, and any vertex $v_2$ coming from $L_3$, say the label of this vertex is $(\rcolor{z}||\rcolor{r'},\bcolor{y'}||\bcolor{w})$. We add an edge from $v_1$ to $v_2$  if the vertices are "corelated", i.e. $\rcolor{\tilde{r}} = \rcolor{r'}$. 
    \item {\textbf{Adding Edges from $R_3$ to $R_2$}:} For any vertices $v_1$ coming from $R_3$, say the label of this vertex is $(\bcolor{x}||\bcolor{w},\rcolor{z}||\rcolor{r})$, and any vertex $v_2$ coming from $R_2$, say the label of this vertex is $(\rcolor{r'}||\bcolor{x'},\rcolor{\tilde{r}'}||\bcolor{y'})$. We add an edge from $v_1$ to $v_2$  if the vertices are "corelated", i.e. $\rcolor{r} = \rcolor{r'}$. (These edges don't arise in the chain we look at in this example, but chains starting from $U^{\dagger}_3$ instead of $U_1$).
    \item {\textbf{Adding Edges from $R_2$ to $R_1$}:} For any vertices $v_1$ coming from $R_2$, say the label of this vertex is $(\rcolor{r}||\bcolor{x},\rcolor{\tilde{r}}||\bcolor{y})$, and any vertex $v_2$ coming from $R_1$, say the label of this vertex is $(\rcolor{z}||\rcolor{r'},\bcolor{y'}||\bcolor{w})$. We add an edge from $v_1$ to $v_2$  if the vertices are "corelated", i.e. $\rcolor{\tilde{r}} = \rcolor{r'}$. (These edges don't arise in the chain we look at in this example, but chains ending from $U^{\dagger}_1$ instead of $U_3$).
\end{itemize}

\noindent Drawing edge structure, we get edges of the following form:

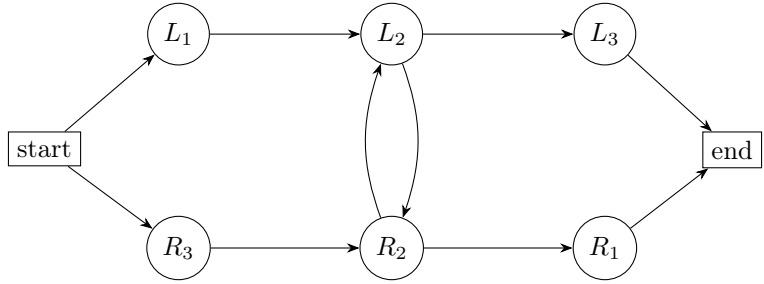
\begin{center}
    
\begin{tikzpicture}[>=Stealth, node distance=2cm, every node/.style={draw,circle}]
  \node (L1) {$L_1$};
  \node (L2) [right=of L1] {$L_2$};
  \node (L3) [right=of L2] {$L_3$};
  
  \node (R1) [below=of L3] {$R_1$};
  \node (R2) [below=of L2] {$R_2$};
  \node (R3) [below=of L1] {$R_3$};

  \node[draw, rectangle, below left=1cm and 1cm of L1] (Start) {start};
  \node[draw, rectangle, below right=1cm and 1cm of L3] (End) {end};

  \draw[->] (L1) -- (L2);
  \draw[->] (L2) -- (L3);
  \draw[->, bend left=20] (L2) to (R2);   
  \draw[->, bend left=20] (R2) to (L2);   
  \draw[->] (R3) -- (R2);
  \draw[->] (R2) -- (R1);
  \draw[->] (Start) -- (L1);
  \draw[->] (Start) -- (R3);
  \draw[->] (R1) -- (End);
  \draw[->] (L3) -- (End);
\end{tikzpicture}
\end{center}

In particular, if we imagine all $\rcolor{r_i}$'s as distinct, we can see that the resulting line graph looks like:

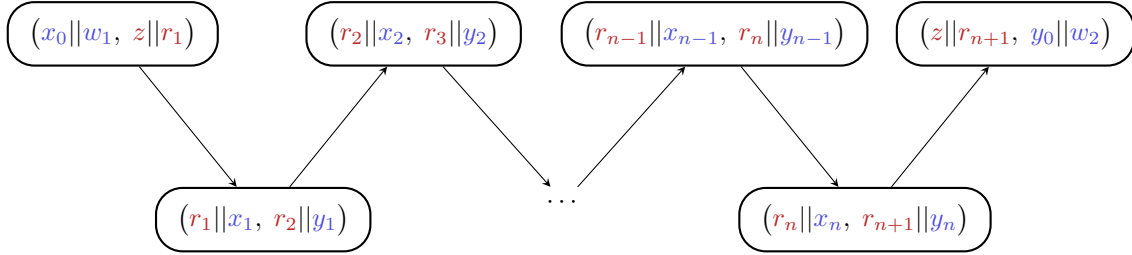
\begin{tikzpicture}[
  >=stealth,
  node distance = 1cm and 1.2cm,
  bubble/.style = {draw, rounded corners=10pt, thick,
                   inner xsep=8pt, inner ysep=6pt}
]
\node[bubble] (A) {$\bigl(\bcolor{x_{0}}|| \bcolor{w_1},\; \rcolor{z}|| \rcolor{r_{1}}\bigr)$};
\node[bubble, below=1.6cm of A, xshift=2cm] (A2) {$\bigl(\rcolor{r_{1}}|| \bcolor{x_{1}},\; \rcolor{r_{2}}|| \bcolor{y_{1}}\bigr)$};
\draw[->] (A) -- (A2);

\node[bubble, above=1.6cm of A2, xshift=2cm] (B)
  {$\bigl(\rcolor{r_{2}}|| \bcolor{x_{2}},\; \rcolor{r_{3}}|| \bcolor{y_{2}}\bigr)$};
\draw[->] (A2) -- (B);

\node (dots) [below=1.6cm of B, xshift=2cm] {$\cdots$};
\draw[->] (B) -- (dots);

\node[bubble, above=1.6cm of dots, xshift=2cm] (D)
  {$\bigl(\rcolor{r_{n-1}}|| \bcolor{x_{n-1}},\; \rcolor{r_{n}}|| \bcolor{y_{n-1}}\bigr)$};
\draw[->] (dots) -- (D);

\node[bubble, below=1.6cm of D, xshift=2cm] (C)
  {$\bigl(\rcolor{r_{n}}|| \bcolor{x_{n}},\; \rcolor{r_{n+1}}|| \bcolor{y_{n}}\bigr)$};
\draw[->] (D) -- (C);

\node[bubble, above=1.6cm of C, xshift=2cm] (C2)
  {$\bigl(\rcolor{z}|| \rcolor{r_{n+1}},\; \bcolor{y_{0}}|| \bcolor{w_{2}}\bigr)$};
\draw[->] (C) -- (C2);
\end{tikzpicture}


\noindent If we imagine any adversary's circuit as some "chains" strung together (recall the example from before), then corresponding to each chain, we get a disjoint line graph, and the database register corresponds to the union of databases corresponding to these disjoint line graphs. Also notice that all the labels in $\rcolor{red}$ are maximally entangled pairs which exist fully in the purification register. Hence, given the $\bcolor{blue}$ labels, we can completely identify the purification states. We refer the reader to~\Cref{sec:struc} to see how we formalise the above.

\subsection{Simulating the Larger Haar Unitary}
\label{sec:tech:sim}

Now that we know that the adversary's query structure can be broken into disjoint chains. To see how to simulate the larger Haar unitary, we will first see what a single disjoint chain looks like and then see what a corresponding database register looks like. To start, we again consider a chain as example:

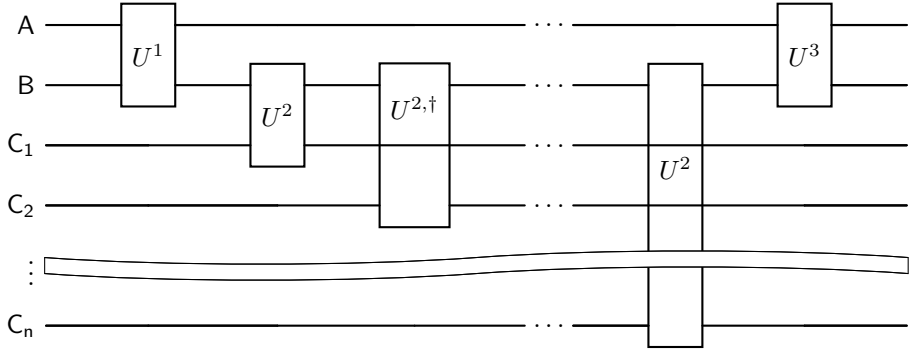
\begin{quantikz}[transparent, column sep=1cm,row sep={0.8cm,between origins}]
\lstick{$\reg{A}$} &
  \gate[wires=2]{U^1} &
  \qw & \qw &
\ \ldots\ & \qw & \gate[wires=2]{U^{3}} & \qw\\
\lstick{$\reg{B}$} & & \gate[wires=2]{U^2} & \gate[wires=3,label
style={yshift=0.5cm}]{U^{2,\dagger}} &\ \ldots\  & \gate[wires=5,label
style={yshift=0.5cm}]{U^{2}} & & \qw \\
\lstick{$\reg{C_1}$} & \qw & & \linethrough &\ \ldots\ & \linethrough &  & \qw \\
\lstick{$\reg{C_2}$}& \qw & \qw & &\ \ldots\ & \linethrough & & \qw  \\
\lstick{$\vdots$} & \wave[draw=black, fill=white, opacity=1] &&&&&&\\ 
\lstick{$\reg{C_n}$}& \qw & \qw & &\ \ldots\ & \qw &  & \qw 
\end{quantikz}

\noindent We know that in the Ideal experiment, we replace $U^{3}_{\reg{AB}} U^{2}_{\reg{BC}} U^{1}_{\reg{AB}}$ and $U^{1,\dagger}_{\reg{AB}} U^{2,\dagger}_{\reg{BC}} U^{3,\dagger}_{\reg{AB}}$ with $O_{\reg{ABC}}$ and $O^{\dagger}_{\reg{ABC}}$, respectively. To do this, we insert dummy unitaries in the abvoe chain. Particularly, we insert a $U^3U^{3,\dagger}$ between $U^2$ and $U^{2,\dagger}$ and we insert a $U^{1,\dagger}U^{1}$ between $U^{2,\dagger}$ and $U^{2}$. Then the above chain looks like: 

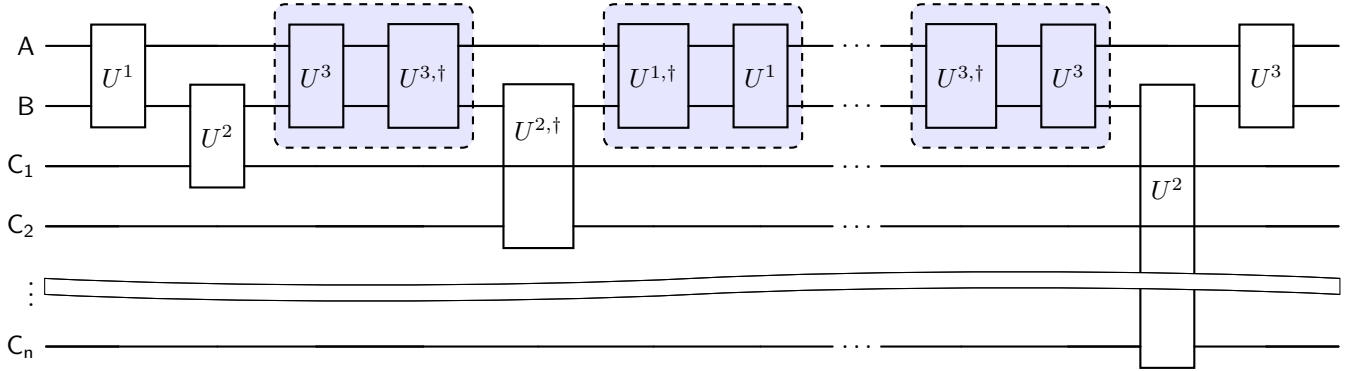
\begin{quantikz}[transparent, column sep=0.6cm,row sep={0.8cm,between origins}]
\lstick{$\reg{A}$} &
  \gate[wires=2]{U^1} &
  \qw & \gate[wires=2]{U^3} 
  \gategroup[wires=2,steps=2,
    style={dashed,rounded corners,inner xsep=2pt,fill=blue!10},
    background]{} & \gate[wires=2]{U^{3,\dagger}} & \qw & \gate[wires=2]{U^{1,\dagger}} \gategroup[wires=2,steps=2,
    style={dashed,rounded corners,inner xsep=2pt,fill=blue!10},
    background]{} & \gate[wires=2]{U^{1}} & 
\ \ldots\ & \gate[wires=2]{U^{3,\dagger}} \gategroup[wires=2,steps=2,
    style={dashed,rounded corners,inner xsep=2pt,fill=blue!10},
    background]{} & \gate[wires=2]{U^{3}} & \qw & \gate[wires=2]{U^{3}} & \qw\\
\lstick{$\reg{B}$} & & \gate[wires=2]{U^2} & & & \gate[wires=3,label
style={yshift=0.5cm}]{U^{2,\dagger}} & & &\ \ldots\ & & & \gate[wires=5,label
style={yshift=0.5cm}]{U^{2}} & & \qw \\
\lstick{$\reg{C_1}$} & \qw & & & & \linethrough & & &\ \ldots\ & & & \linethrough &  & \qw \\
\lstick{$\reg{C_2}$} & \qw & & & \qw & & & &\ \ldots\ & & & \linethrough & & \qw  \\
\lstick{$\vdots$} & \wave[draw=black, fill=white, opacity=1] &&&&&&&&&&&&\\ 
\lstick{$\reg{C_n}$} & \qw & & & \qw & & & &\ \ldots\ & & & \qw &  & \qw 
\end{quantikz}

\noindent Notice that doing this, each component of the chain can be seen as queries to $U^{3}_{\reg{AB}} U^{2}_{\reg{BC}} U^{1}_{\reg{AB}}$ and $U^{1,\dagger}_{\reg{AB}} U^{2,\dagger}_{\reg{BC}} U^{3,\dagger}_{\reg{AB}}$. In particular, the chain looks like alternating queries to $U^{3}_{\reg{AB}} U^{2}_{\reg{BC}} U^{1}_{\reg{AB}}$ and $U^{1,\dagger}_{\reg{AB}} U^{2,\dagger}_{\reg{BC}} U^{3,\dagger}_{\reg{AB}}$. Hence, our example looks as follows:

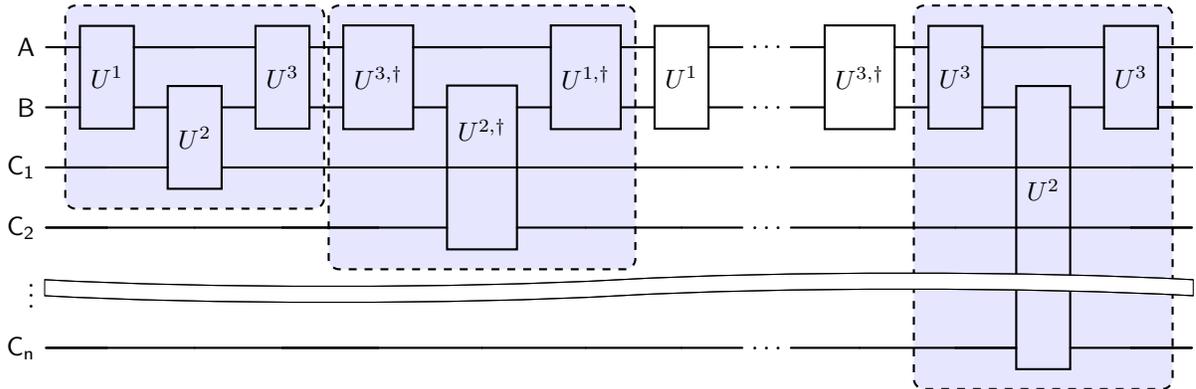
\begin{quantikz}[transparent, column sep=0.45cm,row sep={0.8cm,between origins}]
\lstick{$\reg{A}$} &
  \gate[wires=2]{U^1}\gategroup[wires=3,steps=3,
    style={dashed,rounded corners,inner xsep=2pt,fill=blue!10},
    background]{} &
  \qw & \gate[wires=2]{U^3} 
   & \gate[wires=2]{U^{3,\dagger}} \gategroup[wires=4,steps=3,
    style={dashed,rounded corners,inner xsep=2pt,fill=blue!10},
    background]{} & \qw & \gate[wires=2]{U^{1,\dagger}} & \gate[wires=2]{U^{1}} & 
\ \ldots\ & \gate[wires=2]{U^{3,\dagger}} & \gate[wires=2]{U^{3}} \gategroup[wires=6,steps=3,
    style={dashed,rounded corners,inner xsep=2pt,fill=blue!10},
    background]{} & \qw & \gate[wires=2]{U^{3}} & \qw\\
\lstick{$\reg{B}$} & & \gate[wires=2]{U^2} & & & \gate[wires=3,label
style={yshift=0.5cm}]{U^{2,\dagger}} & & &\ \ldots\ & & & \gate[wires=5,label
style={yshift=0.5cm}]{U^{2}} & & \qw \\
\lstick{$\reg{C_1}$} & \qw & & & & \linethrough & & &\ \ldots\ & & & \linethrough &  & \qw \\
\lstick{$\reg{C_2}$} & \qw & & & \qw & & & &\ \ldots\ & & & \linethrough & & \qw  \\
\lstick{$\vdots$} & \wave[draw=black, fill=white, opacity=1] &&&&&&&&&&&&\\ 
\lstick{$\reg{C_n}$} & \qw & & & \qw & & & &\ \ldots\ & & & \qw &  & \qw 
\end{quantikz}

\noindent In the Ideal experiment, we replace $U^{3}_{\reg{AB}} U^{2}_{\reg{BC}} U^{1}_{\reg{AB}}$ and $U^{1,\dagger}_{\reg{AB}} U^{2,\dagger}_{\reg{BC}} U^{3,\dagger}_{\reg{AB}}$ with $O_{\reg{ABC}}$ and $O^{\dagger}_{\reg{ABC}}$, respectively. Hence, the chain becomes alternating queries to $O_{\reg{ABC}}$ and $O^{\dagger}_{\reg{ABC}}$. Hence, our example looks as follows:

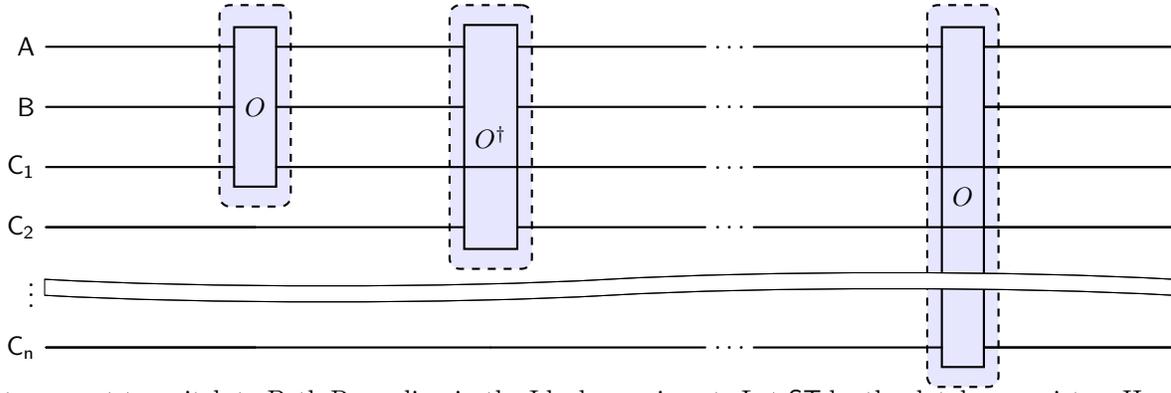
\begin{quantikz}[transparent, column sep=2.5cm,row sep={0.8cm,between origins}]
\lstick{$\reg{A}$} &
  \gate[wires=3]{O}\gategroup[wires=3,steps=1,
    style={dashed,rounded corners,inner xsep=2pt,fill=blue!10},
    background]{} 
   & \gate[wires=4]{O^{\dagger}} \gategroup[wires=4,steps=1,
    style={dashed,rounded corners,inner xsep=2pt,fill=blue!10},
    background]{} & 
\ \ldots\ & \gate[wires=6]{O} \gategroup[wires=6,steps=1,
    style={dashed,rounded corners,inner xsep=2pt,fill=blue!10},
    background]{} & \qw\\
\lstick{$\reg{B}$} & & &\ \ldots\ & & \qw \\
\lstick{$\reg{C_1}$} & & \linethrough &\ \ldots\ & \linethrough & \qw \\
\lstick{$\reg{C_2}$} & \qw & & \ \ldots\ & \linethrough & \qw  \\
\lstick{$\vdots$} & \wave[draw=black, fill=white, opacity=1] &&&&&&&\\ 
\lstick{$\reg{C_n}$} & \qw & &\ \ldots\ & & \qw 
\end{quantikz}

\noindent Next we want to switch to Path Recording in the Ideal experiment. Let $\reg{ST}$ be the database register. Hence, we switch $O_{\reg{ABC}}$ with $V_{L}$ and $O^{\dagger}_{\reg{ABC}}$ with $V_{R}$. Hence, our example looks like: 

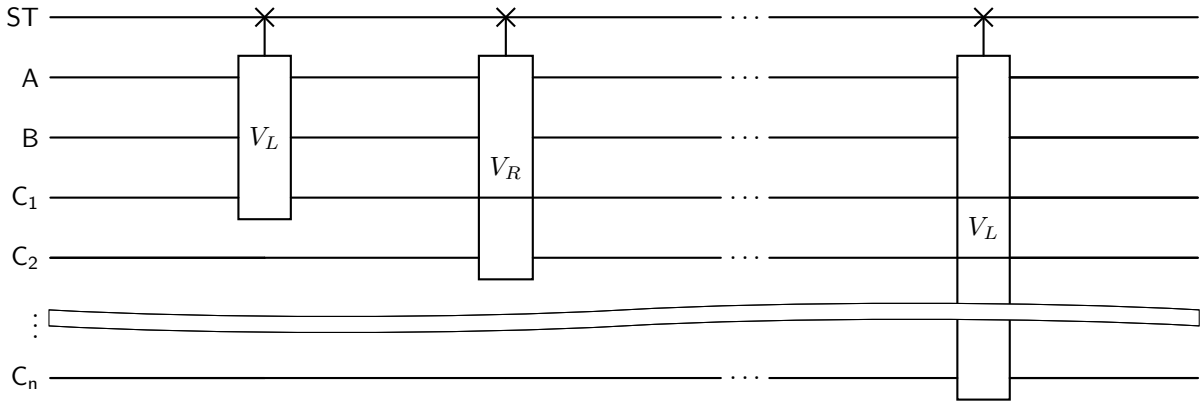
\begin{quantikz}[transparent, column sep=2.5cm,row sep={0.8cm,between origins}]
\lstick{$\reg{ST}$} & \swap{1} & \swap{1} & \ \ldots \ & \swap{1} &\\
\lstick{$\reg{A}$} & \gate[wires=3]{V_L} & \gate[wires=4]{V_R} & \ \ldots\ & \gate[wires=6]{V_L} & \qw\\
\lstick{$\reg{B}$} & & &\ \ldots\ & & \qw \\
\lstick{$\reg{C_1}$} & & \linethrough &\ \ldots\ & \linethrough & \qw \\
\lstick{$\reg{C_2}$} & \qw & & \ \ldots\ & \linethrough & \qw  \\
\lstick{$\vdots$} & \wave[draw=black, fill=white, opacity=1] &&&&&&&\\ 
\lstick{$\reg{C_n}$} & \qw & &\ \ldots\ & & \qw 
\end{quantikz}

\noindent To show that the Ideal experiment is close to the Real experiment, we define a simulator isometry that maps the database register in the Real case to the database register in the Ideal case. To see what this isometry looks like, we first add labels to both the experiments. Recalling the $\bcolor{blue}$ labels for the Real experiment as below:

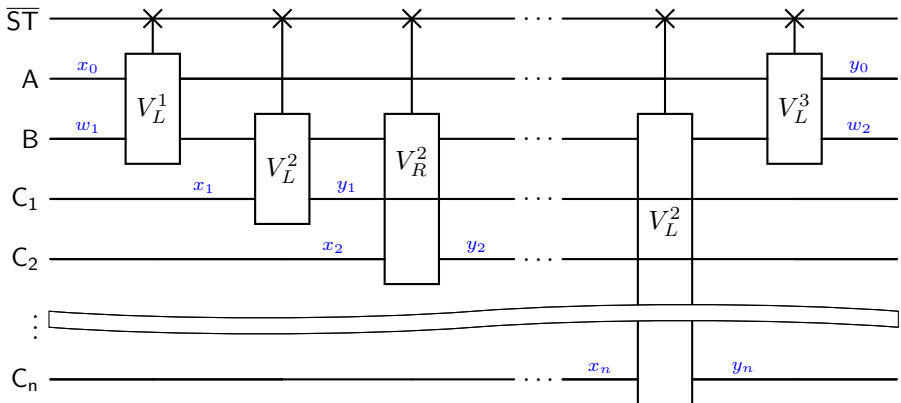
\begin{quantikz}[transparent, column sep=1cm,row sep={0.8cm,between origins}]
\lstick{$\reg{\overline{ST}}$} & \swap{1} & \swap{2} & \swap{2} & \ \ldots \ & \swap{2} & \swap{1} &\\ 
\lstick{$\reg{A}$} \arrow[r,draw=none,"\textcolor{blue}{x_0}"]&
  \gate[wires=2]{V_L^{1}} &
  \qw & \qw &
\ \ldots\ & \qw & \gate[wires=2]{V_L^{3}}\arrow[r,draw=none,"\textcolor{blue}{y_0}"] & \qw\\
\lstick{$\reg{B}$}\arrow[r,draw=none,"\textcolor{blue}{w_1}"] & \qw & \gate[wires=2]{V_L^{2}} & \gate[wires=3,label
style={yshift=0.5cm}]{V_R^{2}} &\ \ldots\ & \gate[wires=5,label
style={yshift=0.5cm}]{V_L^{2}} & \qw\arrow[r,draw=none,"\textcolor{blue}{w_2}"] & \qw \\
\lstick{$\reg{C_1}$} & \qw\arrow[r,draw=none,"\textcolor{blue}{x_1}"]  & \qw\arrow[r,draw=none,"\textcolor{blue}{y_1}"] & \linethrough &\ \ldots\ & \linethrough &  & \qw \\
\lstick{$\reg{C_2}$}& \qw & \qw\arrow[r,draw=none,"\textcolor{blue}{x_2}"] & \qw\arrow[r,draw=none,"\textcolor{blue}{y_2}"] &\ \ldots\ & \linethrough & & \qw  \\
\lstick{$\vdots$} & \wave[draw=black, fill=white, opacity=1] &&&&&&\\ 
\lstick{$\reg{C_n}$}& \qw & \qw & &\ \ldots\ \arrow[r,draw=none,"\textcolor{blue}{x_n}"] & \qw\arrow[r,draw=none,"\textcolor{blue}{y_n}"] & \qw & \qw 
\end{quantikz}

\noindent Recall that the purification state can completely be identified by the $\bcolor{blue}$ labels. Similarly, we add these labels to the Ideal experiment as below: 

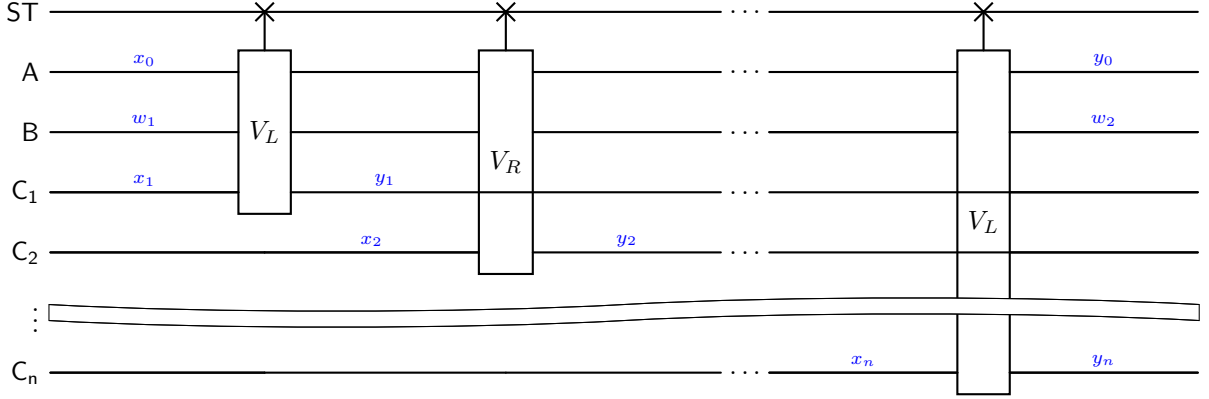
\begin{quantikz}[transparent, column sep=2.5cm,row sep={0.8cm,between origins}]
\lstick{$\reg{ST}$} & \swap{1} & \swap{1} & \ \ldots \ & \swap{1} &\\
\lstick{$\reg{A}$} \arrow[r,draw=none,"\textcolor{blue}{x_0}"]& \gate[wires=3]{V_L} & \gate[wires=4]{V_R} & \ \ldots\ & \gate[wires=6]{V_L}\arrow[r,draw=none,"\textcolor{blue}{y_0}"] & \qw\\
\lstick{$\reg{B}$} \arrow[r,draw=none,"\textcolor{blue}{w_1}"]& & &\ \ldots\ & \qw \arrow[r,draw=none,"\textcolor{blue}{w_2}"]& \qw \\
\lstick{$\reg{C_1}$} \arrow[r,draw=none,"\textcolor{blue}{x_1}"]& \qw\arrow[r,draw=none,"\textcolor{blue}{y_1}"]& \linethrough &\ \ldots\ & \linethrough & \qw \\
\lstick{$\reg{C_2}$} & \qw \arrow[r,draw=none,"\textcolor{blue}{x_2}"]& \qw\arrow[r,draw=none,"\textcolor{blue}{y_2}"]& \ \ldots\ & \linethrough & \qw  \\
\lstick{$\vdots$} & \wave[draw=black, fill=white, opacity=1] &&&&&&&\\ 
\lstick{$\reg{C_n}$} & \qw & &\ \ldots\ \arrow[r,draw=none,"\textcolor{blue}{x_n}"]& \qw \arrow[r,draw=none,"\textcolor{blue}{y_n}"]& \qw 
\end{quantikz}

\noindent Notice that in the above we skip the intermediate labels on the $\reg{AB}$, this is because these are again maximally entangeled pairs completely existing in the database register. Finally, analyzing the database state from the Ideal experiment above and the real experiment from before we can define the simulator isometry (we call this isometry $\Ocomp$). To see how this is formally defined, we refer the reader to~\Cref{sec:gluing:def}.

\subsubsection{Bounding ``Progress Measure''}
\label{sec:tech:prog}

The main challenge in demonstrating that \(\Ocomp\) approximately maps the state in the real case close to the one in the ideal case is the difficulty of obtaining a simple closed-form expression, as was possible in the inverseless setting (see~\cite{MH24}, Appendix C). Instead, we draw inspiration from the query-by-query analysis approach in the literature of the quantum random oracle model~\cite{zhandry2019record,DFMS22}. Specifically, we do query-by-query analysis via defining the \emph{progress measure} as the adversary's distinguishing advantage after each query.

\noindent A key step in our analysis is to show that, for any state \(\ket{\psi}\) (generated using the real oracles), the process of first simulating the ideal database and then making a query to a ideal oracle (e.g., \(V^{\fwd}\)) is close to making a query to a corresponding real oracle (e.g., \(V^{3,\fwd}V^{2,\fwd}V^{1,\fwd}\)) first and then simulating the database. Formally, we show that the following two states are close:
\begin{equation*}
    V \Ocomp \ket{\psi}  \quad \text{and} \quad  \Ocomp V^{\glued} \ket{\psi}\,,
\end{equation*}
which we establish by proving that the operator norm bound 
\[
\|(V \Ocomp - \Ocomp V^{\glued}) \Pi_{\leq t}\|_{\op} = \negl(n)\,,
\]
where $\Pi_{\leq t}$ denotes the projector acting on the database register that checks that the database is of $\poly$-size.
\noindent Similarly, we extend this argument to show inverse queries too, i.e. $$\|(V^{\dagger} \Ocomp - \Ocomp V^{\glued,\dagger}) \Pi_{\leq t}\|_{\op} = \negl(n)$$ 
\noindent To show this, we first show this in each of the  subspaces and combine them to get the final result. Details for this can be found in~\Cref{sec:gluing:fwd}. By establishing these bounds, we can inductively analyze the adversary’s distinguishing advantage after each query (for details, see~\Cref{sec:gluing:induction}). Hence, we show that \(\Ocomp\) approximately maps the state in the real case to the one in the ideal case.

\section{Preliminaries} \label{sec:prelim}

We denote the security parameter by $\secp$. We assume that the reader is familiar with fundamentals of quantum computing, otherwise readers can refer to \cite{nielsen_chuang_2010}. We refer to $\negl(\cdot)$ to be a negligible function. 

\subsection{Notation}

\paragraph{Indexing and sets} We use the notation $[n]$ to refer to the set $\{1, \ldots, n\}$.  For a string $x \in \{0, 1\}^{n+m}$, let $x_{[1:n]}$ to denote the first $n$ bits of $x$. For $N, \ell \in \mathbb{N}$, we let $N^{\downarrow \ell} = \prod_{i = 0}^{\ell-1} (N - i)$. 

\paragraph{Sets and set operators} For two binary strings of the same length $a,b$, we define $a\oplus b$ to be the xor of the two strings. For a set of binary strings $A$ and a binary string $b$, we define the set $A \oplus b := \set{a \oplus b \mid a\in A}$. For two sets of same length binary strings $A$ and $B$, we define the set $A \oplus B := \set{a \oplus b \mid a\in A, b\in B}$.

\paragraph{Set products and the symmetric group}
We use $\Symgp_t$ to refer to the symmetric group over $t$ elements (i.e. the group of all permutations of $t$ elements).  Given a set $A$ and $t \in \mathbb{N}$, we use the notation $A^t$ to denote the $t$-fold Cartesian product of $A$, and the notation $A^t_{\mathrm{dist}}$ to denote distinct subspace of $A^t$, i.e. the vectors in $A^t$, $\vec{y} = (y_1, \ldots, y_t)$, such that for all $i \neq j$, $y_i \neq y_j$. We also define the set $\{\vec{x}\} := \bigcup_{i\in[t]} \{x_i\}$. 

\paragraph{Quantum states and distances}
A register $\reg{R}$ is a named finite-dimensional Hilbert space.  If $\reg{A}$ and $\reg{B}$ are registers, then $\reg{AB}$ denotes the tensor product of the two associated Hilbert spaces.  We denote by $\mathcal{D}(\reg{R})$ the density matrices over register $\reg{R}$.  For $\rho_{\reg{AB}} \in \mathcal{D}(\reg{AB})$, we let $\Tr_\reg{B}(\rho_{\reg{AB}}) \in \mathcal{D}(\reg{A})$ denote the reduced density matrix that results from taking the partial trace over $\reg{B}$.  We denote by $\TD(\rho, \rho') = \frac{1}{2} \norm{\rho - \rho'}_1$ the trace distance between $\rho$ and $\rho'$, where $\norm{X}_1 = \Tr(\sqrt{X^{\dagger} X})$ is the trace norm. For two pure (and possibly subnormalized) states $\ket{\psi}$ and $\ket{\phi}$, we use $\TD\qty(\ket{\psi},\ket{\phi})$ as a shorthand for $\TD\qty(\proj{\psi},\proj{\phi})$. We also say that $A \preceq B$ if $B - A$ is a positive semi-definite matrix.  For positive integers $t, d \in \N$ and a permutation $\sigma \in \Symgp_t$, we let $P_d(\sigma)$ be the $d^t$-dimensional unitary that acts on registers $\reg{R}_1, \ldots, \reg{R}_t$ by permuting the registers according to $\sigma$.  That is,
\begin{equation*}
    P_d(\sigma) \ket{x_1}_{\reg{R}_1} \otimes \dots \otimes \ket{x_t}_{\reg{R}_t} 
    := \ket{x_{\sigma^{-1}(1)}}_{\reg{R}_1} \otimes \dots \otimes \ket{x_{\sigma^{-1}(t)}}_{\reg{R}_t}
\end{equation*}
for all $(x_1, \dots, x_t) \in [d]^t$. We denote by $\haarstates_n$ the Haar distribution over $n$-qubit states, and $\haarunitaries_n$ the Haar measure over $n$-qubit unitaries (i.e. the unique left and right invariant measure).

\paragraph{Relations} Relations are an important part of the path recording framework, here we define relations between sets, as well as what it means to be injective and to take the inverse of a relation.

\begin{definition}[Relation]
    A relation between two finite sets $X$ and $Y$ is a \emph{multiset} of tuples $\{(x_i, y_i)\}_{i\in[t]}$ with $x_i \in X$ and $y_i \in Y$ for all $i \in [t]$.
\end{definition}

\begin{definition}[$\Dom(R)$ and $\Im(R)$]
    For a relation $R = \{(x_i, y_i)\}_{i=1}^{t}$, define $\Dom(R) = \set{x_i}_{i\in [t]}$ and $\Im(R) = \set{y_i}_{i\in [t]}$.
\end{definition}



\begin{definition}[Inverse of a relation]
    The inverse of a relation $R = \{(x_i, y_i)\}_{i = 1}^{t}$ is the relation from $Y$ to $X$ defined by $R^{-1} = \{(y_i, x_i)\}_{i = 1}^{t}$
\end{definition}

\begin{definition}[Substrings]
    Given a string $x\in\bit^{2n+\secp}$, let $x^{\frl(n)}$, $x^{\frm(\secp)}$, and $x^{\frr(n)}$ represent the sub-string on the first $n$, middle $\lambda$, and final $n$ bits respectively, so that $x = x^{\frl(n)} || x^{\frm(\secp)} || x^{\frr(n)}$. Also define $\frl(\cdot)$, $\frr(\cdot)$ and $\frm(\cdot)$ for vectors and sets of strings as follows, let $S = \set{x_i}_{i\in [t]}$, then $S^{\frl(n)} = \set{x^{\frl(n)}_i}_{i\in [t]}$ and let $\vect{x} = (x_1,\ldots,x_t)$, then $\vect{x^{\frl(n)}} = (x^{\frl(n)}_1,\ldots,x^{\frl(n)}_t)$.\footnote{Let $\frl(\cdot)$, $\frr(\cdot)$ and $\frm(\cdot)$ can be defined on strings of other lengths too as first, last and middle substring of some length.}
\end{definition}

\subsection{Cryptographic Primitives}

In this section, we define strong pseudorandom unitaries (strong PRU)~\cite{JLS18}, which are the quantum equivalent of a pseudorandom function, in that an adversary can not distinguish the strong PRU from a truly Haar random unitary, even with inverse access to both.

\begin{definition}[Strong pseudorandom unitaries]
    \label{def:pru}
    We say that a quantum polynomial-size circuit $G$ is a strong pseudorandom unitary if for all quantum polynomial-time adversaries $\mathcal{A}$, there exists a negligible function $\epsilon$ such that for all $\secp$,
    \begin{equation*}
        \left|\mathop{\mathrm{Pr}}_{k \gets \{0, 1\}^{\secp}} \left[1 \gets \mathcal{A}_{\lambda}^{G_\secp(k), {G_\secp(k)}^{\dagger}} \right]- \mathop{\mathrm{Pr}}_{\mathcal{U} \gets \mu_{n(\secp)}} \left[1 \gets \mathcal{A}_{\lambda}^{\mathcal{U}, \mathcal{U}^{\dagger}} \right]\right|  \leq \epsilon(\secp)\,.
    \end{equation*} 
    In the $\mathsf{QHROM}$, both $G_\secp$ and $\mathcal{A}_{\secp}$ have oracle access to an additional family of unitaries $\{U_{\secp}\}_{\secp \in \mathbb{N}}$ sampled from the Haar measure on $\secp$ qubits, and their inverses.
\end{definition}

\subsection{Useful Lemmas}
Here we present useful quantum lemmas that should be familiar to a reader well versed in quantum computation.

\begin{lemma}   \label{lem:op_norm}
For any operator $A$ and vector $\ket{\psi}$, $\norm{A \ket{\psi}}_2 \leq \norm{A}_{\op} \norm{\ket{\psi}}_2$.
\end{lemma}

\begin{lemma}   \label{lem:op_norm_orthogonal}
Let $A$ be an operator and $\cB$ be an orthonormal basis of the domain of $A$. If $A \ket{i}$ is orthogonal to $A \ket{j}$ for all $\ket{i} \neq \ket{j} \in \cB$, then $\norm{A}_{\op} = \max_{\ket{i} \in \cB} \norm{A \ket{i}}_2$.
\end{lemma}

\begin{lemma} \label{lem:proj:commute}
     Let $\Pi_1$ and $\Pi_2$ be two projectors, then $\Pi_1$ and $\Pi_2$ commute if and only if their product is a projector.
\end{lemma}



\section{Path Recording and its Variants}
\label{sec:path}

In this section, we recall the path recording framework from~\cite{MH24}. We give a relatively more general way of looking at path recording and a general theorem for working with path recording. Finally, we define a variant of path recording that we will use throughout this paper.

\subsection{Interpreting the Path Recording Framework}
\label{sec:path:def}

Here we recall the path recording framework from~\cite{MH24}. The path recording framework was defined as a purification of querying a Haar unitary. 

\noindent We will start by stating and interpreting how to "simulate" just forward queries to a Haar unitary. Define the following operator: for any injective relations $R$,
\[
\pr:\ket{x}_{\reg{A}}\ket{R}_{\reg{R}} \mapsto \frac{1}{\sqrt{N - |R|}} \sum_{y \notin \Im(R)} \ket{y}_{\reg{A}} \ket{R\cup\set{(x,y)}}_{\reg{R}}.
\]

\noindent \cite{MH24} gives the following theorem that, in essence, shows that $\pr$ simulates forward queries to a Haar unitary.  
\begin{theorem}[{\cite[Theorem~5]{MH24}}]
\label{thm:MH:fwd}
For any $t$-query algorithm $\Adversary=\left(A_1,\ldots,A_t\right)$, 
\[
\TD \qty( \Ex_{U \sim \haarunitaries_n}\ketbra{\Adversary_t^{U}}{\Adversary_t^{U}}, \Tr_{\reg{R}} \left( \ketbra{\Adversary_t^{\pr}}{\Adversary_t^{\pr}}\right) ) 
\leq O \qty(\frac{t^2}{N^{1/2}}),
\]
where $N = 2^n, \ket{\Adversary_t^{U}} = \prod_{i=1}^{t}\left(U A_i\right) \ket{0}_{\reg{A}} \ket{0}_{\reg{B}}$ and $\ket{\Adversary_t^{\pr}} = \prod_{i=1}^{t}\left( \pr A_i\right) \ket{0}_{\reg{A}} \ket{0}_{\reg{B}} \ket{\varnothing}_{\reg{R}}$.
\end{theorem}

\noindent To understand how $\pr$ works, we must first understand what querying a Haar unitary looks like. In general, querying a Haar unitary on any state gives a maximally mixed state. To understand how $\pr$ does this, we first notice that if you have a maximally entangled state $\frac{1}{\sqrt{N}}\sum_{x\in\bit^n} \ket{x}_{\reg{A}}\ket{x}_{\reg{B}}$ on two registers $\reg{A}$ and $\reg{B}$, then if you only have access to one of these registers $\reg{A}$ or $\reg{B}$, the state looks maximally mixed. $\pr$ uses this property to answer queries to a Haar unitary. In particular, on a query to $\pr$, it saves the input in the purification register and creates a maximally entangled pair, saves one half in the purification register (with the label as the input), and returns the other half as output on the query register. 

\noindent The above idea works almost perfectly, except the operation we define above is not "reversible" (i.e. isn't an isometry or physically realizable). To fix this, we define $\pr$ as only creating an almost maximally entangled state and returns a half in the query register while saving the other in the purification register (labeled by the input) in a way that the operation is still "reversible". 

\noindent Next \cite{MH24} extends this idea to both forward and inverse queries to the Haar unitary. Define the following two operators: for any relations $L,R$, 
\[
V_L:\ket{x}_{\reg{A}}\ket{L}_{\reg{S}}\ket{R}_{\reg{T}} \mapsto \frac{1}{\sqrt{N - |\Im(L\cup R^{-1})|}} \sum_{y \notin \Im(L \cup R^{-1})} \ket{y}_{\reg{A}} \ket{L \cup \{(x,y)\}}_{\reg{S}} \ket{R}_{\reg{T}},
\]
\[
V_R: \ket{x}_{\reg{A}}\ket{L}_{\reg{S}}\ket{R}_{\reg{T}} \mapsto \frac{1}{\sqrt{N - |\Im(R\cup L^{-1})|}} \sum_{y \notin \Im(R \cup L^{-1})} \ket{y}_{\reg{A}} \ket{L}_{\reg{S}} \ket{R \cup \{(x,y)\}}_{\reg{T}}.
\]
\noindent Using $V_L$ and $V_R$, they define the following partial isometry:
$$V = V_L \cdot (I - V_R \cdot V_R^{\dagger}) + (I - V_L \cdot V_L^{\dagger}) \cdot V_R^{\dagger}.$$

\begin{theorem}[{\cite[Theorem~8]{MH24}}]
\label{thm:MH:path}
For any $2t$-query algorithm $\Adversary=\left(A_1,B_1,\ldots,A_t,B_t \right)$, 
\[
\TD \qty( \Ex_{U \sim \haarunitaries_n}\ketbra{\Adversary_t^{U,U^{\dagger}}}{\Adversary_t^{U,U^{\dagger}}}, \Tr_{\reg{ST}} \left( \ketbra{\Adversary_t^{V,V^{\dagger}}}{\Adversary_t^{V,V^{\dagger}}}\right) ) 
\leq O \qty(\frac{t^2}{N^{1/8}}),
\]
where $N = 2^n, \ket{\Adversary_t^{U,U^{\dagger}}} = \prod_{i=1}^{t}\left( U^{\dagger} B_i U A_i\right) \ket{0}_{\reg{A}} \ket{0}_{\reg{B}}$ and $\ket{\Adversary_t^{V,V^{\dagger}}} = \prod_{i=1}^{t}\left( V^{\dagger} B_i V A_i\right) \ket{0}_{\reg{A}} \ket{0}_{\reg{B}} \ket{\varnothing}_{\reg{S}} \ket{\varnothing}_{\reg{T}}$.
\end{theorem}

\noindent To understand how $V$ works, we use the following intuition: given access to the Haar unitary and its inverse, this almost looks like two independent Haar unitaries except if the inverse is queried on the output of a forward query, one must reverse the forward query. To do this, we can think of instantiating two independent forward query oracles $V_L$ and $V_R$, except whenever the inverse oracle is queried, first we check if the input is entangled with something in the purification of $V_L$ (i.e. was the output of $V_L$), if it is, return the associated label, else, just act as a forward query to $V_R$.  Again, to make the above operator "reversible" (i.e. an isometry), we restrict how the "almost" maximally entangled state is defined. 

\noindent Another way of looking at the above operation $V$ that maps well to our intuition is the following: First define the following two projectors (as the projectors that check entanglement between the query register and the database register corresponding to $V_L$ (and $V_R$, respectively)):
\begin{align*} 
    \Pi^{L} &= V_L\cdot V_L^{\dagger} \\
    \Pi^{R} &= V_R\cdot V_R^{\dagger}
\end{align*}
Then we can see that $V$ can be written as 
\begin{align*}
    V &= \Pi^{L}\cdot V_L \cdot (I - \Pi^{R}) + (I - \Pi^{L}) \cdot V_R^{\dagger}\cdot \Pi^{R}\\ 
    V^{\dagger} &= \Pi^{R}\cdot V_R \cdot (I - \Pi^{L}) + (I - \Pi^{R}) \cdot V_L^{\dagger}\cdot \Pi^{L}\\ 
\end{align*}

\noindent Hence operationally, we interpret $V$ as two branches:
\begin{itemize}
    \item If the query register is maximally entangled with the database of $V_R$, apply $V_R^{\dagger}$,
    \item Else apply $V_L$.
\end{itemize}

\noindent In the above, we can think of first checking what subspace a query lies in and then applying the operation depending on this subspace.

\subsection{A Generalization of the Path-Recording Framework}
\label{sec:path:general}

In this section, we will define a generalization of the Path-Recording Framework. The main intuition behind this generalization is that when applying $V_L$ or $V_R$, one doesn't necessarily need a maximally entangled state, just a state with "enough" entanglement. Consider the following operation: Let $V_L$ acts on some query register $\reg{A}$ and some purification register $\reg{ST}$, and added an entry in the database saved in the register $\reg{S}$. Then an operation $V^{f}_L$ action on some query register $\reg{A}$ and some purification register $\reg{ST}$ and an ancilla register $\reg{B}$, is close to $V_L$ where $f$ acts on $\reg{ABST}$ and outputs a "large enough" subset of $[N]\setminus\Im(S)$ over which we create the highly entangled state. 

\noindent Formally, we define the following two operators:
\begin{align*}
    V^{f_L}_L\ket{x}_{\reg{A}}\ket{z}_{\reg{B}}\ket{L}_{\reg{S}}\ket{R}_{\reg{T}} &= \frac{1}{\sqrt{|f_L(x,z,L,R)|}}\sum_{y\in f_L(x,z,L,R)}\ket{y}_{\reg{A}}\ket{z}_{\reg{B}}\ket{L\cup\set{(x,y)}}_{\reg{S}}\ket{R}_{\reg{T}}\\
    V^{f_R}_R\ket{x}_{\reg{A}}\ket{z}_{\reg{B}}\ket{L}_{\reg{S}}\ket{R}_{\reg{T}} &= \frac{1}{\sqrt{|f_R(x,z,L,R)|}}\sum_{y\in f_R(x,z,L,R)}\ket{y}_{\reg{A}}\ket{z}_{\reg{B}}\ket{L}_{\reg{S}}\ket{R\cup\set{(x,y)}}_{\reg{T}}
\end{align*}
where for all $x,z,L,R$, $f_L(x,z,L,R)\subseteq [N]\setminus\Im(L)$ and $f_R(x,z,L,R)\subseteq [N]\setminus\Im(R)$. 

\noindent Using above, we can define $V^{f_L,f_R}$ as $$V^{f_L,f_R} = V^{f_L}_L \cdot (I - V^{f_R}_R \cdot V^{f_R,\dagger}_R) + (I - V^{f_L}_L \cdot V^{f_L,\dagger}_L) \cdot V^{f_R,\dagger}_R.$$

\noindent Then we can show that as long as $f_L$ and $f_R$ give large enough subsets, to a poly query algorithm, $V$ and $V^{f_L,f_R}$ are indistinguishable. Formally, we have the following:

\begin{lemma}[Generalized Path-Recording]
\label{lem:PR:gen}
    Let $f_L$ and $f_R$ be functions such that for all $x,z,L,R$, $|L|+|R|\leq t$, 
    \begin{align*}
        f_L(x,z,L,R)&\subseteq [N]\setminus\Im(L)\\
        f_R(x,z,L,R)&\subseteq [N]\setminus\Im(R)\\
        \frac{\left|N-|f_L(x,z,L,R)|-t\right|}{|f_L(x,z,L,R)|}&\leq \delta\\
        \frac{\left|N-|f_R(x,z,L,R)|-t\right|}{|f_R(x,z,L,R)|}&\leq \delta,
    \end{align*}
    For any $2t$-query algorithm $\Adversary=\left(A_1,B_1,\ldots,A_t,B_t \right)$, 
    \[
    \TD \qty( \Tr_{\reg{ST}} \left( \ketbra{\Adversary_t^{V^{f_L,f_R},V^{f_L,f_R,\dagger}}}{\Adversary_t^{V^{f_L,f_R},V^{f_L,f_R,\dagger}}}\right), \Tr_{\reg{ST}} \left( \ketbra{\Adversary_t^{V,V^{\dagger}}}{\Adversary_t^{V,V^{\dagger}}}\right) ) 
    \leq 16t\sqrt{2\cdot (t+1)\cdot \delta},
    \]
    where,
    \begin{align*}
        \ket{\Adversary_t^{V^{f_L,f_R},V^{f_L,f_R,\dagger}}} &= \prod_{i=1}^{t}\left( V^{f_L,f_R,\dagger} B_i V^{f_L,f_R} A_i\right) \ket{0}_{\reg{A}} \ket{0}_{\reg{B}} \ket{\varnothing}_{\reg{S}} \ket{\varnothing}_{\reg{T}}\\
        \ket{\Adversary_t^{V,V^{\dagger}}} &= \prod_{i=1}^{t}\left( V^{\dagger} B_i V A_i\right) \ket{0}_{\reg{A}} \ket{0}_{\reg{B}} \ket{\varnothing}_{\reg{S}} \ket{\varnothing}_{\reg{T}}
    \end{align*}
\end{lemma}

\noindent The proof of the above lemma is provided in~\Cref{sec:app:PR:gen}.

\subsection{Modified Path Recording}
\label{sec:path:mod}

In this subsection, we define a specific restriction on path recording that we will use later in the proof. We define the following restricted path recording operator as: 
\[
W^{\frm(\secp)}_L:\ket{x}_{\reg{ABC}}\ket{L}_{\reg{S}}\ket{R}_{\reg{T}} \mapsto \frac{1}{\sqrt{2^{2n}(2^{\secp}-|\Im(L\cup R)^{\frm(\secp)}|)}} \sum_{y: y^{\frm(\secp)}\notin \Im(L\cup R)^{\frm(\secp)}} \ket{y}_{\reg{ABC}} \ket{L \cup \{(x,y)\}}_{\reg{S}} \ket{R}_{\reg{T}},
\]
\[
W^{\frm(\secp)}_R:\ket{x}_{\reg{ABC}}\ket{L}_{\reg{S}}\ket{R}_{\reg{T}} \mapsto \frac{1}{\sqrt{2^{2n}(2^{\secp}-|\Im(L\cup R)^{\frm(\secp)}|)}} \sum_{y: y^{\frm(\secp)}\notin \Im(L\cup R)^{\frm(\secp)}} \ket{y}_{\reg{ABC}} \ket{L}_{\reg{S}} \ket{R \cup \{(x,y)\}}_{\reg{T}}.
\]

\noindent Finally, define:
\[
W^{\frm(\secp)} = W_L^{\frm(\secp)}\cdot (I-W_R^{\frm(\secp)}\cdot W_R^{\frm(\secp),\dagger}) + (I-W_L^{\frm(\secp)}\cdot W_L^{\frm(\secp),\dagger})\cdot W_R^{\frm(\secp),\dagger}
\]

\noindent Then from~\Cref{lem:PR:gen}, we have the following lemma.

\begin{lemma}
\label{lem:PR:W}
    For any $2t$-query algorithm $\Adversary=\left(A_1,B_1,\ldots,A_t,B_t \right)$, 
    \[
    \TD \qty( \Tr_{\reg{ST}} \left( \ketbra{\Adversary_t^{W^{\frm(\secp)},W^{\frm(\secp),\dagger}}}{\Adversary_t^{W^{\frm(\secp)},W^{\frm(\secp),\dagger}}}\right), \Tr_{\reg{ST}} \left( \ketbra{\Adversary_t^{V,V^{\dagger}}}{\Adversary_t^{V,V^{\dagger}}}\right) ) 
    \leq O \qty(\sqrt{\frac{t^{3}}{2^{\secp}}}),
    \]
    where $|\reg{ABC}| = 2n+\secp$, $\ket{\Adversary_t^{W^{\frm(\secp)},W^{\frm(\secp),\dagger}}} = \prod_{i=1}^{t}\left( W^{\frm(\secp),\dagger} B_i W^{\frm(\secp)} A_i\right) \ket{0}_{\reg{ABC}} \ket{0}_{\reg{D}} \ket{\varnothing}_{\reg{S}} \ket{\varnothing}_{\reg{T}}$ and $\ket{\Adversary_t^{V,V^{\dagger}}} = \prod_{i=1}^{t}\left( V^{\dagger} B_i V A_i\right) \ket{0}_{\reg{ABC}} \ket{0}_{\reg{D}} \ket{\varnothing}_{\reg{S}} \ket{\varnothing}_{\reg{T}}$.
\end{lemma}

\section{Glued Haar Unitary and its Purification}
\label{sec:glued}

In this section, we present our glued Haar random unitary construction. We study an operation that acts on a purification register and mimics as if querying a glued Haar random unitary construction (similar to how path recording mimics a single Haar unitary). We'll call this operator the Glued Path Recording operator. Similar to how Path Recording has different actions on two different subspaces, Glued Path Recording has different actions on four different subspaces. 

\noindent We start by looking at the naive way to achieve this by just replacing each Haar unitary in the glued construction with an instance of Path Recording Instance. Next, we will define a modification of the Path Recording oracle that is easier to work with in the proof. Finally, we define the four subspaces of interest and, using these subspaces, we define an operator that acts as the Glued Path Recording.

\subsection{Glued Haar Unitary Construction}
\label{sec:glued:cons}

In this section, we present our construction of the Glued Haar Unitary. Let $U^1,U^2$ and $U^3$ be unitaries acting on $n+\secp$ qubits, then we define the glued construction $G(U^1,U^2,U^3)$ acting on $2n+\secp$ qubits as:
$$G(U^1,U^2,U^3)_{\reg{ABC}} = U_{\reg{AB}}^3\cdot U_{\reg{BC}}^2\cdot U_{\reg{AB}}^1$$
where $|\reg{A}|=|\reg{C}|=n$, $|\reg{B}|=\secp$.

\noindent Throughout this paper, we show that an if $U_1,U_2$ and $U_3$ were sampled from the Haar distribution, then no poly query algorithm can distinguish between $G(U^1,U^2,U^3)$ and its inverse from a Haar random untary $O$ and its inverse (where $O$ acts on $2n+\secp$ qubits).

\noindent We start by studying a purification of the construction $G(U^1,U^2,U^3)$. Notice that one can always just purify $U^1$, $U^2$, and $U^3$ individually using Path Recording. While this is a valid purification, it seems hard to work with. We give a more intuitive purification of the above construction in the next section.

\subsection{Glued Path Recording}
\label{sec:glued:PR}

To get a more intuitive purification of the glued Haar unitary construction, we first look at how the construction behaves. Let we make a query to the construction, then we know, from the discussion in~\Cref{sec:path:def}, that if this query is a previous output of $U^1$ (which in the purified view looks like checking entanglement), then $U^1$ inverts this query; else $U^1$ returns an extremely scrambled state. 

\noindent Notice that the output of $U^1$ is (partially) fed into $U^2$. $U^2$ performs a similar check to see if the query is a previous output of $U^2$ or not. Interestingly, notice that if $U^1$ was returning an extremely scrambled state, then this state would almost certainly not be the output of a previous query to $U^2$. 

\noindent To look at this in the purified view, if $U^1$ created a new maximally entangled pair and returns one half in the query register and saves the other half in the first database register, then when $U^2$ checks if this state is maximally entangled with something in the second database, this check almost always fails because by monogamy of entanglement we know that if the query register is maximally entangled with something in the first database, it cannot be maximally entangled with anything in the second database. 

\noindent To formalise this intuition, we start defining some operations similar to path recording. In particular, corresponding to $U^1$ (and $U^2$ and $U^3$, respectively), we define a pair of operations $(V^1_L,V^1_R)$ (and $(V^2_L,V^2_R)$ and $(V^3_L,V^3_R)$, respectively) where $(V^1_L,V^1_R)$ (and $(V^2_L,V^2_R)$ and $(V^3_L,V^3_R)$, respectively) have corresponding purification registers $\reg{S_1T_1}$ (and $\reg{S_2T_2}$ and $\reg{S_3T_3}$, respectively). 

\noindent Now we define some projectors that correspond to checking entanglements (similar to $\Pi^{R}$ in~\Cref{sec:path:def}):
\begin{align*}
    \Pi^{R,1} &= V_R^{1} V_R^{1,\dagger}\\
    \Pi^{R,12} &= V_R^{1} V_R^{2} V_R^{2,\dagger} V_R^{1,\dagger}\\
    \Pi^{R,123} &= V_R^{1} V_R^{2} V_R^{3} V_R^{3,\dagger} V_R^{2,\dagger} V_R^{1,\dagger}
\end{align*}

\noindent Similarly, we define similar projectors in the opposite direction (similar to $\Pi^{L}$ in~\Cref{sec:path:def}):
\begin{align*}
    \Pi^{L,3} &= V_L^{3} V_L^{3,\dagger}\\
    \Pi^{L,32} &= V_L^{3} V_L^{2} V_L^{2,\dagger} V_L^{3,\dagger}\\
    \Pi^{L,321} &= V_L^{3} V_L^{2} V_L^{1} V_L^{1,\dagger} V_L^{2,\dagger} V_L^{3,\dagger}
\end{align*} 

\noindent Then with these in mind, we define the glued purification as follows: 
\begin{align*}
    V^{\glued} = &\left(\Pi^{L,321}\right)\cdot V_L^{3}\cdot V_L^{2}\cdot V_L^{1} \cdot\left(I-\Pi^{R,1}\right) \\
    &+ \left(\Pi^{L,32} -\Pi^{L,321}\right)\cdot V_L^{3}\cdot V_L^{2}\cdot V_R^{1,\dagger}\cdot\left(\Pi^{R,1}-\Pi^{R,12}\right) \\
    &+ \left(\Pi^{L,3} -\Pi^{L,32}\right)\cdot V_L^{3}\cdot V_R^{2,\dagger}\cdot V_R^{1,\dagger}\cdot\left(\Pi^{R,12}-\Pi^{R,123}\right) \\
    &+ \left(I-\Pi^{L,3}\right)\cdot V_R^{3,\dagger}\cdot V_R^{2,\dagger}\cdot V_R^{1,\dagger}\cdot\left(\Pi^{R,123}\right) \\
\end{align*}

\noindent Operationally, the $V^{\glued}$ works as follows:
\begin{itemize}
    \item Check if the query register is in the output of $V_R^{1}\cdot V_R^{2}\cdot V_R^{3}$, if it is, invert these queries.
    \item Else, check if the query register is in the output of $V_R^{1}\cdot V_R^{2}$, if it is, invert these queries and apply $V_L^{3}$.
    \item Else, check if the query register is in the output of $V_R^{1}$, if it is, invert this query and apply $V_L^{3}\cdot V_L^{2}$.
    \item Else, apply $V_L^{3}\cdot V_L^{2}\cdot V_L^{1}$.
\end{itemize}
\noindent One can see that $V^{\glued}$ has a structure that is similar to path recording while also following the intuition stated at the beginning. In fact using the intuition from the discussion above, we formally show the following: 

\begin{lemma}
\label{lem:glued:PR}
    For any $2t$-query algorithm $\Adversary=\left(A_1,B_1,\ldots,A_t,B_t \right)$, 
    \[
    \norm{\ket{\Adversary^{V^{\glued}, V^{\glued,\dagger}}}_{\reg{ABCD\overline{ST}}} 
    - \ket{\Adversary^{V^{3}V^{2}V^{1}, (V^{3}V^{2}V^{1})^{\dagger}}}_{\reg{ABCD\overline{ST}}}}_2
    = O \qty(\frac{t^3}{2^{\secp}}).
    \]
    where $\reg{\overline{ST}} = \reg{S_1S_2S_3T_1T_2T_3}$, $|\reg{ABC}| = 2n+\secp$, $$\ket{\Adversary_t^{V^{\glued},V^{\glued,\dagger}}} = \prod_{i=1}^{t}\left( V^{\glued,\dagger} B_i V^{\glued} A_i\right) \ket{0}_{\reg{ABC}} \ket{0}_{\reg{D}} \ket{\varnothing}_{\reg{S_1}} \ket{\varnothing}_{\reg{T_1}}\ket{\varnothing}_{\reg{S_2}} \ket{\varnothing}_{\reg{T_2}}\ket{\varnothing}_{\reg{S_3}} \ket{\varnothing}_{\reg{T_3}}$$ and $\ket{\Adversary^{V^{3}V^{2}V^{1}, (V^{3}V^{2}V^{1})^{\dagger}}}_{\reg{ABCD\overline{ST}}}$ is defined similarly.
\end{lemma}

\noindent The proof of the above lemma is provided in~\Cref{sec:app:glued}.

\subsection{Modified Glued Path Recording}
\label{sec:glued:PR:modified}

In this subsection, we define a specific restriction on the glued path recording that we will use later in the proof. We define the following restricted glued path recording operator as: 

\noindent We first define the following $$\Im_1^{\midd}(L_1,L_2,L_3,R_1,R_2,R_3) = \Im(L_1\cup R_3)^{\frr(\secp)}\bigcup\Im(L_2\cup R_2)^{\frl(\secp)}\bigcup\Dom(L_2\cup R_2)^{\frl(\secp)},$$
and $$\Im_2^{\midd}(L_1,L_2,L_3,R_1,R_2,R_3) = \Im(L_3\cup R_1)^{\frr(\secp)}.$$

\noindent Next, we define the following partial isometries:
\begin{align*}
    V^{(1),\midd}_L& \ket{x}_{\reg{ABC}} \ket{L_1}_{\reg{S}_1}\ket{R_1}_{\reg{T}_1} \ket{L_2}_{\reg{S}_2}\ket{R_2}_{\reg{T}_2} \ket{L_3}_{\reg{S}_3}\ket{R_3}_{\reg{T}_3} \\
    =& \frac{1}{\sqrt{2^{n}(2^{\secp}-|\Im_1^{\midd}(L_1,L_2,L_3,R_1,R_2,R_3)|)}}\sum_{\substack{y:\\ y^{\frr(\secp)}\notin \Im_1^{\midd}(L_1,L_2,L_3,R_1,R_2,R_3)}}\ket{y}_{\reg{AB}}\ket{x^{\frr(n)}}_{\reg{C}}\\
    &\otimes\ket{L_1\cup\set{(x^{\frl(n+\secp)},y)}}_{\reg{S}_1}\ket{R_1}_{\reg{T}_1} \ket{L_2}_{\reg{S}_2}\ket{R_2}_{\reg{T}_2} \ket{L_3}_{\reg{S}_3}\ket{R_3}_{\reg{T}_3}\\
    V^{(1),\midd}_R& \ket{x}_{\reg{ABC}} \ket{L_1}_{\reg{S}_1}\ket{R_1}_{\reg{T}_1} \ket{L_2}_{\reg{S}_2}\ket{R_2}_{\reg{T}_2} \ket{L_3}_{\reg{S}_3}\ket{R_3}_{\reg{T}_3} \\
    =& \frac{1}{\sqrt{2^{n}(2^{\secp}-|\Im_2^{\midd}(L_1,L_2,L_3,R_1,R_2,R_3)|)}}\sum_{\substack{y:\\ y^{\frr(\secp)}\notin \Im_2^{\midd}(L_1,L_2,L_3,R_1,R_2,R_3)}}\ket{y}_{\reg{AB}}\ket{x^{\frr(n)}}_{\reg{C}}\\
    &\otimes\ket{L_1}_{\reg{S}_1}\ket{R_1\cup\set{(x^{\frl(n+\secp)},y)}}_{\reg{T}_1} \ket{L_2}_{\reg{S}_2}\ket{R_2}_{\reg{T}_2} \ket{L_3}_{\reg{S}_3}\ket{R_3}_{\reg{T}_3} \\
    V^{(2),\midd}_L& \ket{x}_{\reg{ABC}} \ket{L_1}_{\reg{S}_1}\ket{R_1}_{\reg{T}_1} \ket{L_2}_{\reg{S}_2}\ket{R_2}_{\reg{T}_2} \ket{L_3}_{\reg{S}_3}\ket{R_3}_{\reg{T}_3} \\
    =& \frac{1}{\sqrt{2^{n}(2^{\secp}-|\Im_1^{\midd}(L_1,L_2,L_3,R_1,R_2,R_3)|)}}\sum_{\substack{y:\\ y^{\frl(\secp)}\notin \Im_1^{\midd}(L_1,L_2,L_3,R_1,R_2,R_3)}}\ket{x^{\frl(n)}}_{\reg{A}}\ket{y}_{\reg{BC}}\\
    &\otimes\ket{L_1}_{\reg{S}_1}\ket{R_1}_{\reg{T}_1} \ket{L_2\cup\set{(x^{\frr(n+\secp)},y)}}_{\reg{S}_2}\ket{R_2}_{\reg{T}_2} \ket{L_3}_{\reg{S}_3}\ket{R_3}_{\reg{T}_3}\\
    V^{(2),\midd}_R& \ket{x}_{\reg{ABC}} \ket{L_1}_{\reg{S}_1}\ket{R_1}_{\reg{T}_1} \ket{L_2}_{\reg{S}_2}\ket{R_2}_{\reg{T}_2} \ket{L_3}_{\reg{S}_3}\ket{R_3}_{\reg{T}_3} \\
    =& \frac{1}{\sqrt{2^{n}(2^{\secp}-|\Im_1^{\midd}(L_1,L_2,L_3,R_1,R_2,R_3)|)}}\sum_{\substack{y:\\ y^{\frl(\secp)}\notin \Im_1^{\midd}(L_1,L_2,L_3,R_1,R_2,R_3)}}\ket{x^{\frl(n)}}_{\reg{A}}\ket{y}_{\reg{BC}}\\
    &\otimes\ket{L_1}_{\reg{S}_1}\ket{R_1}_{\reg{T}_1} \ket{L_2}_{\reg{S}_2}\ket{R_2\cup\set{(x^{\frr(n+\secp)},y)}}_{\reg{T}_2} \ket{L_3}_{\reg{S}_3}\ket{R_3}_{\reg{T}_3} \\
    V^{(3),\midd}_L& \ket{x}_{\reg{ABC}} \ket{L_1}_{\reg{S}_1}\ket{R_1}_{\reg{T}_1} \ket{L_2}_{\reg{S}_2}\ket{R_2}_{\reg{T}_2} \ket{L_3}_{\reg{S}_3}\ket{R_3}_{\reg{T}_3} \\
    =& \frac{1}{\sqrt{2^{n}(2^{\secp}-|\Im_2^{\midd}(L_1,L_2,L_3,R_1,R_2,R_3)|)}}\sum_{\substack{y:\\ y^{\frr(\secp)}\notin \Im_2^{\midd}(L_1,L_2,L_3,R_1,R_2,R_3)}}\ket{y}_{\reg{AB}}\ket{x^{\frr(n)}}_{\reg{C}}\\
    &\otimes\ket{L_1}_{\reg{S}_1}\ket{R_1}_{\reg{T}_1} \ket{L_2}_{\reg{S}_2}\ket{R_2}_{\reg{T}_2} \ket{L_3\cup\set{(x^{\frl(n+\secp)},y)}}_{\reg{S}_3}\ket{R_3}_{\reg{T}_3}\\
    V^{(3),\midd}_R& \ket{x}_{\reg{ABC}} \ket{L_1}_{\reg{S}_1}\ket{R_1}_{\reg{T}_1} \ket{L_2}_{\reg{S}_2}\ket{R_2}_{\reg{T}_2} \ket{L_3}_{\reg{S}_3}\ket{R_3}_{\reg{T}_3} \\
    =& \frac{1}{\sqrt{2^{n}(2^{\secp}-|\Im_1^{\midd}(L_1,L_2,L_3,R_1,R_2,R_3)|)}}\sum_{\substack{y:\\ y^{\frr(\secp)}\notin \Im_1^{\midd}(L_1,L_2,L_3,R_1,R_2,R_3)}}\ket{y}_{\reg{AB}}\ket{x^{\frr(n)}}_{\reg{C}}\\
    &\otimes\ket{L_1}_{\reg{S}_1}\ket{R_1}_{\reg{T}_1} \ket{L_2}_{\reg{S}_2}\ket{R_2}_{\reg{T}_2} \ket{L_3}_{\reg{S}_3}\ket{R_3\cup\set{(x^{\frl(n+\secp)},y)}}_{\reg{T}_3} \\
\end{align*}

\noindent We know by~\Cref{lem:F_V_close} that for $i\in\set{1,2,3}$ and $X\in\set{L,R}$, \[
\|(V^{i}_X-V^{(i),\midd}_X) \Pi_{\leq t}\|_{\op} = O\left(\sqrt{\frac{t^2}{2^\secp}}\right).
\]

\noindent The reason the above operators are defined this way will become clear in the next section. Basically, querying glued path recording using the above operators results in a mbetter-structuredured purification. We next define the modified path recording with the above operator. 

\noindent Now we define some projectors that correspond to checking entanglements (similar to ~\Cref{sec:glued:PR}):
\begin{align*}
    \Pi^{\cR,1} &= V_R^{(1),\midd} V_R^{(1),\midd,\dagger}\\
    \Pi^{\cR,12} &= V_R^{(1),\midd} V_R^{(2),\midd} V_R^{(2),\midd,\dagger} V_R^{(1),\midd,\dagger}\\
    \Pi^{\cR,123} &= V_R^{(1),\midd} V_R^{(2),\midd} V_R^{(3),\midd} V_R^{(3),\midd,\dagger} V_R^{(2),\midd,\dagger} V_R^{(1),\midd,\dagger}
\end{align*}

\noindent Similarly, we define similar projectors in the opposite direction (similar to ~\Cref{sec:glued:PR}):
\begin{align*}
    \Pi^{\cL,3} &= V_L^{(3),\midd} V_L^{(3),\midd,\dagger}\\
    \Pi^{\cR,32} &= V_L^{(3),\midd} V_L^{(2),\midd} V_L^{(2),\midd,\dagger} V_L^{(3),\midd,\dagger}\\
    \Pi^{\cL,321} &= V_L^{(3),\midd} V_L^{(2),\midd} V_L^{(1),\midd} V_L^{(1),\midd,\dagger} V_L^{(2),\midd,\dagger} V_L^{(3),\midd,\dagger}
\end{align*} 

\noindent Then with these in mind, we define the glued purification as follows: 
\begin{align*}
    W^{\glued} = &\left(\Pi^{\cL,321}\right)\cdot V_L^{(3),\midd}\cdot V_L^{(2),\midd}\cdot V_L^{(1),\midd} \cdot\left(I-\Pi^{\cR,1}\right) \\
    &+ \left(\Pi^{\cL,32} -\Pi^{\cL,321}\right)\cdot V_L^{(3),\midd}\cdot V_L^{(2),\midd}\cdot V_R^{(1),\midd,\dagger}\cdot\left(\Pi^{\cR,1}-\Pi^{\cR,12}\right) \\
    &+ \left(\Pi^{\cL,3} -\Pi^{\cL,32}\right)\cdot V_L^{(3),\midd}\cdot V_R^{(2),\midd,\dagger}\cdot V_R^{(1),\midd,\dagger}\cdot\left(\Pi^{\cR,12}-\Pi^{\cR,123}\right) \\
    &+ \left(I-\Pi^{\cL,3}\right)\cdot V_R^{(3),\midd,\dagger}\cdot V_R^{(2),\midd,\dagger}\cdot V_R^{(1),\midd,\dagger}\cdot\left(\Pi^{\cR,123}\right) \\
\end{align*} 

\noindent Then we have the following:

\begin{lemma}
\label{lem:glued:PR:W}
    For any adversary $\Adversary$ that makes $t$ forward queries and $t$ inverse queries,
    \[
    \norm{\ket{\Adversary^{V^{\glued}, V^{\glued,\dagger}}}_{\reg{ABCD\overline{ST}}} 
    - \ket{\Adversary^{W^{\glued}, W^{\glued,\dagger}}}_{\reg{ABCD\overline{ST}}}}_2
    = O \qty(\sqrt{\frac{t^4}{2^{\secp}}}),
    \]
    where $\reg{\overline{ST}} = \reg{S_1S_2S_3T_1T_2T_3}$.
\end{lemma}

\noindent For ease of notation, we define the following projectors:

\begin{align*}
    \Pi^{\frl,1} &= I-\Pi^{\cR,1}\\
    \Pi^{\frl,2} &= \Pi^{\cR,1}-\Pi^{\cR,12}\\
    \Pi^{\frl,3} &= \Pi^{\cR,12}-\Pi^{\cR,123}\\
    \Pi^{\frl,4} &= \Pi^{\cR,123}
\end{align*}

\section{Structure of Purification for $W^{\glued}$}
\label{sec:struc}

In this section, our main goal would be to understand the structure of purification when using glued path recording. To do this, we start by showing how the purification can be associated to a graph, how this graph is structured and can be parametrized. We then use these graphs to define "good" states. Once we have defined the good states, we study intersection of good states and the subspaces defined with respect to glued path recording. Finally, we show that querying glued path recording only leads to good states. 

\noindent The main intuition behind studying the structure is the following: Looking at the operational definition of $V^{\glued}$ in~\Cref{sec:glued:PR}, we can see that the queries can only be done in certain patterns. In particular, any output of $V^{2}_{L}$ is partially fed as input to a $V^{3}_{L}$. Similarly, whenever we try to invert $V^{3}_{L}$ (i.e. apply $V^{3,\dagger}_{L}$), it is followed by either inverting $V^{2}_{L}$ (i.e. apply $V^{2,\dagger}_{L}$) or applying $V^{2}_{R}$. We associate a graph with the database to keep track of these correlations (i.e. keep track of when the output of some operator is part fed as the input to another operation).

\subsection{Graph associated with the Database}
\label{sec:struc:graph}

\noindent To study the structure of the purification when querying $W^{\glued}$ and $W^{\glued,\dagger}$, we start by associating a graph to the database in the purification. To do this, we associate a vertex for each entry in the database. We label these vertices with the tuple in the database as well as what database it came from. Formally we do the following: 
\noindent Given $(L_1,L_2,L_3,R_1,R_2,R_3)$, we define the following vertex set:
\begin{align*}
    V_{L_1} &= \set{(\frl_1,x,y) | (x,y)\in L_1} \\
    V_{L_2} &= \set{(\frl_2,x,y) | (x,y)\in L_2} \\
    V_{L_3} &= \set{(\frl_3,x,y) | (x,y)\in L_3} \\
    V_{R_1} &= \set{(\frr_1,x,y) | (x,y)\in R_1} \\
    V_{R_2} &= \set{(\frr_2,x,y) | (x,y)\in R_2} \\
    V_{R_3} &= \set{(\frr_3,x,y) | (x,y)\in R_3}\,.
\end{align*}

\noindent Next we want to add edges to the graph. We add edges where we suspect that the output of a given query was partially fed as input to another query. Looking back at the operational definition of $V^{\glued}$ (in~\Cref{sec:glued:PR}), we see that in some operations correlate various enties in this form. We define directed edges between such tuples as follows: 
\begin{align*}
    E_{L_1L_2} &= \set{(v_1,v_2) | v_1 = (\frl_1,x_1,y_1) \in V_{L_1}, v_2 = (\frl_2,x_2,y_2) \in V_{L_2}, {y_1}^{\frr(\secp)} = {x_2}^{\frl(\secp)}} \\
    E_{L_2L_3} &= \set{(v_1,v_2) | v_1 = (\frl_2,x_1,y_1) \in V_{L_2}, v_2 = (\frl_3,x_2,y_2) \in V_{L_3}, {y_1}^{\frl(\secp)} = {x_2}^{\frr(\secp)}} \\
    E_{R_3R_2} &= \set{(v_1,v_2) | v_1 = (\frr_3,x_1,y_1) \in V_{R_3}, v_2 = (\frr_2,x_2,y_2) \in V_{R_2}, {y_1}^{\frr(\secp)} = {x_2}^{\frl(\secp)}} \\
    E_{R_2R_1} &= \set{(v_1,v_2) | v_1 = (\frr_2,x_1,y_1) \in V_{R_2}, v_2 = (\frr_1,x_2,y_2) \in V_{R_1}, {y_1}^{\frl(\secp)} = {x_2}^{\frr(\secp)}} \\
    E_{L_2R_2} &= \set{(v_1,v_2) | v_1 = (\frl_2,x_1,y_1) \in V_{L_2}, v_2 = (\frr_2,x_2,y_2) \in V_{R_2}, {y_1}^{\frl(\secp)} = {x_2}^{\frl(\secp)}} \\
    E_{R_2L_2} &= \set{(v_1,v_2) | v_1 = (\frr_2,x_1,y_1) \in V_{R_2}, v_2 = (\frl_2,x_2,y_2) \in V_{L_2}, {y_1}^{\frl(\secp)} = {x_2}^{\frl(\secp)}} \,.
\end{align*}

\noindent Finally, we this gives us the graph:
\begin{align*}
    V(L_1,L_2,L_3,R_1,R_2,R_3) &= \bigcup_{i=1}^3 \left(V_{L_i} \cup V_{R_i}\right) \\
    E(L_1,L_2,L_3,R_1,R_2,R_3) &= \bigcup_{i=1}^2 \left(E_{L_{i}L_{i+1}} \cup E_{R_{i+1}R_{i}}\right) \cup E_{L_2R_2} \cup E_{R_2L_2} \\
    G(L_1,L_2,L_3,R_1,R_2,R_3) &= (V(L_1,L_2,L_3,R_1,R_2,R_3),E(L_1,L_2,L_3,R_1,R_2,R_3))
\end{align*}

\subsection{Analyzing Structure of the Graph}
\label{sec:struc:good:graph}

\noindent Next we try to study the structure of this graph associated with the database. We know that we added an edge at every instance where we suspect that an output from some operator was partially fed into another operator as the input. Next, we will try to see "chains" of multiple operators where the output of one is fed as input to another. These would look like a path in the associated graph. 

\noindent A path in the graph to a sequence of connected vertices $(v_1,v_2,\ldots,v_n)$ with edges $(v_i,v_{i+1})$. A graph is a line graph if all the vertices in the graph form a path and all edges in the graph are just part of the path. We say a graph is a linear forest if it is a disjoint union of line graphs. For any linear forest, let $\cP(G)$ be the set of disjoint line graphs. For any $p\in\cP(G)$, let $\leng(p)$ denote the number of edges in $p$. 

\noindent Looking back at the operational description of $V^{\glued}$, operations always start with either $V_L^1$ or $V_R^3$. Similarly, operations always end with either $V_L^3$ or $V_R^1$. Hence, we can classify the lines in the linear forest associated to the database. Formally, we define the following classes of Line Graphs in $G$. 

\begin{definition}[Classes of Line Graphs in $G$]
    Given a collection of relations, $\vec{L} = (L_1,L_2,L_3), \vec{R} = (R_1,R_2,R_3)$, define the following sets:
    \begin{align*}
        \cP_{LL}(\vec{L}, \vec{R}) &= \set{p\in \cP(G(L_1,L_2,L_3,R_1,R_2,R_3)) | p_{\sG}\in L_1, p_{\eG}\in L_3} \\
        \cP_{LR}(\vec{L}, \vec{R}) &= \set{p\in \cP(G(L_1,L_2,L_3,R_1,R_2,R_3)) | p_{\sG}\in L_1, p_{\eG}\in R_1} \\
        \cP_{RL}(\vec{L}, \vec{R}) &= \set{p\in \cP(G(L_1,L_2,L_3,R_1,R_2,R_3)) | p_{\sG}\in R_3, p_{\eG}\in L_3} \\
        \cP_{RR}(\vec{L}, \vec{R}) &= \set{p\in \cP(G(L_1,L_2,L_3,R_1,R_2,R_3)) | p_{\sG}\in R_3, p_{\eG}\in R_1}
    \end{align*} 
\end{definition} 

\noindent Next, notice that while part of the output of $V^{1}_L$ is fed into $V^2_L$, the other part of the output is fed to the last operator in the path. We refer the reader to figures in the~\Cref{sec:tech:struc}. We formalise this to define good line in graphs:

\begin{definition}[Good lines in $G$]
    Given a collection of relations, $\vec{L} = (L_1,L_2,L_3), \vec{R} = (R_1,R_2,R_3)$, define the following subsets of $\{\cP_{LL}, \cP_{LR}, \cP_{RL}, \cP_{RR}\}$:
    \begin{align*}
        \cP_{LL}^{\mathrm{good}}(\vec{L}, \vec{R}) &= \{p \in \cP_{LL}(\vec{L}, \vec{R})\ |\ p_{\mathsf{start}} = (\frl_1, x_1, y_1), p_{\mathsf{end}} = (\frl_3, x_2, y_2), y^{\frl(n)}_{1} = x_2^{\frl(n)}\}\\
        \cP_{LR}^{\mathrm{good}}(\vec{L}, \vec{R}) &= \{p \in \cP_{LR}(\vec{L}, \vec{R})\ |\ p_{\mathsf{start}} = (\frl_1, x_1, y_1), p_{\mathsf{end}} = (\frr_1, x_2, y_2), y^{\frl(n)}_{1} = x_2^{\frl(n)}\}\\
        \cP_{RL}^{\mathrm{good}}(\vec{L}, \vec{R}) &= \{p \in \cP_{RL}(\vec{L}, \vec{R})\ |\ p_{\mathsf{start}} = (\frr_3, x_1, y_1), p_{\mathsf{end}} = (\frl_3, x_2, y_2), y^{\frl(n)}_{1} = x_2^{\frl(n)}\}\\
        \cP_{RR}^{\mathrm{good}}(\vec{L}, \vec{R}) &= \{p \in \cP_{RR}(\vec{L}, \vec{R})\ |\ p_{\mathsf{start}} = (\frr_3, x_1, y_1), p_{\mathsf{end}} = (\frr_1, x_2, y_2), y^{\frl(n)}_{1} = x_2^{\frl(n)}\}\,.
    \end{align*}
\end{definition}

\noindent Next we can define a graph as good as graphs which are made of good line graphs.
\begin{definition}[Good graphs]
    Given $L_1,L_2,L_3,R_1,R_2,R_3$, we say $G(L_1,L_2,L_3,R_1,R_2,R_3)$ is "good" if:
    \begin{enumerate}
        \item $G(L_1,L_2,L_3,R_1,R_2,R_3)$ is a linear forest.
        \item All lines in $G(L_1,L_2,L_3,R_1,R_2,R_3)$ are either $\cP_{LL}^{\mathrm{good}}$, $\cP_{LR}^{\mathrm{good}}$, $\cP_{RL}^{\mathrm{good}}$ or $\cP_{RR}^{\mathrm{good}}$. 
    \end{enumerate} 
\end{definition}

\subsection{Parametrizing Good Graphs}

\noindent Now that we have a highly structured graph associated to the database, we try to define some notation to look at these graphs. We start by defining a parametrized form of good line graphs. We know that in good line graphs, part of the output label at any vertex is repeated in the adjacent vertex. We label these by $r_i$. We also know that the first and the last vector also share part of the label, we denote this with $z$. The other labels are made by $x_i$ and $y_i$ along with $w_1,w_2$. We refer the reader to figures in the~\Cref{sec:tech:struc}. We formalise this representation below. 

\begin{definition}[Good line graph parametrization]
    Let $p$ be a line in $\cP_{LL}^{\mathrm{good}}(\vec{L}, \vec{R})$ for some collection $\vec{L}$ and $\vec{R}$, then we can write the line $p$ as follows:
    \begin{equation*}
        p = \set{(\frl_1,x_0||w_1,z||r_1),(\frl_2,r_1||x_1,r_2||y_1),(\frr_2,r_2||x_2,r_3||y_2),\ldots,(\frl_2,r_n||x_n,r_{n+1}||y_n),(\frl_3,z||r_{n+1},y_0||w_2)}\,.
    \end{equation*}
    Then we define the function 
    \begin{equation*}
        \frp(\cL\cL,\vec{x},\vec{y},w_1,w_2,\vec{r},z) = p\,,
    \end{equation*}
    where $\vec{x}$, $\vec{y}$ are $\leng(p)$-length vectors of $n - \lambda$ bit strings, and $\vec{r}$ is a $\leng(p)$-length vector of $\lambda$ bit strings.  
\end{definition}
\noindent We similarly define the functions $\frp$ with the first index $\cL\cR$, $\cR\cL$ and $\cR\cR$ for paths in $\cP_{LR}^{\mathrm{good}}(\vec{L}, \vec{R})$, $\cP_{RL}^{\mathrm{good}}(\vec{L}, \vec{R})$ and $\cP_{RR}^{\mathrm{good}}(\vec{L}, \vec{R})$, respectively. 

\noindent We extend the above formalization to a "good" graph (i.e. a disjoint union of good line graphs). We define the following parametrized representations of "good" graphs:

\begin{definition}[Good graph parametrization]
    Given any "good" graph $G$, we define the following representation: Let 
    $$G = \bigcup_{X,Y\in\set{\cL,\cR}}\left(\bigcup_{i} p^{XY}_i\right)$$
    where $p_i^{\cL\cL} = \frp(\cL\cL,\vect{x^{\cL\cL,i}},\vect{y^{\cL\cL,i}},w^{\cL\cL,i}_1,w^{\cL\cL,i}_2,\vect{r^{\cL\cL,i}},z^{\cL\cL,i}),$ and similarly $p_i^{\cL\cR}$, $p_i^{\cR\cL}$ and $p_i^{\cR\cR}$. Then we define $\overline{\cG}$ as:
    \begin{align*}
    G = \overline{\cG} \left(\bigcup_{X,Y\in\set{\cL,\cR}}\left(\bigcup_{i}\set{(XY,\vect{x^{XY,i}},\vect{y^{XY,i}},w^{XY,i}_1,w^{XY,i}_2,\vect{r^{XY,i}},z^{XY,i})}\right)\right).
    \end{align*}
\end{definition}

\subsection{Defining Good Auxiliary States}
\label{sec:struc:aux}

Next we associate even more structure with the database state. In particular, we notice that the internal labels for each line graph (i.e. the $\vect{r}$'s and $\vect{z}$'s) exist as maximally entangled pairs, with one half in the output labels with respect to some operators and the other half in the input labels with respect to some other operator. These maximally entangled pairs can be sampled later and added into the database state. Formally, given some partial description of line graphs (i.e. all parameters except the $\vect{r}$'s and $\vect{z}$'s), we can sample the rest of the parameters and create a superposition over these. We will call these partial descriptions of line graphs a "state structure parameter". 

\noindent We start by defining some notation to help us formalize the above intuition. We define "state structure parameter" notation as the following four sets:
\begin{itemize}
    \item $S_{\cL\cL} = \set{\frq_{i}^{\cL\cL}}_{i}$, where $\frq_{i}^{\cL\cL} = (\cL\cL,\vect{x^{\cL\cL,i}},\vect{y^{\cL\cL,i}},w^{\cL\cL,i}_1,w^{\cL\cL,i}_2)$ and $|\vect{x^{\cL\cL,i}}| = |\vect{y^{\cL\cL,i}}|$.
    \item $S_{\cL\cR} = \set{\frq_{i}^{\cL\cR}}_{i}$, where $\frq_{i}^{\cL\cR} = (\cL\cR,\vect{x^{\cL\cR,i}},\vect{y^{\cL\cR,i}},w^{\cL\cR,i}_1,w^{\cL\cR,i}_2)$ and $|\vect{x^{\cL\cR,i}}| = |\vect{y^{\cL\cR,i}}|$.
    \item $S_{\cR\cL} = \set{\frq_{i}^{\cR\cL}}_{i}$, where $\frq_{i}^{\cR\cL} = (\cR\cL,\vect{x^{\cR\cL,i}},\vect{y^{\cR\cL,i}},w^{\cR\cL,i}_1,w^{\cR\cL,i}_2)$ and $|\vect{x^{\cR\cL,i}}| = |\vect{y^{\cR\cL,i}}|$.
    \item $S_{\cR\cR} = \set{\frq_{i}^{\cR\cR}}_{i}$, where $\frq_{i}^{\cR\cR} = (\cR\cR,\vect{x^{\cR\cR,i}},\vect{y^{\cR\cR,i}},w^{\cR\cR,i}_1,w^{\cR\cR,i}_2)$ and $|\vect{x^{\cR\cR,i}}| = |\vect{y^{\cR\cR,i}}|$.
\end{itemize}

\noindent Before defining the notation to study "good states", we define the following helper functions for "state structure parameter":
\begin{definition}
    Given a state structure parameter $\overline{S}$, with $\overline{S} = \bigcup_{X,Y\in\set{\cL,\cR}}S_{XY}$ and $S_{\cL\cL} = \set{\frq_{i}^{\cL\cL}}_{i}$, $S_{\cL\cR} = \set{\frq_{i}^{\cL\cR}}_{i}$, $S_{\cR\cL} = \set{\frq_{i}^{\cR\cL}}_{i}$ and $S_{\cR\cR} = \set{\frq_{i}^{\cR\cR}}_{i}$. Define:
    \begin{itemize}
        \item For $X,Y\in\set{\cL,\cR}$, $\leng(S_{XY}) = \sum_{i\in |S_{XY}|} \leng(\vect{x^{XY,i}})$ and $\leng(\overline{S}) = \sum_{X,Y\in\set{\cL,\cR}} \leng(S_{XY})$.
        \item For $X,Y\in\set{\cL,\cR}$, $\coun(S_{XY}) = |S_{XY}|$ and $\coun(\overline{S}) = \sum_{X,Y\in\set{\cL,\cR}} \coun(S_{XY})$.
        \item $\Im(\overline{S}) = \set{w_2^{XY,i}| X,Y\in\set{\cL,\cR}, i}$.
    \end{itemize}
\end{definition}

\noindent Before defining "Good States", we first impose an extra condition on state structure parameters.

\begin{definition}[Good State Structure Parameter]
    Given a state parameter $\overline{S}$, with $\overline{S} = \overline{S} = \bigcup_{X,Y\in\set{\cL,\cR}}S_{XY}$ and $S_{\cL\cL} = \set{\frq_{i}^{\cL\cL}}_{i}$, $S_{\cL\cR} = \set{\frq_{i}^{\cL\cR}}_{i}$, $S_{\cR\cL} = \set{\frq_{i}^{\cR\cL}}_{i}$ and $S_{\cR\cR} = \set{\frq_{i}^{\cR\cR}}_{i}$. We say $\overline{S}$ is a good state parameter if $|\Im(\overline{S})|=\coun(\overline{S})$.
\end{definition}

\noindent Finally, we define "good states":

\begin{definition}[Good States]
    Given $\overline{S} = \bigcup_{X,Y\in\set{\cL,\cR}}S_{XY}$ with $S_{\cL\cL} = \set{(\cL\cL,\vect{x^{\cL\cL,i}},\vect{y^{\cL\cL,i}},w^{\cL\cL,i}_1,w^{\cL\cL,i}_2)}_i$ and similarly $S_{\cL\cR}$, $S_{\cR\cL}$ and $S_{\cR\cR}$ with $a = \coun(\overline{S})$ and $b = \leng(\overline{S})$. Let, for $X,Y\in\set{\cL,\cR}$, $z^{XY,i}\in\bit^{\secp}$ and $\set{\vect{r^{XY,i}}}_{X,Y,i}\in\bit^{bn}_{\dist}$. Say $Z = \set{z^{XY,i}}_{X,Y,i}$ and $R = \set{\vect{r^{XY,i}}}_{X,Y,i}$. Define $\G$ as the following notation: 
    \begin{align*}
        \G(\overline{S},R,Z) =&  \overline{\cG} \bigg(\bigcup_{X,Y\in\set{\cL,\cR}}\set{(XY,\vect{x^{XY,i}},\vect{y^{XY,i}},w^{XY,i}_1,w^{XY,i}_2,\vect{r^{XY,i}},z^{XY,i})}_i\bigg)
    \end{align*}
    Then we define the following state (defining notation $\frG$):
    \begin{align*}
        \ket{\frG(\overline{S})}_{\reg{\overline{ST}}} = \frac{1}{\sqrt{2^{an}(2^\secp)\ldots(2^\secp-b+1) }}\sum_{\substack{Z\in\bit^{an}\\ R\in\left(\bit^{\secp}\right)^{b}_{\dist}}} \ket{\G \left(\overline{S},R,Z\right)}_{\reg{\overline{ST}}},
    \end{align*}
    \noindent where $\ket{\G \left(\overline{S},R,Z\right)}_{\reg{\overline{ST}}}$ denotes the $\ket{L_1}_{\reg{S}_1}\ket{L_2}_{\reg{S}_2}\ket{L_3}_{\reg{S}_3}\ket{R_1}_{\reg{T}_1}\ket{R_2}_{\reg{T}_2}\ket{R_3}_{\reg{T}_3}$ corresponding to the $\G \left(\overline{S},R,Z\right)$.
\end{definition}

\noindent Next, we define the projector on these good state as below:

\begin{definition}[Good Projector]
    Define "good" projector as follows:
    \begin{align*}
        \Pi^{\good} = \sum_{\substack{\overline{S}\\ \overline{S}\text{ is good}}}\ketbra{\frG(\overline{S})}{\frG(\overline{S})}
    \end{align*}
\end{definition}

\subsection{Subspaces of $\Pi^{\good}$}
\label{sec:struc:proj}

Now that we have defined this projector $\Pi^{\good}$, we want to say that any algorithm querying $V^{\glued}$, the purification mostly lies in the subspace defined by $\Pi^{\good}$. Before we approach this, know from~\Cref{sec:glued:PR}, $V^{\glued}$ has different operations on different subspaces. We show how these subspaces behave with $\Pi^{\good}$. In particular, we know that $V^{\glued}$ has different operations on subspaces defined by projectors $\Pi^{\cR, 1}$, $\Pi^{\cR, 12}$ and $\Pi^{\cR, 123}$. In particular, we show that $\Pi^{\cR, 1}$, $\Pi^{\cR, 12}$ and $\Pi^{\cR, 123}$ commute with $\Pi^{\good}$.

\noindent From~\Cref{lem:proj:commute}, we know that two projectors commute if and only if their product is a projector. Hence, to show that $\Pi^{\cR, 1}$, $\Pi^{\cR, 12}$ and $\Pi^{\cR, 123}$ commute with $\Pi^{\good}$, we show that their product is a projector. To define these projectors, we first define the following vectors:

\begin{itemize}
    \item Let $\overline{S'}$ be a good state parameter, let $X\in\set{\cL,\cR}$, $t\in\N$, $\vect{x}\in\bit^{(t+1)n}$ and $\vect{y}\in\bit^{tn}$ and $w_1\in\bit^{\secp}$, let $a=\coun(\overline{S'})$, define 
    $$\ket{\chi^{\frl,1}_{\overline{S'},X,\vect{x},\vect{y},w_1}} = \frac{1}{\sqrt{2^{n}(2^\secp-a+1)}}\sum_{\substack{y'_0\in\set{0,1}^n\\ w'_2\in\left(\bit^{\secp}\setminus\Im(\overline{S'})\right)}} \ket{y_0',w_2',\frG(\overline{S'}\cup\set{(X\cR,\vect{x},y'_0||\vect{y},w_1,w'_2)})}_{\reg{AB\overline{ST}}}.$$
    \item Let $\overline{S'}$ be a good state parameter, let $t\in\N$, $X\in\set{\cL,\cR}$, $\vect{x}\in\bit^{(t+2)n}$ and $\vect{y}\in\bit^{tn}$ and $w\in\bit^{\secp}$, let $a=\coun(\overline{S'})$, define 
    $$\ket{\chi^{\frl,2}_{\overline{S'},X,\vect{x},\vect{y},w_1}} = \frac{1}{2^{n}\sqrt{(2^\secp-a+1)}}\sum_{\substack{y_0'\in\set{0,1}^n\\ y_1'\in\set{0,1}^n\\ w_2'\in\left(\bit^{\secp}\setminus\Im(\overline{S'})\right)}} \ket{y_0',w_2',y_1',\frG(\overline{S'}\cup\set{(X\cR,\vect{x},y_0'||\vect{y}||y_1',w_1,w_2')})}_{\reg{ABC\overline{ST}}}.$$
    \item Let $\overline{S'}$ be a good state parameter, let $\vect{x}\in\bit^{2n}$ and $w\in\bit^{\secp}$, let $a=\coun(\overline{S'})$, define 
    $$\ket{\chi^{\frl,3}_{\overline{S'},\vect{x},w_1}} = \frac{1}{2^{n}\sqrt{(2^\secp-a+1)}}\sum_{\substack{y_0'\in\set{0,1}^n\\ y_1'\in\set{0,1}^n\\ w_2'\in\left(\bit^{\secp}\setminus\Im(\overline{S'})\right)}} \ket{y_0',w_2',y_1',\frG(\overline{S'}\cup\set{(\cR\cR,\vect{x},(y_0',y_1'),w_1,w_2')})}_{\reg{ABC\overline{ST}}}.$$
\end{itemize}

\noindent Notice that the above states defined are norm $1$ and orthogonal.

\noindent The way to think about these states is the following:
\begin{itemize}
    \item The space spanned by $\ket{\chi^{\frl,1}_{\overline{S'},X,\vect{x},\vect{y},w_1}}$ are the "Good" states in the image of $V^{(1),\midd}_R$ (i.e. in $\Pi^{\cR, 1}$). 
    \item The space spanned by $\ket{\chi^{\frl,2}_{\overline{S'},X,\vect{x},\vect{y},w_1}}$ are the "Good" states in the image of $V^{(1),\midd}_RV^{(2),\midd}_R$ (i.e. in $\Pi^{\cR, 12}$).
    \item The space spanned by $\ket{\chi^{\frl,3}_{\overline{S'},\vect{x},w_1}}$ are the "Good" states in the image of $V^{(1),\midd}_RV^{(2),\midd}_RV^{(3),\midd}_R$ (i.e. in $\Pi^{\cR, 123}$).
\end{itemize}

\noindent Similar to above, we also define $\ket{\chi^{\frr,i}}$ as:
\begin{itemize}
    \item Let $\overline{S'}$ be a good state parameter, let $X\in\set{\cL,\cR}$, $t\in\N$, $\vect{x}\in\bit^{(t+1)n}$ and $\vect{y}\in\bit^{tn}$ and $w_1\in\bit^{\secp}$, let $a=\coun(\overline{S'})$, define 
    $$\ket{\chi^{\frr,1}_{\overline{S'},X,\vect{x},\vect{y},w_1}} = \frac{1}{\sqrt{2^{n}(2^\secp-a+1)}}\sum_{\substack{y'_0\in\set{0,1}^n\\ w'_2\in\left(\bit^{\secp}\setminus\Im(\overline{S'})\right)}} \ket{y_0',w_2',\frG(\overline{S'}\cup\set{(X\cL,\vect{x},y'_0||\vect{y},w_1,w'_2)})}_{\reg{AB\overline{ST}}}.$$
    \item Let $\overline{S'}$ be a good state parameter, let $t\in\N$, $X\in\set{\cL,\cR}$, $\vect{x}\in\bit^{(t+2)n}$ and $\vect{y}\in\bit^{tn}$ and $w\in\bit^{\secp}$, let $a=\coun(\overline{S'})$, define 
    $$\ket{\chi^{\frr,2}_{\overline{S'},X,\vect{x},\vect{y},w_1}} = \frac{1}{2^{n}\sqrt{(2^\secp-a+1)}}\sum_{\substack{y_0'\in\set{0,1}^n\\ y_1'\in\set{0,1}^n\\ w_2'\in\left(\bit^{\secp}\setminus\Im(\overline{S'})\right)}} \ket{y_0',w_2',y_1',\frG(\overline{S'}\cup\set{(X\cL,\vect{x},y_0'||\vect{y}||y_1',w_1,w_2')})}_{\reg{ABC\overline{ST}}}.$$
    \item Let $\overline{S'}$ be a good state parameter, let $\vect{x}\in\bit^{2n}$ and $w\in\bit^{\secp}$, let $a=\coun(\overline{S'})$, define 
    $$\ket{\chi^{\frr,3}_{\overline{S'},\vect{x},w_1}} = \frac{1}{2^{n}\sqrt{(2^\secp-a+1)}}\sum_{\substack{y_0'\in\set{0,1}^n\\ y_1'\in\set{0,1}^n\\ w_2'\in\left(\bit^{\secp}\setminus\Im(\overline{S'})\right)}} \ket{y_0',w_2',y_1',\frG(\overline{S'}\cup\set{(\cL\cL,\vect{x},(y_0',y_1'),w_1,w_2')})}_{\reg{ABC\overline{ST}}}.$$
\end{itemize}

\noindent Finally, we have the following lemmas that formalize that $\ket{\chi^i}$'s span $\Pi^{\good}\Pi^{\cR,i}$:

\begin{lemma}
\label{lem:proj:com1}
    We have the following:
    $$\Pi^{\good}\Pi^{\cR,1} = \sum_{\overline{S'},X,\vect{x},\vect{y},w_1} \ketbra{\chi^{\frl,1}_{\overline{S'},X,\vect{x},\vect{y},w_1}}{\chi^{\frl,1}_{\overline{S'},X,\vect{x},\vect{y},w_1}}$$
\end{lemma}

\begin{lemma}
\label{lem:proj:com2}
    We have the following:
    $$\Pi^{\good}\Pi^{\cR,12} = \sum_{\overline{S'},X,\vect{x},\vect{y},w_1} \ketbra{\chi^{\frl,2}_{\overline{S'},X,\vect{x},\vect{y},w_1}}{\chi^{\frl,2}_{\overline{S'},X,\vect{x},\vect{y},w_1}}$$
\end{lemma}

\begin{lemma}
\label{lem:proj:com3}
    We have the following:
    $$\Pi^{\good}\Pi^{\cR,123} = \sum_{\overline{S'},\vect{x},w_1} \ketbra{\chi^{\frl,3}_{\overline{S'},\vect{x},w_1}}{\chi^{\frl,3}_{\overline{S'},\vect{x},w_1}}$$
\end{lemma}

\noindent Similar to above, we have $\Pi^{\good}\Pi^{\cL,i}$ as a projector on space spanned by $\ket{\chi^{\frr,i}}$. We give proofs for the above lemmas in~\Cref{sec:app:proj}.

\subsection{Action of $W^{\glued}$ on states in $\Pi^{\good}$}
\label{sec:struc:action}

Now finally we can start talking about the action of $W^{\glued}$ on states in $\Pi^{\good}$. In particular, we want to show that querying $W^{\glued}$ on a state in $\Pi^{\good}$ gives us states in $\Pi^{\good}$. Recall that:
\begin{align*}
    W^{\glued} = &\left(\Pi^{\cL,321}\right)\cdot V_L^{(3),\midd}\cdot V_L^{(2),\midd}\cdot V_L^{(1),\midd} \cdot\left(I-\Pi^{\cR,1}\right) \\
    &+ \left(\Pi^{\cL,32} -\Pi^{\cL,321}\right)\cdot V_L^{(3),\midd}\cdot V_L^{(2),\midd}\cdot V_R^{(1),\midd,\dagger}\cdot\left(\Pi^{\cR,1}-\Pi^{\cR,12}\right) \\
    &+ \left(\Pi^{\cL,3} -\Pi^{\cL,32}\right)\cdot V_L^{(3),\midd}\cdot V_R^{(2),\midd,\dagger}\cdot V_R^{(1),\midd,\dagger}\cdot\left(\Pi^{\cR,12}-\Pi^{\cR,123}\right) \\
    &+ \left(I-\Pi^{\cL,3}\right)\cdot V_R^{(3),\midd,\dagger}\cdot V_R^{(2),\midd,\dagger}\cdot V_R^{(1),\midd,\dagger}\cdot\left(\Pi^{\cR,123}\right) \\
\end{align*} 

\noindent Recall that from the previous section, we know exactly how to characterize good states in the four branches. In particular, we know that these states in various branches are spanned by $\ket{\chi^{\frl,i}}$. We show the action of $W^{\glued}$ on $\ket{\chi^{\frl,i}}$. In particular, we show the following:

\begin{lemma}
    \label{lem:LtoR:1}
    Let $\overline{S}$ be some good state parameter and $x_0,x_1\in\bit^{n}$, $w_1\in\bit^{\secp}$. Then we have the following: 
    \begin{align*}
        V^{(1),\midd}_RV^{(2),\midd}_RV^{(3),\midd}_R\ket{x_0}\ket{w_1}\ket{x_1}\ket{\frG(\overline{S})} &= \ket{\chi^{\frl,3}_{\overline{S},(x_0,x_1),w_1}} \\
        V^{(3),\midd}_LV^{(2),\midd}_LV^{(1),\midd}_L\ket{x_0}\ket{w_1}\ket{x_1}\ket{\frG(\overline{S})} &= \ket{\chi^{\frr,3}_{\overline{S},(x_0,x_1),w_1}} 
    \end{align*}
\end{lemma}

\begin{lemma}
    \label{lem:LtoR:2}
    Let $\overline{S'}$ be a good state parameter, let $X\in\set{\cL,\cR}$, $t\in\N$, $\vect{x}\in\bit^{(t+1)n}$ and $\vect{y}\in\bit^{tn}$, $x_0\in\bit^{n}$ and $w_1\in\bit^{\secp}$. Then we have the following: 
    \begin{align*}
         V_R^{(1),\midd}V_R^{(2),\midd}V_L^{(3),\midd,\dagger}\ket{\chi^{\frr,1}_{\overline{S},X,\vect{x},\vect{y},w_1}}_{\reg{AB\overline{ST}}}\ket{x'}_{\reg{C}} &= \ket{\chi^{\frl,2}_{\overline{S},X,\vect{x}||x',\vect{y},w_1}}_{\reg{ABC\overline{ST}}} \\
         V_L^{(3),\midd}V_L^{(2),\midd}V_R^{(1),\midd,\dagger}\ket{\chi^{\frl,1}_{\overline{S},X,\vect{x},\vect{y},w_1}}_{\reg{AB\overline{ST}}}\ket{x'}_{\reg{C}} &= \ket{\chi^{\frr,2}_{\overline{S},X,\vect{x}||x',\vect{y},w_1}}_{\reg{ABC\overline{ST}}}
    \end{align*}
\end{lemma}

\noindent We provide the proofs of the above lemmas in~\Cref{sec:app:action}. Finally, now since we understand the action of $W^{\glued}$ on basis states of the subspaces, we show that applying $W^{\glued}$ on good states gives good states. Formally, we show the following lemma:

\begin{lemma}
\label{lem:struc:good:fwd}
    Let $\ket{\phi}$ be some state such that $\Pi^{\good}\ket{\phi} = \ket{\phi}$, then $(I-\Pi^{\good}) W^{\glued}\ket{\phi} = 0$.
\end{lemma}

\noindent The proof of the above lemma is in~\Cref{sec:app:action}. Symmetrically, we can also get the lemma below.

\begin{lemma}
\label{lem:struc:good:inv}
    Let $\ket{\phi}$ be some state such that $\Pi^{\good}\ket{\phi} = \ket{\phi}$, then $(I-\Pi^{\good}) W^{\glued,\dagger}\ket{\phi}=0$.
\end{lemma}

\subsection{Purification of $W^{\glued}$ is Good}
\label{sec:struc:induction}

\noindent Let $\Adversary$ denote a $2t$-query algorithm. For any unitaries $U$, define 
\[
\ket{\Adversary^{U, U^{\dagger}}} 
= \prod_{i=1}^{t} \left(U^{\dagger}_{\reg{ABC}} B_i U_{\reg{ABC}} A_i\right) \ket{0}_{\reg{ABC}} \ket{0}_{\reg{D}}.
\]

\begin{lemma}
\label{lem:struc:good}
    Let $\Adversary$ denote a $2t$-query algorithm. Define 
    \begin{align*}
        \ket{\psi} &= \ket{\Adversary^{W^{\glued},W^{\glued,\dagger}}} \\
        \ket{\phi} &= \ket{\Adversary^{\Pi^{\good}W^{\glued},\Pi^{\good}W^{\glued,\dagger}}} \\
    \end{align*} Then
    $$\ket{\psi} = \ket{\phi}$$
\end{lemma}

\noindent We get the above by~\Cref{lem:struc:good:fwd,lem:struc:good:inv}.

\section{Strong Gluing Theorem}
\label{sec:gluing}

In this section, we finally prove the Strong Gluing Theorem. We start by reducing the large Haar Unitary to the modified path recording and the glued construction to the modified glued path recording. Next we define an operator that simulates the purification in the large path recording from the purification in the gluing construction. To do this, we do a query-by-query analysis. 

\noindent We now state the main result of this section.
\begin{theorem}[Strong gluing of random unitaries]
\label{thm:gluing}
    Let $\reg{A}, \reg{B}, \reg{C}$ be registers, and $U^1_{\reg{AB}}$, $U^2_{\reg{BC}}$, and $U^3_{\reg{AB}}$ be Haar random unitaries on $n+\secp$ qubits, with $\reg{B}$ being $\secp$ qubits.  Then for any $t$-query adversary $\mathcal{A}^{(\cdot)}$, the following holds
    \begin{multline*}
        \TD\left(\avg_{U^1, U^2, U^3 \leftarrow \haarunitaries_n}\left[\proj{\mathcal{A}^{U^3_{\reg{AB}}U^2_{\reg{BC}}U^1_{\reg{AB}}, (U^3_{\reg{AB}}U^2_{\reg{BC}}U^1_{\reg{AB}})^\dagger}}\right], \avg_{O \leftarrow \haarunitaries_{2n - \lambda}} \left[\proj{\mathcal{A}^{O_{\reg{ABC}}, O^{\dagger}_{\reg{ABC}}}}\right]\right) \\= O\left(\frac{t^2}{2^{\secp/2}} + \frac{t^3}{2^{\secp}} + \frac{t^3}{2^{(n+\secp)/8}}\right).
    \end{multline*}
\end{theorem}

\subsection{Proof of~\Cref{thm:gluing}}
\label{sec:strong:proof:pru:qhrom}

\noindent Let $\Adversary$ denote a strong PRU adversary. For any unitaries $U$, define 
\[
\ket{\Adversary^{U, U^{\dagger}}} 
= \prod_{i=1}^{t} \left(U^{\dagger}_{\reg{ABC}} B_i U_{\reg{ABC}} A_i\right) \ket{0}_{\reg{ABC}} \ket{0}_{\reg{D}}.
\]
We define the following hybrids (changes are denoted in \rcolor{red}):

\paragraph{Hybrid $\hyb_1$:} Define:
\[
\ket{u_1(O)} = \ket{\Adversary^{O_{\reg{ABC}}, O^{\dagger}_{\reg{ABC}}}}.
\]
Output 
\[
\Ex_{O \sim \haarunitaries_{2n+\secp}} \left[\ketbra{u_1(O)}{u_1(O)}\right].
\]

\paragraph{Hybrid $\hyb_2$:} Define:
\[
\ket{u_2} = \ket{\Adversary^{\rcolor{V}, \rcolor{V^{\dagger}}}},
\]
where $V, V^{\dagger}$ acts on the registers $\reg{ABCST}$ and registers $\reg{S}$ and $\reg{T}$ are initialised as $\ket{\varnothing}$. Output 
\[
\rcolor{\Tr_{\reg{ST}}}\left(\ketbra{u_2}{u_2}\right).
\]

\paragraph{Hybrid $\hyb_3$:} Define \[\ket{u_3} = \ket{\Adversary^{\rcolor{W^{\frm(\secp)}}, \rcolor{W^{\frm(\secp),\dagger}}}},
\]
where $W^{\fwd}, W^{\inv}$ acts on the registers $\reg{ABCST}$ and registers $\reg{S}$ and $\reg{T}$ are initialised as $\ket{\varnothing}$. Output $$\Tr_{\reg{ST}}\left(\rcolor{\ketbra{u_3}{u_3}}\right).$$

\paragraph{Hybrid $\hyb_4$:} Define \[\ket{u_4} = \ket{\Adversary^{\rcolor{\Pi^{\good}W^{\glued}}, \rcolor{\Pi^{\good}W^{\glued,\dagger}}}},
\]
where $\Pi^{\good}W^{\glued}, \Pi^{\good}W^{\glued,\dagger}$ acts on the registers $\reg{ABCS_1T_1S_2T_2S_3T_3}$ and registers $\reg{S}_i$ and $\reg{T}_i$ are initialised as $\ket{\varnothing}$ for $i\in [3]$. Output $$\Tr_{\rcolor{\reg{S_1S_2S_3T_1T_2T_3}}}\left(\rcolor{\ketbra{u_4}{u_4}}\right).$$

\paragraph{Hybrid $\hyb_5$:} Define \[\ket{u_5} = \ket{\Adversary^{\rcolor{W^{\glued}}, \rcolor{W^{\glued,\dagger}}}},
\]
where $W^{\glued}, W^{\glued,\dagger}$ acts on the registers $\reg{ABCS_1T_1S_2T_2S_3T_3}$ and registers $\reg{S}_i$ and $\reg{T}_i$ are initialised as $\ket{\varnothing}$ for $i\in [3]$. Output $$\Tr_{\rcolor{\reg{S_1S_2S_3T_1T_2T_3}}}\left(\rcolor{\ketbra{u_5}{u_5}}\right).$$

\paragraph{Hybrid $\hyb_6$:} Define \[\ket{u_6} = \ket{\Adversary^{\rcolor{V^{\glued}}, \rcolor{V^{\glued,\dagger}}}},
\]
where $V^{\glued}, V^{\glued,\dagger}$ acts on the registers $\reg{ABCS_1T_1S_2T_2S_3T_3}$ and registers $\reg{S}_i$ and $\reg{T}_i$ are initialised as $\ket{\varnothing}$ for $i\in [3]$. Output $$\Tr_{\reg{S_1S_2S_3T_1T_2T_3}}\left(\rcolor{\ketbra{u_6}{u_6}}\right).$$

\paragraph{Hybrid $\hyb_7$:} Define 
\[\ket{u_7(U^1,U^2,U^3)} = \ket{\Adversary^{U_{\reg{AB}}^3U_{\reg{BC}}^2U_{\reg{AB}}^1, (U_{\reg{AB}}^3U_{\reg{BC}}^2U_{\reg{AB}}^1)^{\dagger}}}.
\]
Output $$\rcolor{\Ex_{U_1,U_2,U_3 \sim \haarunitaries_{n+\secp}}} \left[\ketbra{u_7(U^1,U^2,U^3)}{u_7(U^1,U^2,U^3)}\right].$$

\paragraph{Statistical Indistinguishability of Hybrids.}

\noindent We prove the closeness as follows:

\begin{myclaim} 
    The trace distance between $\hyb_1$ and $\hyb_2$ is $O \qty(\frac{t^3}{N^{1/8}})$.
\end{myclaim}
\begin{proof}
    By~\Cref{thm:MH:path}.
\end{proof}

\begin{myclaim} 
    The trace distance between $\hyb_2$ and $\hyb_3$ is $O \qty(\sqrt{\frac{t^{4}}{2^{\secp}}})$.
\end{myclaim}
\begin{proof}
    By~\Cref{lem:PR:W}.  
\end{proof}

\begin{myclaim} \label{thm:gluing:induction}
    The trace distance between $\hyb_3$ and $\hyb_4$ is $O \qty(\frac{t^3}{2^{\secp}})$. 
\end{myclaim}

\noindent Proving~\Cref{thm:gluing:induction} is the main technical step of this section. We begin by defining $\Ocomp$ in~\Cref{sec:gluing:def}, which we then use to prove~\Cref{thm:gluing:induction} in~\Cref{sec:gluing:induction}.

\begin{myclaim} 
    The trace distance between $\hyb_4$ and $\hyb_5$ is $0$.
\end{myclaim}
\begin{proof}
     By~\Cref{lem:struc:good}.
\end{proof}

\begin{myclaim} 
    The trace distance between $\hyb_5$ and $\hyb_6$ is $O \qty(\sqrt{\frac{t^4}{2^{\secp}}})$.
\end{myclaim}
\begin{proof}
     By~\Cref{lem:glued:PR:W}.
\end{proof}

\begin{myclaim} 
    The trace distance between $\hyb_6$ and $\hyb_5$ is $O \qty(\frac{t^3}{2^{\secp}}+\frac{t^3}{2^{(n+\secp)/8}})$.
\end{myclaim}
\begin{proof}
     By~\Cref{lem:glued:PR} and~\Cref{thm:MH:path}.
\end{proof}

\subsection{Defining $\Ocomp$}
\label{sec:gluing:def}

In this section, we will define $\Ocomp$, which maps the purification from querying $\Pi^{\good}W^{\glued}$ to purification of $W^{\frm(\secp)}$. We know that the purification of $\Pi^{\good}W^{\glued}$ is in the good subspace, hence is in the span of $\ket{\frG(\overline{S})}$. We start by thinking about what $\Ocomp$ should do for a single $\ket{\frG(\overline{S})}$.

\noindent The intuition towards defining $\Ocomp$ is the following: Given any path $\frq\in \overline{S}$, we know that it signifies multiple interweaving queries to $V_X^{(i),\midd}$ with outputs from one being partially fed as input into the next. We define $\Ocomp$ to convert these interweaving queries to $W^{\frm(\secp)}$ and $W^{\frm(\secp),\dagger}$. We refer the reader to figures in the~\Cref{sec:tech:sim}. 

\noindent Again, when we have interweaving queries to $W^{\frm(\secp)}$ and $W^{\frm(\secp),\dagger}$, we see that the output of one feeding into the exists as a maximally entangled pair between the two databases (similar to~\Cref{sec:struc:aux}). Hence, we can sample these maximally entangled pairs and append them to a state structure parameter. Let's denote these by $\vect{u}$ and $\vect{v}$. 

\noindent Formally, we define $\comp$ as below: Given $\frq = (\cL\cL,\vect{x^{LL}},\vect{y^{LL}},w^{LL}_1,w^{LL}_2)\in\overline{S}$ with $\leng(\frq)>2$, then define for $\vect{u^{LL}}\in\bit^{(\leng(\frq)-1)n},\vect{v^{LL}}\in\bit^{(\leng(\frq)-1)\secp}$
\begin{align*}
    &\ket{\comp(\frq,\vect{u^{LL}},\vect{v^{LL}})} = \\ 
    &\qquad\big|\set{(x_0^{LL}||w^{LL}_1||x_1^{LL},u_1^{LL}||v_1^{LL}||y_1^{LL}), (u_2^{LL}||v_2^{LL}||x_3^{LL},u_3^{LL}||v_3^{LL}||y_3^{LL}),\\
    &\qquad\qquad\qquad\ldots,(u_{n-3}^{LL}||v_{n-3}^{LL}||x_{n-2}^{LL},u_{n-2}^{LL}||v_{n-2}^{LL}||y_{n-2}^{LL}),(u_{n-1}^{LL}||v_{n-1}^{LL}||x_{n-1}^{LL},y_0^{LL}||w^{LL}_2||y_{n-1}^{LL})}\big\rangle\\
    &\qquad\qquad\otimes \ket{\set{(u_1^{LL}||v_1^{LL}||x_2^{LL},u_2^{LL}||v_2^{LL}||y_2^{LL}),\ldots,(u_{n-2}^{LL}||v_{n-2}^{LL}||x_{n-1}^{LL},u_{n-1}^{LL}||v_{n-1}^{LL}||y_{n-1}^{LL})}}
\end{align*}

\noindent On any $\frq = (\cL\cL,\vect{x^{LL}},\vect{y^{LL}},w^{LL}_1,w^{LL}_2)\in \overline{S}$ with $\leng(\frq)=2$, $\vect{u^{LL}} = (),\vect{v^{LL}} = ()$, then define 
\begin{align*}
    \ket{\comp(\frq,\vect{u^{LL}},\vect{v^{LL}})} = \ket{\set{(x_0^{LL}||w^{LL}_1||x_1^{LL},y_0^{LL}||w^{LL}_2||y_{1}^{LL})}}\ket{\emptyset}
\end{align*}

\noindent We think of $\vect{u^{LL}}$ and $\vect{v^{LL}}$ the simulated $\reg{A}$ and $\reg{B}$. 

\noindent Similarly, define $\comp$ on $\frq$ with first element $\cL\cR,\cR\cL$ and $\cR\cR$. Next, we define an operation that takes $\overline{S}$, a set of $\vect{u}$'s and $\vect{v}$'s, and give a combined database: 

\noindent Formally, let $\overline{S}$ with $a = \coun(\overline{S})$ and $b = \leng(\overline{S})$. Let $\cU = \set{\vect{u^{i}}}_{i}\in\bit^{(b-a)n}$ and $\cV = \set{\vect{v^{i}}}_{i}\in\bit^{(b-a)\secp}$, then define 
\begin{align*}
    \ket{\F(\overline{S},\cU,\cV)} = & \ket{\bigcup_{\frq_i\in \overline{S}} \comp(\frq_i,\vect{u^{i}},\vect{v^{i}})}
\end{align*}

\noindent Finally, we define $\Ocomp$
\begin{align*}
    \Ocomp\ket{\frG(\overline{S})} = \frac{1}{\sqrt{2^{(b-a)n}\left((2^{\secp}-a)\ldots(2^{\secp}-b+1)\right)}}& \sum_{\substack{\cU\in\bit^{(b-a)n}\\ \cV\in(\bit^{\secp}\setminus\Im(\overline{S}))^{b-a}_{\dist}}}\ket{\F(\overline{S},\cU,\cV)}
\end{align*}

\noindent We define the notation $\frF$ as $$\Ocomp\ket{\frG(\overline{S})} = \ket{\frF(\overline{S})}$$

\noindent Then we prove that for each query, $\Ocomp$ simulates the database, i.e. we prove the following:
\begin{lemma}
\label{lem:gluing:fwd}
    For any integer $t \geq 0$,
    \begin{itemize}
        \item Forward query: 
        $$\|\left(\Ocomp \Pi^{\good}W^{\gluedfwd} - W^{\frm(\secp)}\Ocomp\right)\Pi^{\good}\Pi_{\leq t}\|_{\op} = O(t^2/2^{\secp})$$
        \item Inverse query:
        $$\|\left(\Ocomp \Pi^{\good}W^{\glued,\dagger} - W^{\frm(\secp),\dagger}\Ocomp\right)\Pi^{\good}\Pi_{\leq t}\|_{\op} = O(t^2/2^{\secp})$$
    \end{itemize}
\end{lemma}

\noindent We prove the above in~\Cref{sec:gluing:fwd} but before that we will finish proving~\Cref{thm:gluing:induction} in~\Cref{sec:gluing:induction} using~\Cref{lem:gluing:fwd}.

\subsection{Proof of~\Cref{thm:gluing:induction}:~Closeness between $\hyb_3$ and $\hyb_4$}
\label{sec:gluing:induction}

\noindent Denote the initial joint state in $\hyb_3$ by
\[
\ket{\psi_0} := \ket{0}_{\reg{ABC}}\ket{0}_{\reg{D}}\ket{\varnothing}_{\reg{S}}\ket{\varnothing}_{\reg{T}}.
\]
For $i \in [2t]$, denote the joint state right after the $i$-th query by
\[
\ket{\psi_i} := 
\begin{cases}
    W^{\frm(\secp)}A_{(i+1)/2}\ket{\psi_{i-1}} \quad & \text{, if $i\equiv 1\bmod{2}$} \\
    W^{\frm(\secp),\dagger}B_{i/2}\ket{\psi_{i-1}} \quad & \text{, if $i\equiv 0\bmod{2}$}.
\end{cases}
\]
The output of $\hyb_3$ is 
\[
\Tr_{\reg{ST}}\left( \ketbra{\psi_{2t}}{\psi_{2t}} \right).
\]
Similarly, denote the initial joint state in $\hyb_4$ by
\[
\ket{\phi_0} := 
\ket{0}_{\reg{ABC}} \ket{0}_{\reg{D}} \ket{\varnothing}_{\reg{S_1}} \ket{\varnothing}_{\reg{T_1}} \ket{\varnothing}_{\reg{S_2}} \ket{\varnothing}_{\reg{T_2}} \ket{\varnothing}_{\reg{S_3}} \ket{\varnothing}_{\reg{T_3}} .
\]
For $i \in [2t]$, denote the joint state right after the $i$-th query by
\[
\ket{\phi_i} := 
\begin{cases}
    \Pi^{\good}W^{\gluedfwd}A_{(i+1)/2}\ket{\phi_{i-1}} \quad & \text{, if $i\equiv 1\bmod{2}$} \\
    \Pi^{\good}W^{\gluedinv}B_{i/2}\ket{\phi_{i-1}} \quad & \text{, if $i\equiv 0\bmod{2}$}.
\end{cases}
\]
The output of $\hyb_4$ is $$\Tr_{\reg{S_1S_2S_3T_1T_2T_3}}\left(\ketbra{\phi_{2t}}{\phi_{2t}}\right).$$

\noindent We prove the following claim by induction: for $i \in [2t]$, $\|\Ocomp\ket{\phi_i}-\ket{\psi_i}\|_2 = O(i^3/2^{\secp})$.

\begin{itemize}
    \item Base case: $\Ocomp\ket{\phi_0}=\ket{\psi_0}$.
    \item Induction hypothesis: Suppose $\|\Ocomp\ket{\phi_{i-1}}-\ket{\psi_{i-1}}\|_2 = O((i-1)^3/2^{\secp})$.
\end{itemize}

\noindent Consider the following two cases: 

\paragraph{Case 1: $i$ is odd:} 

\begin{align*}
& \|\Ocomp\ket{\phi_{i}}-\ket{\psi_{i}}\|_2 \\
= & \|\Ocomp \Pi^{\good}W^{\gluedfwd}A_{(i+1)/2}\ket{\phi_{i-1}}-W^{\frm(\secp)}A_{(i+1)/2}\ket{\psi_{i-1}}\|_2
\tag{by expanding the definition of $\ket{\psi_{i}}$ and $\ket{\phi_{i}}$} \\
\leq & \|\Ocomp \Pi^{\good}W^{\gluedfwd}A_{(i+1)/2}\ket{\phi_{i-1}} \rcolor{- W^{\frm(\secp)}A_{(i+1)/2} \Ocomp\ket{\phi_{i-1}}} \|_2 \\
& + \| \rcolor{W^{\frm(\secp)} A_{(i+1)/2}\Ocomp\ket{\phi_{i-1}}} - W^{\frm(\secp)}A_{(i+1)/2}\ket{\psi_{i-1}}\|_2 
\tag{by the triangle inequality} \\
= & \|(\Ocomp \Pi^{\good}W^{\gluedfwd}A_{(i+1)/2} - W^{\frm(\secp)}\Ocomp) A_{(i+1)/2} \ket{\phi_{i-1}}\|_2
+ \|W^{\frm(\secp)}A_{(i+1)/2}(\Ocomp\ket{\phi_{i-1}} - \ket{\psi_{i-1}})\|_2 
\tag{since $\Ocomp$ and $A_{(i+1)/2}$ commute} \\
\leq & \|(\Ocomp \Pi^{\good} W^{\gluedfwd}A_{(i+1)/2} - W^{\frm(\secp)}\Ocomp)\Pi^{\good} \Pi_{\leq t}\|_{\op}
+ \norm{\Ocomp\ket{\phi_{i-1}}-\ket{\psi_{i-1}}}_2 
\tag{by~\Cref{lem:op_norm}} \\
= & O(i^2/2^{\secp}) +O((i-1)^3/2^{\secp}) .
\tag{by~\Cref{lem:gluing:fwd} and the induction hypothesis} \\
=& O(i^3/2^{\secp})
\end{align*}

\paragraph{Case 2: $i$ is even:} 

\begin{align*}
& \|\Ocomp\ket{\phi_{i}}-\ket{\psi_{i}}\|_2 \\
= & \|\Ocomp \Pi^{\good} W^{\gluedinv}B_{i/2}\ket{\phi_{i-1}}-W^{\frm(\secp),\dagger}B_{i/2}\ket{\psi_{i-1}}\|_2
\tag{by expanding the definition of $\ket{\psi_{i}}$ and $\ket{\phi_{i}}$} \\
\leq & \|\Ocomp \Pi^{\good}W^{\gluedinv}B_{i/2}\ket{\phi_{i-1}} \rcolor{- W^{\frm(\secp),\dagger}B_{i/2} \Ocomp\ket{\phi_{i-1}}} \|_2 \\
& + \| \rcolor{W^{\frm(\secp),\dagger} B_{i/2}\Ocomp\ket{\phi_{i-1}}} - W^{\frm(\secp),\dagger}B_{i/2}\ket{\psi_{i-1}}\|_2 
\tag{by the triangle inequality} \\
= & \|(\Ocomp\Pi^{\good} W^{\gluedinv}B_{i/2} - W^{\frm(\secp),\dagger}\Ocomp) B_{i/2} \ket{\phi_{i-1}}\|_2
+ \|W^{\frm(\secp),\dagger}B_{i/2}(\Ocomp\ket{\phi_{i-1}} - \ket{\psi_{i-1}})\|_2 
\tag{since $\Ocomp$ and $B_{i/2}$ commute} \\
\leq & \|(\Ocomp \Pi^{\good}W^{\gluedinv}B_{i/2} - W^{\frm(\secp),\dagger}\Ocomp)\Pi^{\good} \Pi_{\leq t}\|_{\op}
+ \norm{\Ocomp\ket{\phi_{i-1}}-\ket{\psi_{i-1}}}_2 
\tag{by~\Cref{lem:op_norm}} \\
= & O(i^2/2^{\secp}) +O((i-1)^3/2^{\secp}) .
\tag{by~\Cref{lem:gluing:fwd} and the induction hypothesis} \\
=& O(i^3/2^{\secp})
\end{align*}

\subsection{Proof of~\Cref{lem:gluing:fwd}:~Closeness of the  Oracle Queries}
\label{sec:gluing:fwd}

To prove~\Cref{lem:gluing:fwd}, we start by recalling that
\begin{align*}
    W^{\glued} = &\left(\Pi^{\cL,321}\right)\cdot V_L^{(3),\midd}\cdot V_L^{(2),\midd}\cdot V_L^{(1),\midd} \cdot\left(I-\Pi^{\cR,1}\right) \\
    &+ \left(\Pi^{\cL,32} -\Pi^{\cL,321}\right)\cdot V_L^{(3),\midd}\cdot V_L^{(2),\midd}\cdot V_R^{(1),\midd,\dagger}\cdot\left(\Pi^{\cR,1}-\Pi^{\cR,12}\right) \\
    &+ \left(\Pi^{\cL,3} -\Pi^{\cL,32}\right)\cdot V_L^{(3),\midd}\cdot V_R^{(2),\midd,\dagger}\cdot V_R^{(1),\midd,\dagger}\cdot\left(\Pi^{\cR,12}-\Pi^{\cR,123}\right) \\
    &+ \left(I-\Pi^{\cL,3}\right)\cdot V_R^{(3),\midd,\dagger}\cdot V_R^{(2),\midd,\dagger}\cdot V_R^{(1),\midd,\dagger}\cdot\left(\Pi^{\cR,123}\right) \\
\end{align*} 
And that from~\Cref{sec:struc:proj}, we know a nice basis for "good" states in the four subspaces on which we have different operations of $W^{\glued}$. Hence, we start by proving that $\Ocomp$ works in this four subspaces. Formally, we show the following lemmas:

\begin{lemma}
\label{lem:gluing:fwd_1}
For any integer $t \geq 0$,
    $$\|\left(\Ocomp \Pi^{\good} W^{\gluedfwd} - W^{\frm(\secp)}\Ocomp\right)\Pi^{\good}\Pi^{\frl,1}\Pi_{\leq t}\|_{\op} = O(t^2/2^{\secp})$$
\end{lemma}

\begin{lemma}
\label{lem:gluing:fwd_2}
For any integer $t \geq 0$,
    $$\|\left(\Ocomp \Pi^{\good} W^{\gluedfwd} - W^{\frm(\secp)}\Ocomp\right)\Pi^{\good}\Pi^{\frl,2}\Pi_{\leq t}\|_{\op} = O(t^2/2^{\secp})$$
\end{lemma}

\begin{lemma}
\label{lem:gluing:fwd_3}
For any integer $t \geq 0$,
    $$\|\left(\Ocomp \Pi^{\good} W^{\gluedfwd} - W^{\frm(\secp)}\Ocomp\right)\Pi^{\good}\Pi^{\frl,3}\Pi_{\leq t}\|_{\op} = O(t^2/2^{\secp})$$
\end{lemma}

\begin{lemma}
\label{lem:gluing:fwd_4}
For any integer $t \geq 0$,
    $$\|\left(\Ocomp \Pi^{\good} W^{\gluedfwd} - W^{\frm(\secp)}\Ocomp\right)\Pi^{\good}\Pi^{\frl,4}\Pi_{\leq t}\|_{\op} = O(t^2/2^{\secp})$$
\end{lemma}

\noindent We give proofs of the above lemmas in~\Cref{sec:app:fwd}.

\noindent We restate~\Cref{lem:gluing:fwd} for convenience.
\begin{lemma}[\Cref{lem:gluing:fwd}, restated]
For any integer $t \geq 0$,
\begin{itemize}
    \item Forward query: 
    $$\|\left(\Ocomp \Pi^{\good} W^{\gluedfwd} - W^{\frm(\secp)}\Ocomp\right)\Pi^{\good}\Pi_{\leq t}\|_{\op} = O(t^2/2^{\secp})$$
    \item Inverse query:
    $$\|\left(\Ocomp \Pi^{\good} W^{\gluedinv} - W^{\frm(\secp),\dagger}\Ocomp\right)\Pi^{\good}\Pi_{\leq t}\|_{\op} = O(t^2/2^{\secp})$$
\end{itemize}
\end{lemma}

\begin{proof}
    We prove the lemma for forward queries and we get it for inverse queries symmetrically. We want to show 
    $$\|\left(\Ocomp \Pi^{\good} W^{\gluedfwd} - W^{\frm(\secp)}\Ocomp\right)\Pi^{\good}\Pi_{\leq t}\|_{\op} = O(t^2/2^{\secp}).$$
    
    \noindent First, notice that $\sum_{i=1}^{4}\Pi^{\frl,i} = I$. Hence, we get 
    \begin{align*}
        \gamma =& \|\left(\Ocomp \Pi^{\good} W^{\gluedfwd} - W^{\frm(\secp)}\Ocomp\right)\Pi^{\good}\Pi_{\leq t}\|_{\op}\\
        =& \|\left(\Ocomp \Pi^{\good} W^{\gluedfwd} - W^{\frm(\secp)}\Ocomp\right)\Pi^{\good}\left(\sum_{i=1}^{4}\Pi^{\frl,i}\right)\Pi_{\leq t}\|_{\op}\\
        =& \|\sum_{i=1}^{4}\left(\left(\Ocomp \Pi^{\good} W^{\gluedfwd} - W^{\frm(\secp)}\Ocomp\right)\Pi^{\good}\Pi^{\frl,i}\right)\Pi_{\leq t}\|_{\op}\\
        \leq & \sum_{i=1}^{4}\|\left(\left(\Ocomp \Pi^{\good} W^{\gluedfwd} - W^{\frm(\secp)}\Ocomp\right)\Pi^{\good}\Pi^{\frl,i}\right)\Pi_{\leq t}\|_{\op}\\
        =& O(t^2/2^{\secp})
    \end{align*}
    \noindent Where the fouth line is by triangle inequality, and fifth line is by~\Cref{lem:gluing:fwd_1,lem:gluing:fwd_2,lem:gluing:fwd_3,lem:gluing:fwd_4}.
\end{proof}

\section{Stretching Strong Pseudorandom Unitaries}
\label{sec:stretch}

Now we show how to apply our results to get nearly linear depth and to stretch the length of \emph{any} strong pseudorandom unitary, relative to its key size.  

To prove that we can stretch the keys in a PRU family, we start by recalling the following result from~\cite{CRYPTO:ABGL25},
\begin{theorem}
    \label{thm:stretch_pru}
    For any $f(n) = \omega(\log n)$, $k_1,k_2,k_3\in\bit^{f(n)}$, define
    \[
    G^U(k_1||k_2||k_3) := (X^{k_3}\otimes I_{n-f(n)}) U (X^{k_2}\otimes I_{n-f(n)}) U (X^{k_1}\otimes I_{n-f(n)}),
    \]
    where $U$ is an $n$-qubit unitary. Then $\set{G_U(k_1||k_2||k_3)}_{k_1,k_2,k_3\in\bit^{f(n)}}$ is a strong PRU in the QHROM.
\end{theorem}

\noindent At a high level, from a single instance of a strong pseudorandom unitary for a key $k$, and $9$ random strings of length $O(\log^{2}(n))$, we can create three additional instances of a strong pseudorandom unitaries, that are random even relative to the original instance of the pseudorandom unitary, by applying the construction above thrice.  Then we can apply the strong gluing theorem to join these three pseudorandom unitaries into a strong pseudorandom unitary acting on a larger input. Formally, we have the following theorem.

\begin{theorem}[Stretching a strong PRU]
\label{thm:stretch}
    Let $\{\mathsf{PRU}_{\lambda, k}\}_{\lambda \in \mathbb{N}, k \in \{0, 1\}^{\lambda}}$ be a strong pseudorandom unitary family with keys of size $\lambda$ acting on $t(\lambda)$ many qubits.  Then there exists a family of strong pseudorandom unitaries $\{\mathsf{StretchPRU}_{\lambda, k \in \{0, 1\}^{\lambda + 9 \log^2(\lambda)}}\}$ with keys of length $\lambda + 9\log^2(\lambda)$ that acts on $2t(\lambda) - \log^2(\lambda)$ qubits.
\end{theorem}

\begin{proof}[Proof of~\Cref{thm:stretch}]
    Let $k_1||\ldots||k_9$ be a string of length $9\log^2(\lambda)$ where each $k_i$ is length $\log^2(\lambda)$.  Then consider the following construction of $\mathsf{StretchPRU}$ on registers $\reg{ABC}$, where $\reg{A}$ is $t(\lambda) - \log^2(\lambda)$ qubits, $\reg{B}$ is $\log^2(\lambda)$ qubits, and $\reg{C}$ is $t(\lambda) - \log^2(\lambda)$ qubits.
    \begin{multline*}
        \mathsf{StretchPRU}_{k||k_1||\ldots||k_9} = \\(X^{k_1} \mathsf{PRU}_k X^{k_2} \mathsf{PRU}_k X^{k_3})_{\reg{AB}} (X^{k_4} \mathsf{PRU}_k X^{k_5} \mathsf{PRU}_k X^{k_6})_{\reg{BC}} (X^{k_7} \mathsf{PRU}_k X^{k_8} \mathsf{PRU}_k X^{k_9})_{\reg{AB}}\,.
    \end{multline*}
    Let $\mathsf{Stretch}_{k_1||\ldots||k_9}(U)$ be the same construction, except that $\mathsf{PRU}_k$ is replaced with a unitary $U$.  Then by the definition of a strong pseudo-random unitary, we have the following for all polynomial-time adversaries $\mathcal{A}$.
    \begin{multline*}
        \Bigg|\Pr_{k||k_1||\ldots||k_9 \gets \{0, 1\}^{\lambda + 9 \log^2(\lambda)}} \left[\top \leftarrow \mathcal{A}^{\mathsf{StretchPRU}_{k||k_1||\ldots||k_9}, \mathsf{StretchPRU}_{k||k_1||\ldots||k_9}^{\dagger}}\right] \\- \Pr_{\substack{U \gets \mu_{t(\lambda)} \\ k_1||\ldots||k_9 \gets \{0, 1\}^{9\log^2(\lambda)}}}\left[\top \leftarrow \mathcal{A}^{\mathsf{Stretch}_{k_1||\ldots||k_9}(U), \left(\mathsf{Stretch}_{k_1||\ldots||k_9}(U)\right)^{\dagger}}\right]\Bigg| \leq \negl(\lambda)\,.
    \end{multline*}
    From \Cref{thm:stretch_pru}, applied three times, we have the following bound:
    \begin{multline*}
        \Bigg| \Pr_{\substack{U \gets \mu_{t(\lambda)} \\ k_1||\ldots||k_9 \gets \{0, 1\}^{9\log^2(\lambda)}}}\left[\top \leftarrow \mathcal{A}^{\mathsf{Stretch}_{k_1||\ldots||k_9}(U), \left(\mathsf{Stretch}_{k_1||\ldots||k_9}(U)\right)^{\dagger}}\right] \\- \Pr_{U', V, W \gets \mu_{t(\lambda)}}\left[\top \leftarrow \mathcal{A}^{U'VW, (U'VW)^{\dagger}}\right]\Bigg| \leq \negl(\lambda)\,.
    \end{multline*}

    \noindent Finally, applying \Cref{thm:gluing}, we have the following:
    \begin{equation*}
        \Bigg|\Pr_{U', V, W \gets \mu_{t(\lambda)}}\left[\top \leftarrow \mathcal{A}^{U'VW, (U'VW)^{\dagger}}\right] - \Pr_{O \gets \mu_{2t(\lambda) - \log^2(\lambda)}}\left[\top \leftarrow \mathcal{A}^{O, O^{\dagger}}\right] \Bigg| \leq \negl(\lambda)\,.
    \end{equation*}
    Applying the triangle inequality, the construction of $\mathsf{StretchPRU}$ is indistinguishable from a large Haar random unitary on $2t(\lambda) - \log^2(\lambda)$ qubits.  
\end{proof}

\begin{corollary}[Strong pseudorandom unitaries with small keys]
\label{thm:short:keys}
    If there exists a family of strong pseudorandom unitaries, then for every constant $c$ there exists a strong pseudorandom unitary family such that
    \begin{enumerate}
        \item The key size is $\lambda + 9c \log^3(\lambda)$.
        \item The pseudorandom unitary family acts on $\lambda^c (t(\lambda) - \log^2(\lambda)) + \log^2(\lambda)$ qubits.
    \end{enumerate}
\end{corollary}

\begin{proof}[Proof of~\Cref{thm:short:keys}]
    We recursively apply the previous theorem $c\log(\lambda)$ many times.  Each time, we need $9\log^2(\lambda)$ additional bits of randomness, and we double (minus $9\log^2(\lambda)$) the output length of the strong pseudorandom unitary.  Thus, after performing this transformation recursively $n$ times, our output length is
    \begin{equation*}
        2^n t(\lambda) - 2^{n-1} \cdot 9\log^2(\lambda) - 2^{n-2} \cdot 9\log^2(\lambda) - \ldots - 9\log^2(\lambda) = 2^{n} (t(\lambda) - 9\log^2(\lambda))\,.
    \end{equation*}
    This setting $n = c\log(\lambda)$, we get the desired key length and output length.  Note that this requires running the original strong pseudorandom unitary $O(\lambda^c)$ times, which is polynomial in $\lambda$ for constant $c$.
\end{proof}

Rescaling so that $\lambda \gets \lambda + 9c\log^3(\lambda)$, we have that there is a family of strong pseudo-random unitaries with keys of length $\lambda$ and output size roughly $\lambda^c$.

Next, we prove that our strong gluing theorem implies that strong PRUs exist at near linear depth.
\begin{corollary}[Shortening a super-linear depth PRU]
    If there exists a family of strong pseudorandom unitaries, then for every constant $c$ there exists a family of strong pseudorandom unitaries with depth $O(n^{1 + 1/c})$.  
\end{corollary}
\begin{proof}
    Let $\cG = \set{\cG^{n}}_{n\in\N}$ denote a strong pseudorandom unitary family with $\cG^{n}$ denoting the unitaries with input length $n$. Let $\cK^{n}$ denote the set of keys assoicated with $\cG^{n}$. By the definition of strong pseudorandom unitaries, the depth of any circuit in $\cG^{n}$ is asymptotically bounded by a polynomial in $n$, the input size. Let $c_1$ be a constant such that the depth of the family of strong pseudorandom unitaries is asymptotically bounded by $O(n^{c_1})$. 

    Then circuits in $\cG^{n^{1 / (c_1 \cdot c)}}$ are strong PRUs on input length $n^{1 / (c_1 \cdot c)}$ qubits, whose depth is bounded by $O(n^{1/c})$. Then for any $t\in\poly(n)$, we sample some $k_1,\ldots,k_{2t+1}\gets\cK^{n^{1 / (c_1 \cdot c)}}$. We arrange $U^{k_1},\ldots,U^{k_{2t+1}}$ in the following circuit:

    
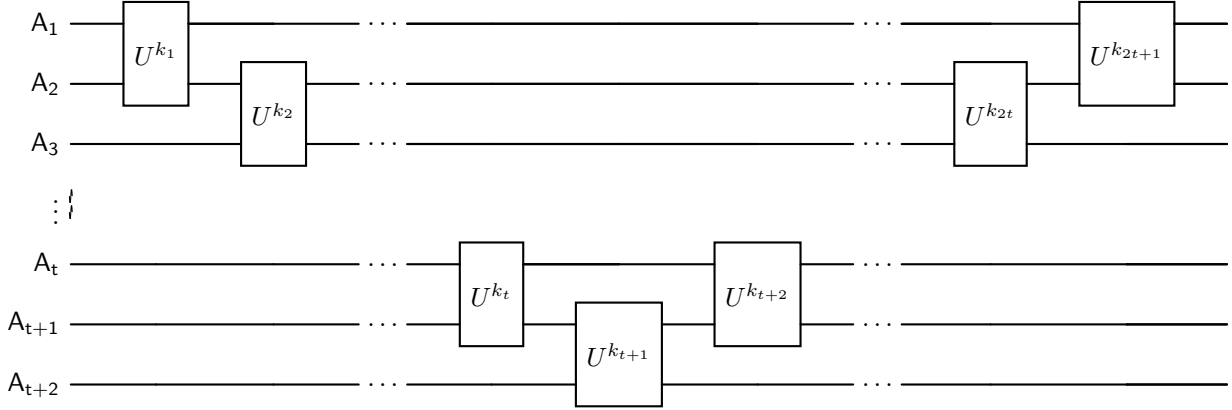
\begin{quantikz}[transparent, column sep=0.7cm,row sep={0.8cm,between origins}]
    \lstick{$\reg{A_1}$} & \gate[wires=2]{U^{k_1}} & \qw & \ \ldots\ & \qw & \qw & \qw & \ \ldots\ & \qw & \gate[wires=2]{U^{k_{2t+1}}} & \qw\\
    \lstick{$\reg{A_2}$} &  & \gate[wires=2]{U^{k_2}} & \ \ldots\ & \qw & \qw & \qw & \ \ldots\ & \gate[wires=2]{U^{k_{2t}}} &  & \qw\\
    \lstick{$\reg{A_3}$} & & & \ \ldots\ & \qw & \qw & \qw & \ \ldots\ & & & \qw\\
    \lstick{$\vdots$}\wave&&&&&&&&&&\\ 
    \lstick{$\reg{A_{t}}$} &  &  & \ \ldots\ & \gate[wires=2]{U^{k_{t}}} & \qw & \gate[wires=2]{U^{k_{t+2}}} & \ \ldots\ &  &  & \qw\\
    \lstick{$\reg{A_{t+1}}$} &  &  & \ \ldots\ & & \gate[wires=2]{U^{k_{t+1}}} & & \ \ldots\ &  &  & \qw\\
    \lstick{$\reg{A_{t+2}}$} &  &  & \ \ldots\ & & & & \ \ldots\ &  &  & \qw\\
    \end{quantikz}

    In the above, $|A_i| = n^{1/(c_1 \cdot c)} / 2$. Notice that the above circuit is on input size $(t + 2) \cdot n^{1/(c_1 \cdot c)} /2$ that has depth $(2t+1) \cdot O(n^{1/c})$. Let $t = O(n)$, we have a family of circuits on input length $O(n)$ that has depth $O(n^{1 + 1/c})$.
    
    We prove sampling $k_1,\ldots,k_{2t+1}\gets\cK^{n^{1 / (c_1 \cdot c)}}$ and arranging as above gives us a family of strong PRUs with input length $O(n)$ that has depth $O(n^{1 + 1/c})$. To prove that the above circuit is a PRU, we start by applying the strong gluing theorem on the middle three unitaries (i.e. $U^{k_{t}}U^{k_{t+1}}U^{k_{t+2}}$) and replacing it with a larger Haar unitary (say $V_1$). Next, we apply the strong gluing theorem on the new middle three unitaries (i.e. $U^{k_{t-1}}V_1U^{k_{t+3}}$) and replacing it with a larger Haar unitary (say $V_2$). Repeating this process a total of $t$ times gives us a single large Haar unitary. 
\end{proof}

\section*{Acknowledgments}
PA, AG and YTL are supported by the National Science Foundation under the grants FET-2329938, CAREER-2341004 and, FET-2530160.

\printbibliography
\newpage
\appendix

\section{Proof of~\Cref{lem:PR:gen}}
\label{sec:app:PR:gen}

We start by proving that $V_L^{f_L}$ is close to $V_L$ and $V_R^{f_R}$ is close to $V_R$. 

\begin{lemma}   \label{lem:FL_VL_FR_VR_close}
For any integer $t \geq 0$,
\[
\|(V_L-V_L^{f_L}) \Pi_{\leq t}\|_{\op} \leq \sqrt{2\cdot(t+1)\cdot\delta}
\quad \text{and} \quad
\|(V_R-V_R^{f_R}) \Pi_{\leq t}\|_{\op} \leq \sqrt{2\cdot(t+1)\cdot\delta}.
\]
\end{lemma}
\begin{proof}
Consider an arbitrary (normalized) state in the support of $\Pi_{\leq t}$
\[
\ket{\psi}_{\reg{ABST}} = \sum_{x,z,L,R} \alpha_{x,z,L,R} \ket{x}_{\reg{A}} \ket{z}_{\reg{B}} \ket{L}_{\reg{S}} \ket{R}_{\reg{T}},
\]
where $\alpha_{x,z,L,R} = 0$ whenever $|L \cup R| > t$. Then 
\[
V_L \ket{\psi}_{\reg{ABST}} = \sum_{x,z,L,R} \frac{\alpha_{x,z,L,R}}{\sqrt{2^{n}-|\Im(L\cup R^{-1})|}} \sum_{y \notin \Im(L\cup R^{-1})} \ket{y}_{\reg{A}} \ket{z}_{\reg{B}} \ket{L \cup \set{(x,y)}}_{\reg{S}} \ket{R}_{\reg{T}},
\]
and 
\[
V_L^{f_L} \ket{\psi}_{\reg{ABST}} = \sum_{x,z,L,R} \frac{\alpha_{x,z,L,R}}{\sqrt{|f_L(x,z,L,R)|}} \sum_{y\in f_L(x,z,L,R)} \ket{y}_{\reg{A}}\ket{z}_{\reg{B}} \ket{L \cup \set{(x,y)}}_{\reg{S}} \ket{R}_{\reg{T}}.
\]

\noindent Subtracting, 
\begin{align*}
(V_L & - V_L^{f_L}) \ket{\psi}_{\reg{ABST}} \\
& = \sum_{x,z,L,R} \alpha_{x,z,L,R} \sum_{\substack{y: y\in f_L(x,z,L,R)\\ y\not\in\Im(L\cup R^{-1})}} \ket{y}_{\reg{A}} \ket{z}_{\reg{B}} \ket{L \cup \set{(x,y)}}_{\reg{S}} \ket{R}_{\reg{T}} \\
&\underbrace{\qquad\qquad\qquad\times\left(\frac{1}{\sqrt{2^{n}-|\Im(L\cup R^{-1})|}}-\frac{1}{\sqrt{|f_L(x,z,L,R)|}}\right)}_{\ket{v}} \\
& + \underbrace{\sum_{x,z,L,R} \alpha_{x,z,L,R} \sum_{\substack{y: y\not\in f_L(x,z,L,R)\\ y\not\in\Im(L\cup R^{-1})}} \ket{y}_{\reg{A}} \ket{z}_{\reg{B}} \ket{L \cup \set{(x,y)}}_{\reg{S}} \ket{R}_{\reg{T}} \left(\frac{1}{\sqrt{2^{n}-|\Im(L\cup R^{-1})|}}\right)}_{\ket{w_1}}\\
& + \underbrace{\sum_{x,z,L,R} \alpha_{x,z,L,R} \sum_{\substack{y: y\in f_L(x,z,L,R)\\ y\in\Im(L\cup R^{-1})}} \ket{y}_{\reg{A}} \ket{z}_{\reg{B}} \ket{L \cup \set{(x,y)}}_{\reg{S}} \ket{R}_{\reg{T}} \left(-\frac{1}{\sqrt{|f_L(x,z,L,R)|}}\right)}_{\ket{w_2}}.
\end{align*}
Note that $\ket{w_1}$, $\ket{w_2}$ and $\ket{v}$ are orthogonal. Thus,
\[
\norm{(V_L - V_L^{f_L})\ket{\psi}_{\reg{ABST}}}^2_2 = \braket{v}{v} + \braket{w_1}{w_1}  + \braket{w_2}{w_2} 
\]

\myparagraph{Bounding $\braket{v}{v}$} Similar to~\cite{MH24}, by changing the order of summation, we can rewrite $\ket{v}$ as
    \begin{align*}
        \ket{v} = \sum_{\substack{y,z,L',R}} \ket{y}\ket{z} \ket{L'} \ket{R} \left( \sum_{\substack{(x,L): \\ L' = L \cup \{(x,y)\},\\ y\in f_L(x,z,L,R)\\ y\not\in\Im(L\cup R^{-1})}} \alpha_{x,z,L,R} \left(\frac{1}{\sqrt{2^{n}-|\Im(L\cup R^{-1})|}}-\frac{1}{\sqrt{|f_L(x,z,L,R)|}}\right) \right),
    \end{align*}
    and thus
    \begin{align*}
        \braket{v}{v} = \sum_{\substack{y,z,L',R}}& \left( \sum_{\substack{(x,L): \\ L' = L \cup \{(x,y)\},\\ y\in f_L(x,z,L,R)\\ y\not\in\Im(L\cup R^{-1})}} \alpha_{x,z,L,R} \left(\frac{1}{\sqrt{2^{n}-|\Im(L\cup R^{-1})|}}-\frac{1}{\sqrt{|f_L(x,z,L,R)|}}\right) \right)^2\\
        \leq \sum_{\substack{y,z,L',R}}& \left( \sum_{\substack{(x,L): \\ L' = L \cup \{(x,y)\},\\ y\in f_L(x,z,L,R)\\ y\not\in\Im(L\cup R^{-1})}} |\alpha_{x,z,L,R}|^2 \right) \\
        &\times \left( \sum_{\substack{(x,L): \\ L' = L \cup \{(x,y)\},\\ y\in f_L(x,z,L,R)\\ y\not\in\Im(L\cup R^{-1})}} \left(\frac{1}{\sqrt{2^{n}-|\Im(L\cup R^{-1})|}}-\frac{1}{\sqrt{|f_L(x,z,L,R)|}}\right)^2 \right),
    \end{align*}
    where the last inequality is by Cauchy-Schwarz. We can bound the summand by writing
    \begin{align*}
         \sum_{\substack{(x,L): \\ L' = L \cup \{(x,y)\},\\ y\in f_L(x,z,L,R)\\ y\not\in\Im(L\cup R^{-1})}} &\left(\frac{1}{\sqrt{2^{n}-|\Im(L\cup R^{-1})|}}-\frac{1}{\sqrt{|f_L(x,z,L,R)|}}\right)^2 \\
        &= \sum_{\substack{(x,L): \\ L' = L \cup \{(x,y)\},\\ y\in f_L(x,z,L,R)\\ y\not\in\Im(L\cup R^{-1})}} \left(\frac{\sqrt{|f_L(x,z,L,R)|} - \sqrt{2^{n}-|\Im(L\cup R^{-1})|}}{\sqrt{|f_L(x,z,L,R)|(2^{n}-|\Im(L\cup R^{-1})|)}}\right)^2\\
        & \leq \sum_{\substack{(x,L): \\ L' = L \cup \{(x,y)\},\\ y\in f_L(x,z,L,R)\\ y\not\in\Im(L\cup R^{-1})}} \left(\frac{\sqrt{\left|2^{n}-|\Im(L\cup R^{-1})|-|f_L(x,z,L,R)|\right|}}{\sqrt{|f_L(x,z,L,R)|(2^{n}-|\Im(L\cup R^{-1})|)}}\right)^2 \tag{since $\sqrt{a} - \sqrt{b} \leq \sqrt{a-b}$ when $a\geq b \geq 0$}\\
        & \leq \frac{(\abs{L} + 1) \cdot \left|2^{n}-|\Im(L\cup R^{-1})|-|f_L(x,z,L,R)|\right|}{|f_L(x,z,L,R)|(2^{n}-|\Im(L\cup R^{-1})|)} \\
        & \leq \delta\cdot\frac{(\abs{L} + 1)}{(2^{n}-|\Im(L\cup R^{-1})|)} 
    \end{align*}
    where the last inequality uses the fact that for any fixed $L'$, there are at most $\abs{L} +1$ choices of $(x,L)$ that can satisfy $L' = L \cup \{(x,y)\}$. Thus,
    \begin{align*}
        \braket{v}{v} &\leq \delta\cdot\frac{(\abs{L} + 1)}{(2^{n}-|\Im(L\cup R^{-1})|)} \cdot \sum_{\substack{y,z,L',R}} \left(\sum_{\substack{(x,L): \\ L' = L \cup \{(x,y)\},\\ y\in f_L(x,z,L,R)\\ y\not\in\Im(L\cup R^{-1})}} \abs{\alpha_{x,z,L,R}}^2 \right) \\
        &=\delta\cdot\frac{(\abs{L} + 1)}{(2^{n}-|\Im(L\cup R^{-1})|)} \cdot \sum_{\substack{x,z,L,R}} \abs{\alpha_{x,z,L,R}}^2 \cdot \left( \sum_{y \in \bit^{2n+\secp}} \indic(y\in f_L(x,z,L,R))\indic(y\notin \Im(L\cup R^{-1})) \right) \\
        &\leq \delta\cdot (\abs{L} + 1) \cdot \sum_{\substack{x,z,L,R}} \abs{\alpha_{x,z,L,R}}^2 = \delta\cdot (\abs{L} + 1).
    \end{align*}

\myparagraph{Bounding $\braket{w_1}{w_1}$} We know that 
    \begin{align*}
        \ket{w_1} =& \sum_{x,z,L,R} \alpha_{x,z,L,R} \sum_{\substack{y: y\not\in f_L(x,z,L,R)\\ y\not\in\Im(L\cup R^{-1})}} \ket{y}_{\reg{A}} \ket{z}_{\reg{B}} \ket{L \cup \set{(x,y)}}_{\reg{S}} \ket{R}_{\reg{T}} \left(\frac{1}{\sqrt{2^{n}-|\Im(L\cup R^{-1})|}}\right) \\
        =& \sum_{y,z,(L',R)}\ket{y}\ket{z}\ket{L'}\ket{R}\sum_{\substack{(x,L): \\ L' = L \cup \{(x,y)\},\\ y\not\in f_L(x,z,L,R)\\ y\not\in\Im(L\cup R^{-1})}} \left(\frac{\alpha_{x,z,L,R}}{\sqrt{2^{n}-|\Im(L\cup R^{-1})|}}\right)
    \end{align*}
    \noindent Then 
    \begin{align*}
        \braket{w_1}{w_1} &= \sum_{y,z,(L',R)}\left|\sum_{\substack{(x,L): \\ L' = L \cup \{(x,y)\},\\ y\not\in f_L(x,z,L,R)\\ y\not\in\Im(L\cup R^{-1})}} \frac{\alpha_{x,z,L,R}}{\sqrt{2^{n}-|\Im(L\cup R^{-1})|}}\right|^{2} \leq \sum_{y,z,(L',R)}\sum_{\substack{(x,L): \\ L' = L \cup \{(x,y)\},\\ y\not\in f_L(x,z,L,R)\\ y\not\in\Im(L\cup R^{-1})}} \frac{\left|\alpha_{x,z,L,R}\right|^{2}}{2^{n}-|\Im(L\cup R^{-1})|} \\
        &= \sum_{x,z,L,R} \frac{\left|\alpha_{x,z,L,R}\right|^{2}}{2^{n}-|\Im(L\cup R^{-1})|}\left(\sum_{\substack{y:y\not\in f_L(x,z,L,R)\\ y\not\in\Im(L\cup R^{-1})}}1\right) \leq \sum_{x,z,L,R}\frac{(2^n-|f_L(x,z,L,R)|)\left|\alpha_{x,z,L,R}\right|^{2}}{2^{n}-|\Im(L\cup R^{-1})|} \\
        &\leq \frac{\delta}{\delta + 1} \sum_{x,z,L,R} \left|\alpha_{x,z,L,R}\right|^{2} = \frac{\delta}{\delta + 1}\leq \delta
    \end{align*}

\noindent Similarly, we also have $$\braket{w_2}{w_2} \leq \sum_{x,z,L,R}\frac{t\left|\alpha_{x,z,L,R}\right|^{2}}{|f_L(x,z,L,R)|}\leq t\cdot\delta.$$

\noindent Hence, it holds that
\[
\|(V_L - W^{\frm(\secp)}_L) \Pi_{\leq t}\|_{\op}
\leq \sqrt{2\cdot(t+1)\cdot\delta}.
\]
By a symmetric argument, we have
\[
\|(V_R - W^{\frm(\secp)}_R) \Pi_{\leq t}\|_{\op}
\leq \sqrt{2\cdot(t+1)\cdot\delta}. \qedhere
\]
\end{proof}

\noindent Then by applying triangle inequality multiple times, we have:
\begin{lemma}   \label{lem:F_V_close}
For any integer $t \geq 0$,
\[
\|(V-V^{f_L,f_R}) \Pi_{\leq t}\|_{\op} \leq 8\sqrt{2\cdot(t+1)\cdot\delta}.
\]
\end{lemma}

\noindent Finally, by bounding each query distance by $\delta$, we get the following:
\begin{lemma}[Generalized Path-Recording (restated)]
    Let $f_L$ and $f_R$ be functions such that for all $x,z,L,R$, $|L|+|R|\leq t$, 
    \begin{align*}
        f_L(x,z,L,R)&\subseteq [N]\setminus\Im(L)\\
        f_R(x,z,L,R)&\subseteq [N]\setminus\Im(R)\\
        \frac{\left|N-|f_L(x,z,L,R)|-t\right|}{|f_L(x,z,L,R)|}&\leq \delta\\
        \frac{\left|N-|f_R(x,z,L,R)|-t\right|}{|f_R(x,z,L,R)|}&\leq \delta,
    \end{align*}
    For any $2t$-query algorithm $\Adversary=\left(A_1,B_1,\ldots,A_t,B_t \right)$, 
    \[
    \TD \qty( \Tr_{\reg{ST}} \left( \ketbra{\Adversary_t^{V^{f_L,f_R},V^{f_L,f_R,\dagger}}}{\Adversary_t^{V^{f_L,f_R},V^{f_L,f_R,\dagger}}}\right), \Tr_{\reg{ST}} \left( \ketbra{\Adversary_t^{V,V^{\dagger}}}{\Adversary_t^{V,V^{\dagger}}}\right) ) 
    \leq 16t\sqrt{2\cdot (t+1)\cdot \delta},
    \]
    where,
    \begin{align*}
        \ket{\Adversary_t^{V^{f_L,f_R},V^{f_L,f_R,\dagger}}} &= \prod_{i=1}^{t}\left( V^{f_L,f_R,\dagger} B_i V^{f_L,f_R} A_i\right) \ket{0}_{\reg{A}} \ket{0}_{\reg{B}} \ket{\varnothing}_{\reg{S}} \ket{\varnothing}_{\reg{T}}\\
        \ket{\Adversary_t^{V,V^{\dagger}}} &= \prod_{i=1}^{t}\left( V^{\dagger} B_i V A_i\right) \ket{0}_{\reg{A}} \ket{0}_{\reg{B}} \ket{\varnothing}_{\reg{S}} \ket{\varnothing}_{\reg{T}}
    \end{align*}
\end{lemma}

\section{Glued Path Recording}
\label{sec:app:glued}
\noindent We want to show that querying $V^{3}V^{2}V^{1}$ is close to querying $V^{\glued}$. We start by formalising the intuition we had about monogamy of entanglement by proving the following lemma:

\begin{lemma}
\label{lem:V1V2dagger}
    Let $V^1_L$ be such that it acts on $\reg{ABS_1T_1}$ and $V_{R}^2$ be such that it acts on $\reg{BCS_2T_2}$, then $$\|V_L^{1,\dagger}V_{R}^{2}\Pi_{\leq t}\|_{\op}\leq O\left(\frac{t^2}{2^{\secp}}\right).$$
\end{lemma}

\begin{proof}
    Let $$\ket{\psi} = \sum_{x,L_1,L_2,R_1,R_2}\alpha_{xL_1L_2R_1R_2} \ket{x}_{\reg{ABC}}\ket{L_1}_{\reg{S}_1}\ket{R_1}_{\reg{T}_1}\ket{L_2}_{\reg{S}_2}\ket{R_2}_{\reg{T}_2}.$$

    \noindent Then let $\ket{\chi} = V_L^{1,\dagger}V_{R}^{2}\ket{\psi}$
    \begin{align*}
        \ket{\chi} &= \sum_{y,L_1,L_2,R_1,R_2}\alpha_{yL_1L_2R_1R_2} V_L^{1,\dagger}V_{R}^{2}\ket{y}_{\reg{ABC}}\ket{L_1}_{\reg{S}_1}\ket{R_1}_{\reg{T}_1}\ket{L_2}_{\reg{S}_2}\ket{R_2}_{\reg{T}_2} \\
        &= \frac{1}{\sqrt{2^{n+\secp}}} \sum_{\substack{y,L_1,L_2,R_1,R_2\\ x\not\in\Dom(L\cup R^{-1})}}\alpha_{yL_1L_2R_1R_2} V_L^{1,\dagger}\ket{y^{\frl(n)}}_{\reg{A}}\ket{x}_{\reg{BC}}\ket{L_1}_{\reg{S}_1}\ket{R_1}_{\reg{T}_1}\ket{L_2}_{\reg{S}_2}\ket{R_2\cup\set{(y^{\frr(n+\secp)},x)}}_{\reg{T}_2} \\
        &= \frac{1}{2^{n+\secp}} \sum_{\substack{y,L_1,L_2,R_1,R_2\\ x\not\in\Dom(L_2\cup R_2^{-1})\\ z: (z,y^{\frl(n)}||x^{\frl(\secp)})\in L_1}}\alpha_{yL_1L_2R_1R_2} \ket{z}_{\reg{AB}}\ket{x^{\frr(n)}}_{\reg{C}}\ket{L_1\setminus\set{(z,y^{\frl(n)}||x^{\frl(\secp)})}}_{\reg{S}_1}\ket{R_1}_{\reg{T}_1}\ket{L_2}_{\reg{S}_2}\ket{R_2\cup\set{(y^{\frr(n+\secp)},x)}}_{\reg{T}_2} \\
        &= \frac{1}{2^{n+\secp}} \sum_{\substack{y,L_1,L_2,R_1,R_2\\ x\not\in\Dom(L_2\cup R_2)\\ z: (z,y^{\frl(n)}||x^{\frl(\secp)})\in L_1}}\alpha_{yL_1L_2R_1R_2} \ket{z}_{\reg{AB}}\ket{x^{\frr(n)}}_{\reg{C}}\ket{L_1\setminus\set{(z,y^{\frl(n)}||x^{\frl(\secp)})}}_{\reg{S}_1}\ket{R_1}_{\reg{T}_1}\ket{L_2}_{\reg{S}_2}\ket{R_2\cup\set{(y^{\frr(n+\secp)},x)}}_{\reg{T}_2} \\
        &= \frac{1}{2^{n+\secp}} \sum_{\substack{x,y,z,L'_1,L_2,R_1,R'_2}}\alpha_{yL'_1\cup\set{(z,y^{\frl(n)}||x^{\frl(\secp)})}L_2R_1R'_2\setminus\set{(y^{\frr(n+\secp)},x)}} \ket{z}_{\reg{AB}}\ket{x^{\frr(n)}}_{\reg{C}}\ket{L'_1}_{\reg{S}_1}\ket{R_1}_{\reg{T}_1}\ket{L_2}_{\reg{S}_2}\ket{R'_2}_{\reg{T}_2} \\
    \end{align*}

    \noindent Then $\|\ket{\chi}\|^2_2 = \braket{\chi}{\chi}$,
    \begin{align*}
        \|\ket{\chi}\|^2_2 &= \frac{1}{2^{2(n+\secp)}}\sum_{z,x^{\frr(n)},L'_1,R_1,L_2,R'_2} \left|\sum_{y^{\frl(n)},y^{\frr(n+\secp)},x^{\frl(\secp)}}\alpha_{yL'_1\cup\set{(z,y^{\frl(n)}||x^{\frl(\secp)})}L_2R_1R'_2\setminus\set{(x,y^{\frr(n+\secp)})}}\right|^2 \\
        &\leq \frac{t^2 2^{2n}}{2^{2(n+\secp)}} \sum_{z,x^{\frr(n)},L'_1,R_1,L_2,R'_2} \sum_{y^{\frl(n)},y^{\frr(n+\secp)},x^{\frl(\secp)}}\left|\alpha_{yL'_1\cup\set{(z,y^{\frl(n)}||x^{\frl(\secp)})}L_2R_1R'_2\setminus\set{(x,y^{\frr(n+\secp)})}}\right|^2 \\
        &\leq \frac{t^4 }{2^{2\secp}}
    \end{align*}

    \noindent Hence $$\|V_L^{1,\dagger}V_{R}^{2}\Pi_{\leq t}\|_{\op}\leq \frac{t^2}{2^{\secp}}.$$
    
\end{proof}

\begin{lemma}
\label{lem:glued:PR:close}
    Let $V^{\glued}$ be defined as before, then $$\|(V^{\glued} - V^{3}V^{2}V^{1})\Pi_{\leq t}\|_{\op}\leq O\left(\frac{t^2}{2^{\secp}}\right).$$
\end{lemma}

\noindent The above lemma gives us the following:

\begin{lemma}[\Cref{lem:glued:PR} restated]
    For any $2t$-query algorithm $\Adversary=\left(A_1,B_1,\ldots,A_t,B_t \right)$, 
    \[
    \norm{\ket{\Adversary^{V^{\glued}, V^{\glued,\dagger}}}_{\reg{ABCD\overline{ST}}} 
    - \ket{\Adversary^{V^{3}V^{2}V^{1}, (V^{3}V^{2}V^{1})^{\dagger}}}_{\reg{ABCD\overline{ST}}}}_2
    = O \qty(\frac{t^3}{2^{\secp}}).
    \]
    where $\reg{\overline{ST}} = \reg{S_1S_2S_3T_1T_2T_3}$, $|\reg{ABC}| = 2n+\secp$, $$\ket{\Adversary_t^{V^{\glued},V^{\glued,\dagger}}} = \prod_{i=1}^{t}\left( V^{\glued,\dagger} B_i V^{\glued} A_i\right) \ket{0}_{\reg{ABC}} \ket{0}_{\reg{D}} \ket{\varnothing}_{\reg{S_1}} \ket{\varnothing}_{\reg{T_1}}\ket{\varnothing}_{\reg{S_2}} \ket{\varnothing}_{\reg{T_2}}\ket{\varnothing}_{\reg{S_3}} \ket{\varnothing}_{\reg{T_3}}$$ and $\ket{\Adversary^{V^{3}V^{2}V^{1}, (V^{3}V^{2}V^{1})^{\dagger}}}_{\reg{ABCD\overline{ST}}}$ is defined similarly.
\end{lemma}

\section{Proofs of~\Cref{sec:struc:proj}}
\label{sec:app:proj}

\begin{proof}[Proof of~\Cref{lem:proj:com1}]
    Fix some $y,w,\overline{S}$ where $\overline{S}$ is good and $a = \coun(\overline{S})$ and $b = \leng(\overline{S})$. We start by looking at what $$\ket{\psi_{y,w,\overline{S}}} = \Pi^{\cR,1} \ket{y}_{\reg{A}}\ket{w}_{\reg{B}}\ket{\frG(\overline{S})}_{\reg{\overline{ST}}},$$
    \noindent Then we have the following:
    \begin{align*}
        \ket{\psi_{y,w,\overline{S}}} =& \Pi^{\cR,1}\ket{y}_{\reg{A}}\ket{w}_{\reg{B}}\ket{\frG(\overline{S})}_{\reg{\overline{ST}}} \\
        =& \frac{1}{\sqrt{2^{an}\cdot\Pi_{i=1}^{b}(2^\secp-i+1) }}\sum_{\substack{Z\in\bit^{an}\\ R\in\left(\bit^{\secp}\right)^{b}_{\dist}}} \Pi^{\cR,1}  \ket{y}_{\reg{A}}\ket{w}_{\reg{B}}\ket{\G \left(\overline{S},R,Z\right)}_{\reg{\overline{ST}}}
    \end{align*}

    \noindent Notice that the above is zero if there's no line in $\G \left(\overline{S},R,Z\right)$ of the form $(X\cR,\vect{x},y||\vect{y},w_1,w,\vect{r},z)$ for some $X\in\set{\cL,\cR}$ and $\vect{x},\vect{y},w_1,w_2,\vect{r},z$. That is that $\overline{S} = \overline{S'}\cup\set{(X\cR,\vect{x},y||\vect{y},w_1,w)}$.Then the above looks like:
    \begin{align*}
        \ket{\psi_{y,w,\overline{S}}} =& \frac{1}{\sqrt{2^{an}\cdot\Pi_{i=1}^{b}(2^\secp-i+1) }}\sum_{\substack{Z'\in\bit^{(a-1)n}\\ R'\in\left(\bit^{\secp})\cup\set{w}\right)^{b-|\vect{x}|}_{\dist}\\ z\in\set{0,1}^n\\ \vect{r}\in\left(\bit^{\secp}\setminus R'\right)^{|\vect{x}|}_{\dist}}} \Pi^{\cR,1} \ket{y}_{\reg{A}}\ket{w}_{\reg{B}}\\
        &\times\ket{\G \left(\overline{S'},R',Z'\right)\cup\set{(X\cR,\vect{x},y||\vect{y},w_1,w,\vect{r},z)}}_{\reg{\overline{ST}}}\\
        =& \frac{1}{2^n(2^\secp-a+1)\sqrt{2^{an}\cdot\Pi_{i=1}^{b}(2^\secp-i+1) }}\sum_{\substack{Z'\in\bit^{(a-1)n}\\ R'\in\left(\bit^{\secp}\right)^{b-|\vect{x}|}_{\dist}\\ z\in\set{0,1}^n\\ \vect{r}\in\left(\bit^{\secp}\setminus R'\right)^{|\vect{x}|}_{\dist}\\ y'\in\set{0,1}^n\\ w'\in\left(\bit^{\secp}\setminus\Im(\overline{S'})\right)}} \ket{y'}_{\reg{A}}\ket{w'}_{\reg{B}}\\
        &\times \ket{\G \left(\overline{S'},R',Z'\right)\cup\set{(X\cR,\vect{x},y'||\vect{y},w_1,w',\vect{r},z)}}_{\reg{\overline{ST}}}\\
        =& \frac{1}{2^n(2^\secp-a+1)\sqrt{2^{an}\cdot\Pi_{i=1}^{b}(2^\secp-i+1) }}\sum_{\substack{y'\in\set{0,1}^n\\ w'\in\left(\bit^{\secp}\setminus\Im(\overline{S'})\right)\\ Z\in\bit^{an}\\ R\in\left(\bit^{\secp}\right)^{b}_{\dist}}} \ket{y'}_{\reg{A}}\ket{w'}_{\reg{B}}\\
        &\times \ket{\G \left(\overline{S'}\cup\set{(X\cR,\vect{x},y'||\vect{y},w_1,w')},R,Z\right)}_{\reg{\overline{ST}}}\\
        =& \frac{1}{\sqrt{2^{n}(2^\secp-a+1)}}\underbrace{\frac{1}{\sqrt{2^{n}(2^\secp-a+1)}}\sum_{\substack{y'\in\set{0,1}^n\\ w'\in\left(\bit^{\secp}\setminus\Im(\overline{S'})\right)}} \ket{y',w',\frG(\overline{S'}\cup\set{(X\cR,\vect{x},y'||\vect{y},w_1,w')})}_{\reg{AB\overline{ST}}}}_{\ket{\chi^{\frl,1}_{\overline{S'},X,\vect{x},\vect{y},w_1}}}\\
    \end{align*}

    \noindent Finally, to understand $\Pi^{\cR,1}\Pi^{\good}$, we expanding the projector $\Pi^{\good}$: 
    \begin{align*}
        \Pi^{\cR,1}\Pi^{\good} =& \Pi^{\cR,1}\sum_{\substack{y,w,\overline{S}\\ \overline{S}\text{ is good}}} \ketbra{y,w,\frG(\overline{S})}{y,w,\frG(\overline{S})}_{\reg{AB\overline{ST}}}\\
        =& \sum_{\substack{y,w,\overline{S}\\ \overline{S}\text{ is good}}}  \Pi^{\cR,1}\ketbra{y,w,\frG(\overline{S})}{y,w,\frG(\overline{S})}_{\reg{AB\overline{ST}}}\\
        =& \sum_{\substack{y,w\\ X,\vect{x},\vect{y},w_1,
        \overline{S'}\\ w\in\bit^{\secp}\setminus\Im(\overline{S'})}}  \Pi^{\cR,1} \ketbra{y,w,\frG(\overline{S'}\cup\set{(X\cR,\vect{x},y||\vect{y},w_1,w)})}{y,w,\frG(\overline{S'}\cup\set{(X\cR,\vect{x},y||\vect{y},w_1,w)})}_{\reg{AB\overline{ST}}}\\
        =& \sum_{\substack{y,w\\X,\vect{x},\vect{y},w_1,
        \overline{S'}\\ w\in\bit^{\secp}\setminus\Im(\overline{S'})}}\frac{1}{\sqrt{2^{n}(2^\secp-a+1)}} \ket{\chi^{\frl,1}_{\overline{S'},X,\vect{x},\vect{y},w_1}}\bra{y,w,\frG(\overline{S'}\cup\set{(X\cR,\vect{x},y||\vect{y},w_1,w)})}_{\reg{AB\overline{ST}}} \\
        =& \sum_{\substack{X,\vect{x},\vect{y},w_1,
        \overline{S'}}}\ket{\chi^{\frl,1}_{\overline{S'},X,\vect{x},\vect{y},w_1}}\bra{\chi^{\frl,1}_{\overline{S'},X,\vect{x},\vect{y},w_1}}_{\reg{AB\overline{ST}}} \\
    \end{align*}       

    \noindent Hence, $$\Pi^{\good}\Pi^{\cR,1} = \sum_{\overline{S'},X,\vect{x},\vect{y},w_1} \ketbra{\chi^{\frl,1}_{\overline{S'},X,\vect{x},\vect{y},w_1}}{\chi^{\frl,1}_{\overline{S'},X,\vect{x},\vect{y},w_1}}$$
\end{proof}

\begin{proof}[Proof of~\Cref{lem:proj:com2}]
    Fix some $y_0,w,y_1,\overline{S}$ where $\overline{S}$ is good and $a = \coun(\overline{S})$ and $b = \leng(\overline{S})$. We start by looking at what $$\ket{\psi_{y_0,w,y_1,\overline{S}}} = \Pi^{\cR,12} \ket{y_0}_{\reg{A}}\ket{w}_{\reg{B}}\ket{y_1}_{\reg{C}}\ket{\frG(\overline{S})}_{\reg{\overline{ST}}},$$
    \noindent Then we have the following:
    \begin{align*}
        \ket{\psi_{y_0,w,y_1,\overline{S}}} =& \Pi^{\cR,12}\ket{y_0}_{\reg{A}}\ket{w}_{\reg{B}}\ket{y_1}_{\reg{C}}\ket{\frG(\overline{S})}_{\reg{\overline{ST}}} \\
        =& \frac{1}{\sqrt{2^{an}\cdot\Pi_{i=1}^{b}(2^\secp-i+1) }}\sum_{\substack{Z\in\bit^{an}\\ R\in\left(\bit^{\secp}\right)^{b}_{\dist}}} \Pi^{\cR,12}  \ket{y_0}_{\reg{A}}\ket{w}_{\reg{B}}\ket{y_1}_{\reg{C}}\ket{\G \left(\overline{S},R,Z\right)}_{\reg{\overline{ST}}}
    \end{align*}

    \noindent Notice that the above is zero if there's no line in $\G \left(\overline{S},R,Z\right)$ of the form $(X\cR,\vect{x},y_0||\vect{y}||y_1,w_1,w,\vect{r},z)$ for some $X\in\set{\cL,\cR}$ and $\vect{x},\vect{y},w_1,\vect{r},z$. That is that $\overline{S} = \overline{S'}\cup\set{(X\cR,\vect{x},y_0||\vect{y}||y_1,w_1,w)}$.Then the above looks like:
    \begin{align*}
        \ket{\psi_{y_0,w,y_1,\overline{S}}} =& \frac{1}{\sqrt{2^{an}\cdot\Pi_{i=1}^{b}(2^\secp-i+1) }}\sum_{\substack{Z'\in\bit^{(a-1)n}\\ R'\in\left(\bit^{\secp})\cup\set{w}\right)^{b-|\vect{x}|}_{\dist}\\ z\in\set{0,1}^n\\ \vect{r}\in\left(\bit^{\secp}\setminus R'\right)^{|\vect{x}|}_{\dist}}} \Pi^{\cR,12} \ket{y_0}_{\reg{A}}\ket{w}_{\reg{B}}\ket{y_1}_{\reg{C}}\\
        &\times\ket{\G \left(\overline{S'},R',Z'\right)\cup\set{(X\cR,\vect{x},y_0||\vect{y}||y_1,w_1,w,\vect{r},z)}}_{\reg{\overline{ST}}}\\
        =& \frac{1}{2^{2n}(2^\secp-a+1)\sqrt{2^{an}\cdot\Pi_{i=1}^{b}(2^\secp-i+1) }}\sum_{\substack{Z'\in\bit^{(a-1)n}\\ R'\in\left(\bit^{\secp}\right)^{b-|\vect{x}|}_{\dist}\\ z\in\set{0,1}^n\\ \vect{r}\in\left(\bit^{\secp}\setminus R'\right)^{|\vect{x}|}_{\dist}\\ y_0'\in\set{0,1}^n\\ w'\in\left(\bit^{\secp}\setminus\Im(\overline{S'})\right)\\ y_1'\in\set{0,1}^n}} \ket{y_0'}_{\reg{A}}\ket{w'}_{\reg{B}}\ket{y_1'}_{\reg{A}}\\
        &\times \ket{\G \left(\overline{S'},R',Z'\right)\cup\set{(X\cR,\vect{x},y'_0||\vect{y}||y_1',w_1,w',\vect{r},z)}}_{\reg{\overline{ST}}}\\
        =& \frac{1}{2^{2n}(2^\secp-a+1)\sqrt{2^{an}\cdot\Pi_{i=1}^{b}(2^\secp-i+1) }}\sum_{\substack{y_0'\in\set{0,1}^n\\ w'\in\left(\bit^{\secp}\setminus\Im(\overline{S'})\right)\\ y_1'\in\set{0,1}^n\\ Z\in\bit^{an}\\ R\in\left(\bit^{\secp}\right)^{b}_{\dist}}} \ket{y_0'}_{\reg{A}}\ket{w'}_{\reg{B}}\ket{y_1'}_{\reg{C}}\\
        &\times \ket{\G \left(\overline{S'}\cup\set{(X\cR,\vect{x},y_0'||\vect{y}||y_1',w_1,w')},R,Z\right)}_{\reg{\overline{ST}}}\\
        =& \frac{1}{2^{n}\sqrt{(2^\secp-a+1)}}\underbrace{\frac{1}{2^{n}\sqrt{(2^\secp-a+1)}}\sum_{\substack{y_0'\in\set{0,1}^n\\ w'\in\left(\bit^{\secp}\setminus\Im(\overline{S'})\right)\\ y_1'\in\set{0,1}^n}} \ket{y_0',w',y_1',\frG(\overline{S'}\cup\set{(X\cR,\vect{x},y_0'||\vect{y}||y_1',w_1,w')})}_{\reg{ABC\overline{ST}}}}_{\ket{\chi^{\frl,2}_{\overline{S'},X,\vect{x},\vect{y},w_1}}}\\
    \end{align*}

    \noindent Finally, to understand $\Pi^{\cR,12}\Pi^{\good}$, we expanding the projector $\Pi^{\good}$: 
    \begin{align*}
        \Pi^{\cR,12}\Pi^{\good} =& \Pi^{\cR,12}\sum_{\substack{y_0,w,y_1,\overline{S}\\ \overline{S}\text{ is good}}} \ketbra{y_0,w,y_1,\frG(\overline{S})}{y_0,w,y_1,\frG(\overline{S})}_{\reg{ABC\overline{ST}}}\\
        =& \sum_{\substack{y_0,w,y_1,\overline{S}\\ \overline{S}\text{ is good}}}  \Pi^{\cR,12}\ketbra{y_0,w,y_1,\frG(\overline{S})}{y_0,w,y_1,\frG(\overline{S})}_{\reg{ABC\overline{ST}}}\\
        =& \sum_{\substack{y_0,w,y_1\\ X,\vect{x},\vect{y},w_1,
        \overline{S'}\\ w\in\bit^{\secp}\setminus\Im(\overline{S'})}}  \Pi^{\cR,12} \ket{y_0,w,y_1,\frG(\overline{S'}\cup\set{(X\cR,\vect{x},y_0||\vect{y}||y_1,w_1,w)})}\\
        &\times\bra{y_0,w,y_1,\frG(\overline{S'}\cup\set{(X\cR,\vect{x},y_0||\vect{y}||y_1,w_1,w)})}_{\reg{ABC\overline{ST}}}\\
        =& \sum_{\substack{y_0,w,y_1\\X,\vect{x},\vect{y},w_1,
        \overline{S'}\\ w\in\bit^{\secp}\setminus\Im(\overline{S'})}}\frac{1}{2^{n}\sqrt{(2^\secp-a+1)}} \ket{\chi^{\frl,2}_{\overline{S'},X,\vect{x},\vect{y},w_1}}\bra{y_0,w,y_1,\frG(\overline{S'}\cup\set{(X\cR,\vect{x},y_0||\vect{y}||y_1,w_1,w)})}_{\reg{ABC\overline{ST}}} \\
        =& \sum_{\substack{X,\vect{x},\vect{y},w_1,
        \overline{S'}}}\ket{\chi^{\frl,2}_{\overline{S'},X,\vect{x},\vect{y},w_1}}\bra{\chi^{\frl,2}_{\overline{S'},X,\vect{x},\vect{y},w_1}}_{\reg{ABC\overline{ST}}} \\
    \end{align*}       

    \noindent Hence, $$\Pi^{\good}\Pi^{\cR,12} = \sum_{\overline{S'},X,\vect{x},\vect{y},w_1} \ketbra{\chi^{\frl,2}_{\overline{S'},X,\vect{x},\vect{y},w_1}}{\chi^{\frl,2}_{\overline{S'},X,\vect{x},\vect{y},w_1}}$$
\end{proof}

\begin{proof}[Proof of~\Cref{lem:proj:com3}]
    Fix some $y_0,w,y_1,\overline{S}$ where $\overline{S}$ is good and $a = \coun(\overline{S})$ and $b = \leng(\overline{S})$. We start by looking at what $$\ket{\psi_{y_0,w,y_1,\overline{S}}} = \Pi^{\cR,123} \ket{y_0}_{\reg{A}}\ket{w}_{\reg{B}}\ket{y_1}_{\reg{C}}\ket{\frG(\overline{S})}_{\reg{\overline{ST}}},$$
    \noindent Then we have the following:
    \begin{align*}
        \ket{\psi_{y_0,w,y_1,\overline{S}}} =& \Pi^{\cR,123}\ket{y_0}_{\reg{A}}\ket{w}_{\reg{B}}\ket{y_1}_{\reg{C}}\ket{\frG(\overline{S})}_{\reg{\overline{ST}}} \\
        =& \frac{1}{\sqrt{2^{an}\cdot\Pi_{i=1}^{b}(2^\secp-i+1) }}\sum_{\substack{Z\in\bit^{an}\\ R\in\left(\bit^{\secp}\right)^{b}_{\dist}}} \Pi^{\cR,123}  \ket{y_0}_{\reg{A}}\ket{w}_{\reg{B}}\ket{y_1}_{\reg{C}}\ket{\G \left(\overline{S},R,Z\right)}_{\reg{\overline{ST}}}
    \end{align*}

    \noindent Notice that the above is zero if there's no line in $\G \left(\overline{S},R,Z\right)$ of the form $(\cR\cR,\vect{x},(y_0,y_1),w_1,w,\vect{r},z)$ for some $\vect{x},w_1,\vect{r},z$. That is that $\overline{S} = \overline{S'}\cup\set{(\cR\cR,\vect{x},(y_0,y_1),w_1,w)}$.Then the above looks like:
    \begin{align*}
        \ket{\psi_{y_0,w,y_1,\overline{S}}} =& \frac{1}{\sqrt{2^{an}\cdot\Pi_{i=1}^{b}(2^\secp-i+1) }}\sum_{\substack{Z'\in\bit^{(a-1)n}\\ R'\in\left(\bit^{\secp})\cup\set{w}\right)^{b-2}_{\dist}\\ z\in\set{0,1}^n\\ \vect{r}\in\left(\bit^{\secp}\setminus R'\right)^{2}_{\dist}}} \Pi^{\cR,123} \ket{y_0}_{\reg{A}}\ket{w}_{\reg{B}}\ket{y_1}_{\reg{C}}\\
        &\times\ket{\G \left(\overline{S'},R',Z'\right)\cup\set{(\cR\cR,\vect{x},(y_0,y_1),w_1,w,\vect{r},z)}}_{\reg{\overline{ST}}}\\
        =& \frac{1}{2^{2n}(2^\secp-a+1)\sqrt{2^{an}\cdot\Pi_{i=1}^{b}(2^\secp-i+1) }}\sum_{\substack{Z'\in\bit^{(a-1)n}\\ R'\in\left(\bit^{\secp}\right)^{b-2}_{\dist}\\ z\in\set{0,1}^n\\ \vect{r}\in\left(\bit^{\secp}\setminus R'\right)^{2}_{\dist}\\ y_0'\in\set{0,1}^n\\ w'\in\left(\bit^{\secp}\setminus\Im(\overline{S'})\right)\\ y_1'\in\set{0,1}^n}} \ket{y_0'}_{\reg{A}}\ket{w'}_{\reg{B}}\ket{y_1'}_{\reg{A}}\\
        &\times \ket{\G \left(\overline{S'},R',Z'\right)\cup\set{(\cR\cR,\vect{x},(y'_0,y_1'),w_1,w',\vect{r},z)}}_{\reg{\overline{ST}}}\\
        =& \frac{1}{2^{2n}(2^\secp-a+1)\sqrt{2^{an}\cdot\Pi_{i=1}^{b}(2^\secp-i+1) }}\sum_{\substack{y_0'\in\set{0,1}^n\\ w'\in\left(\bit^{\secp}\setminus\Im(\overline{S'})\right)\\ y_1'\in\set{0,1}^n\\ Z\in\bit^{an}\\ R\in\left(\bit^{\secp}\right)^{b}_{\dist}}} \ket{y_0'}_{\reg{A}}\ket{w'}_{\reg{B}}\ket{y_1'}_{\reg{C}}\\
        &\times \ket{\G \left(\overline{S'}\cup\set{(\cR\cR,\vect{x},(y_0',y_1'),w_1,w')},R,Z\right)}_{\reg{\overline{ST}}}\\
        =& \frac{1}{2^{n}\sqrt{(2^\secp-a+1)}}\underbrace{\frac{1}{2^{n}\sqrt{(2^\secp-a+1)}}\sum_{\substack{y_0'\in\set{0,1}^n\\ w'\in\left(\bit^{\secp}\setminus\Im(\overline{S'})\right)\\ y_1'\in\set{0,1}^n}} \ket{y_0',w',y_1',\frG(\overline{S'}\cup\set{(\cR\cR,\vect{x},(y_0',y_1'),w_1,w')})}_{\reg{ABC\overline{ST}}}}_{\ket{\chi^{\frl,3}_{\overline{S'},\vect{x},w_1}}}\\
    \end{align*}

    \noindent Finally, to understand $\Pi^{\cR,123}\Pi^{\good}$, we expanding the projector $\Pi^{\good}$: 
    \begin{align*}
        \Pi^{\cR,123}\Pi^{\good} =& \Pi^{\cR,123}\sum_{\substack{y_0,w,y_1,\overline{S}\\ \overline{S}\text{ is good}}} \ketbra{y_0,w,y_1,\frG(\overline{S})}{y_0,w,y_1,\frG(\overline{S})}_{\reg{ABC\overline{ST}}}\\
        =& \sum_{\substack{y_0,w,y_1,\overline{S}\\ \overline{S}\text{ is good}}}  \Pi^{\cR,123}\ketbra{y_0,w,y_1,\frG(\overline{S})}{y_0,w,y_1,\frG(\overline{S})}_{\reg{ABC\overline{ST}}}\\
        =& \sum_{\substack{y_0,w,y_1\\ \vect{x},w_1,
        \overline{S'}\\ w\in\bit^{\secp}\setminus\Im(\overline{S'})}}  \Pi^{\cR,123} \ket{y_0,w,y_1,\frG(\overline{S'}\cup\set{(\cR\cR,\vect{x},(y_0,y_1),w_1,w)})}\\
        &\times\bra{y_0,w,y_1,\frG(\overline{S'}\cup\set{(\cR\cR,\vect{x},(y_0,y_1),w_1,w)})}_{\reg{ABC\overline{ST}}}\\
        =& \sum_{\substack{y_0,w,y_1\\ \vect{x},w_1,
        \overline{S'}\\ w\in\bit^{\secp}\setminus\Im(\overline{S'})}}\frac{1}{2^{n}\sqrt{(2^\secp-a+1)}} \ket{\chi^{\frl,3}_{\overline{S'},\vect{x},w_1}}\bra{y_0,w,y_1,\frG(\overline{S'}\cup\set{(\cR\cR,\vect{x},(y_0,y_1),w_1,w)})}_{\reg{ABC\overline{ST}}} \\
        =& \sum_{\substack{\vect{x},w_1,
        \overline{S'}}}\ket{\chi^{\frl,3}_{\overline{S'},\vect{x},w_1}}\bra{\chi^{\frl,3}_{\overline{S'},\vect{x},w_1}}_{\reg{ABC\overline{ST}}} \\
    \end{align*}       

    \noindent Hence, $$\Pi^{\good}\Pi^{\cR,123} = \sum_{\overline{S'},\vect{x},w_1} \ketbra{\chi^{\frl,3}_{\overline{S'},\vect{x},w_1}}{\chi^{\frl,3}_{\overline{S'},\vect{x},w_1}}$$
\end{proof}

\section{Proofs of~\Cref{sec:struc:action}}
\label{sec:app:action}

\begin{proof}[Proof of~\Cref{lem:LtoR:1}]
Recall that we want to analyse: 
$$\ket{\phi} = V^{(1),\midd}_RV^{(2),\midd}_RV^{(3),\midd}_R\ket{x_0}\ket{w_1}\ket{x_1}\ket{\frG(\overline{S})}$$
\noindent In the below calculation, we don't explicitly write the normalisation, and can be verified. 
    \begin{align*}
         \ket{\phi} &= V^{(1),\midd}_RV^{(2),\midd}_RV^{(3),\midd}_R\ket{x_0}\ket{w_1}\ket{x_1}\ket{\frG(\overline{S})}\\
         &= \sum_{\substack{Z\in\bit^{an}\\ R\in(\bit^{\secp})^{b}_{\dist}}}V^{(1),\midd}_RV^{(2),\midd}_RV^{(3),\midd}_R\ket{x_0}\ket{w_1}\ket{x_1}\ket{\G(\overline{S},R,Z)}\\
         &= \sum_{\substack{Z\in\bit^{an}\\ R\in(\bit^{\secp})^{b}_{\dist}\\ (r_1,r_2)\in (\bit^{\secp}\setminus R)^{2}_{\dist}\\ z\in\bit^{n}\\ y_0,y_1\in\bit^n\\ w_2\in\bit^{\secp}\setminus\Im(\overline{S})}}\ket{y_0}\ket{w_2}\ket{y_1}\ket{\G(\overline{S},R,Z)\cup\frp(\cR\cR,(x_0,x_1),(y_0,y_1),w_1,w_2,(r_1,r_2),z)}\\
         &= \sum_{\substack{Z\cup\set{z}\in\bit^{(a+1)n}\\ R\cup\set{(r_1,r_2)}\in(\bit^{\secp})^{b_2}_{\dist}\\ y_0,y_1\in\bit^n\\ w_2\in\bit^{\secp}\setminus\Im(\overline{S})}}\ket{y_0}\ket{w_2}\ket{y_1}\ket{\G(\overline{S}\cup\set{(\cR\cR,(x_0,x_1),(y_0,y_1),w_1,w_2)},R\cup\set{(r_1,r_2)},Z\cup\set{z})}\\
         &= \sum_{\substack{y_0,y_1\in\bit^n\\ w_2\in\bit^{\secp}\setminus\Im(\overline{S})}}\ket{y_0}\ket{w_2}\ket{y_1}\ket{\frG(\overline{S}\cup\set{(\cR\cR,(x_0,x_1),(y_0,y_1),w_1,w_2)})}\\
         &= \ket{\chi^{\frl,3}_{\overline{S},(x_0,x_1),w_1}} 
    \end{align*}

    \noindent Hence, we have 
    \begin{align*}
          V^{(1),\midd}_RV^{(2),\midd}_RV^{(3),\midd}_R\ket{x_0}\ket{w_1}\ket{x_1}\ket{\frG(\overline{S})} = \ket{\chi^{\frl,3}_{\overline{S},(x_0,x_1),w_1}} 
    \end{align*}
\end{proof}

\begin{proof}[Proof of~\Cref{lem:LtoR:2}]
Recall that we want to analyse:
$$\ket{\phi} = V_R^{(1),\midd}V_R^{(2),\midd}V_L^{(3),\midd,\dagger}\ket{\chi^{\frr,1}_{\overline{S},X,\vect{x},\vect{y},w_1}}_{\reg{AB\overline{ST}}}\ket{x'}_{\reg{C}}$$
\noindent In the below calculation, we don't explicitly write the normalisation, and can be verified. 
    \begin{align*}
         \ket{\phi} &= V_R^{(1),\midd}V_R^{(2),\midd}V_L^{(3),\midd,\dagger}\ket{\chi^{\frr,1}_{\overline{S},X,\vect{x},\vect{y},w_1}}_{\reg{AB\overline{ST}}}\ket{x'}_{\reg{C}}\\
         &= V_R^{(1),\midd}V_R^{(2),\midd}V_L^{(3),\midd,\dagger}\sum_{\substack{y_0\in\bit^{n}\\ w_2\in\bit^{\secp}\setminus\Im(\overline{S})}}\ket{y_0,w_2,\frG(\overline{S}\cup\set{(XL,\vect{x},y_0||\vect{y},w_1,w_2)})}_{\reg{AB\overline{ST}}}\ket{x'}_{\reg{C}}\\
         &= V_R^{(1),\midd}V_R^{(2),\midd}V_L^{(3),\midd,\dagger}\sum_{\substack{y_0\in\bit^{n}\\ w_2\in\bit^{\secp}\setminus\Im(\overline{S})\\ Z\in\bit^{an}\\ R\in(\bit^{\secp})^{b}_{\dist}\\ z\in\bit^{n}\\ \vect{r}\in(\bit^{\secp}\setminus R)^{a'}_{\dist}}}\ket{y_0,w_2,\G(\overline{S}\cup\set{(XL,\vect{x},y_0||\vect{y},w_1,w_2)},R\cup\set{\vect{r}},Z\cup\set{z})}_{\reg{AB\overline{ST}}}\ket{x'}_{\reg{C}}\\
         &= \sum_{\substack{y_0\in\bit^{n}\\ w_2\in\bit^{\secp}\setminus\Im(\overline{S})\\ y_1\in\bit^{n}\\ Z\in\bit^{an}\\ R\in(\bit^{\secp})^{b}_{\dist}\\ z\in\bit^{n}\\ \vect{r}\in(\bit^{\secp}\setminus R)^{a'}_{\dist}\\ r\in\bit^{\secp}\setminus R\cup\set{\vect{r}}}}\ket{y_0,w_2,y_1\G(\overline{S}\cup\set{(XR,\vect{x}||x',y_0||\vect{y}||y_1,w_1,w_2)},R\cup\set{\vect{r}||r},Z\cup\set{z})}_{\reg{ABC\overline{ST}}}\\
         &= \sum_{\substack{y_0\in\bit^{n}\\ w_2\in\bit^{\secp}\setminus\Im(\overline{S})\\ y_1\in\bit^{n}}}\ket{y_0,w_2,y_1\frG(\overline{S}\cup\set{(XR,\vect{x}||x',y_0||\vect{y}||y_1,w_1,w_2)})}_{\reg{ABC\overline{ST}}}\\
         &= \ket{\chi^{\frl,2}_{\overline{S},X,\vect{x}||x',\vect{y},w_1}}_{\reg{ABC\overline{ST}}} 
    \end{align*}

    \noindent Hence, we have 
    \begin{align*}
          V_R^{(1),\midd}V_R^{(2),\midd}V_L^{(3),\midd,\dagger}\ket{\chi^{\frr,1}_{\overline{S},X,\vect{x},\vect{y},w_1}}_{\reg{AB\overline{ST}}}\ket{x'}_{\reg{C}} = \ket{\chi^{\frl,2}_{\overline{S},X,\vect{x}||x',\vect{y},w_1}}_{\reg{ABC\overline{ST}}}
    \end{align*}
\end{proof}

\begin{proof}[Proof of~\Cref{lem:struc:good:fwd}]
    Recall that:
    \begin{align*}
        W^{\glued} = &\left(\Pi^{\cL,321}\right)\cdot V_L^{(3),\midd}\cdot V_L^{(2),\midd}\cdot V_L^{(1),\midd} \cdot\left(I-\Pi^{\cR,1}\right) \\
        &+ \left(\Pi^{\cL,32} -\Pi^{\cL,321}\right)\cdot V_L^{(3),\midd}\cdot V_L^{(2),\midd}\cdot V_R^{(1),\midd,\dagger}\cdot\left(\Pi^{\cR,1}-\Pi^{\cR,12}\right) \\
        &+ \left(\Pi^{\cL,3} -\Pi^{\cL,32}\right)\cdot V_L^{(3),\midd}\cdot V_R^{(2),\midd,\dagger}\cdot V_R^{(1),\midd,\dagger}\cdot\left(\Pi^{\cR,12}-\Pi^{\cR,123}\right) \\
        &+ \left(I-\Pi^{\cL,3}\right)\cdot V_R^{(3),\midd,\dagger}\cdot V_R^{(2),\midd,\dagger}\cdot V_R^{(1),\midd,\dagger}\cdot\left(\Pi^{\cR,123}\right) \\
    \end{align*} 

    \noindent We know $\ket{\phi} = \Pi^{\good}\ket{\phi}$. Then we also have the following:
    \begin{align*}
        \left(I-\Pi^{\cR,1}\right)\ket{\phi} =& \left(I-\Pi^{\cR,1}\right)\Pi^{\good}\ket{\phi}\\
        =& \Pi^{\good}\left(I-\Pi^{\cR,1}\right)\ket{\phi}
    \end{align*}
    \noindent Similarly, we have:
    \begin{align*}
        \underbrace{\left(I-\Pi^{\cR,1}\right)\ket{\phi}}_{\ket{\phi_1}} =& \Pi^{\good}\left(I-\Pi^{\cR,1}\right)\ket{\phi}\\
        \underbrace{\left(\Pi^{\cR,1}-\Pi^{\cR,12}\right)\ket{\phi}}_{\ket{\phi_2}} =& \Pi^{\good}\left(\Pi^{\cR,1}-\Pi^{\cR,12}\right)\ket{\phi}\\
        \underbrace{\left(\Pi^{\cR,12}-\Pi^{\cR,123}\right)\ket{\phi}}_{\ket{\phi_3}} =& \Pi^{\good}\left(\Pi^{\cR,12}-\Pi^{\cR,123}\right)\ket{\phi}\\
        \underbrace{\left(\Pi^{\cR,123}\right)\ket{\phi}}_{\ket{\phi_4}} =& \Pi^{\good}\left(\Pi^{\cR,123}\right)\ket{\phi}\\
    \end{align*}
    \noindent And we have $\ket{\phi} = \sum_{i\in [4]}\ket{\phi_i}$.
    Then for $i\in [4]$, we compute that $\Pi^{\good} W^{\glued}\ket{\phi_i} - W^{\glued}\ket{\phi_i}$.

    \noindent We start by computing the above for $\ket{\phi_1}$. 
    \begin{align*}
        W^{\glued}\ket{\phi_1} &= W^{\glued}\left(I-\Pi^{\cR,1}\right)\ket{\phi_1}\\
        &= \left(\Pi^{\cL,321}\right)\cdot V_L^{(3),\midd}\cdot V_L^{(2),\midd}\cdot V_L^{(1),\midd}\left(I-\Pi^{\cR,1}\right)\ket{\phi_1}
    \end{align*}
    Notice that $\Pi^{\good}$ commutes with $\Pi^{\cL,321}$. Also, since $\ket{\phi_1}$ is in the subspace $\Pi^{\good}$, we can write it as a superposition over $\ket{x_0}\ket{w_1}\ket{x_1}\ket{\frG(\overline{S})}$. Then by~\Cref{lem:LtoR:1}, we get that $V_L^{(3),\midd}\cdot V_L^{(2),\midd}\cdot V_L^{(1),\midd}$ on $\ket{\phi_1}$ is in $\Pi^{\good}$. Hence, we get $(I-\Pi^{\good})W^{\glued}\ket{\phi_1} = 0$.

    \noindent Next, we compute the above for $\ket{\phi_2}$. 
    \begin{align*}
        W^{\glued}\ket{\phi_2} &= W^{\glued}\left(\Pi^{\cR,1}-\Pi^{\cR,12}\right)\ket{\phi_2}\\
        &= \left(\Pi^{\cL,32}-\Pi^{\cL,321}\right)\cdot V_L^{(3),\midd}\cdot V_L^{(2),\midd}\cdot V_R^{(1),\midd,\dagger}\left(\Pi^{\cR,1}-\Pi^{\cR,12}\right)\ket{\phi_2}
    \end{align*}
    Notice that $\Pi^{\good}$ commutes with $\Pi^{\cL,32}-\Pi^{\cL,321}$. Also, since $\ket{\phi_1}$ is in the subspace $\Pi^{\cR,1}\Pi^{\good}$, by~\Cref{lem:proj:com1}, we can write it as a superposition over $\ket{\chi^1_{\overline{S'},\vect{x},\vect{y},w_1}}\ket{x'}$. Then by~\Cref{lem:LtoR:2}, we get that $V_L^{(3),\midd}\cdot V_L^{(2),\midd}\cdot V_R^{(1),\midd,\dagger}$ on $\ket{\phi_2}$ is in $\Pi^{\good}$. Hence, we get $(I-\Pi^{\good})W^{\glued}\ket{\phi_2} = 0$.

    \noindent Similarly by, we have for $i\in \set{3,4}$, $(I-\Pi^{\good})W^{\glued}\ket{\phi_i} = 0$. Hence, we get $(I-\Pi^{\good})W^{\glued}\ket{\phi} = 0$. 
\end{proof}

\section{Proofs of~\Cref{sec:gluing:fwd}}
\label{sec:app:fwd}

\begin{lemma}
    \label{lem:gluing:proj}
    We have the following:
    $$\|\Ocomp^{\dagger}W_R^{\frm(\secp)}W_R^{\frm(\secp),\dagger}\Ocomp\Pi^{\good}\Pi_{\leq t} - \left(\sum_{\overline{S'},X,\vect{x},\vect{y},w_1} \ketbra{\chi^{\frl,2}_{\overline{S'},X,\vect{x},\vect{y},w_1}}{\chi^{\frl,2}_{\overline{S'},X,\vect{x},\vect{y},w_1}}\right)\Pi_{\leq t}\|_{\op} = O(t^{2}/2^{\secp})$$
\end{lemma}
\noindent The proof of the above is similar to~\Cref{lem:proj:com1}.

\noindent Notice that~\Cref{lem:proj:com2}, we have $$\sum_{\overline{S'},X,\vect{x},\vect{y},w_1} \ketbra{\chi^{\frl,2}_{\overline{S'},X,\vect{x},\vect{y},w_1}}{\chi^{\frl,2}_{\overline{S'},X,\vect{x},\vect{y},w_1}} = \Pi^{\good}\Pi^{cR,12}.$$
Notice that $$\Pi^{cR,12}\Pi^{\frl,1} = \Pi^{cR,12}\Pi^{\frl,2} = 0,$$ $$\Pi^{cR,12}\Pi^{\frl,3} = \Pi^{\frl,3}$$ and $$\Pi^{cR,12}\Pi^{\frl,4} = \Pi^{\frl,4}$$

\noindent Then we can see how $\Pi^{\frl,i}$ behaves with the above projector. In particular, we have the following:
\begin{lemma}
    \label{lem:gluing:com_1}
    We have the following:
    $$\|\Ocomp^{\dagger}W_R^{\frm(\secp)}W_R^{\frm(\secp),\dagger}\Ocomp\Pi^{\good}\Pi^{\frl,1}\Pi_{\leq t}\|_{\op} = O(t^{2}/2^{\secp})$$
\end{lemma}
\begin{lemma}
    \label{lem:gluing:com_2}
    We have the following:
    $$\|\Ocomp^{\dagger}W_R^{\frm(\secp)}W_R^{\frm(\secp),\dagger}\Ocomp\Pi^{\good}\Pi^{\frl,2}\Pi_{\leq t}\|_{\op} = O(t^{2}/2^{\secp})$$
\end{lemma}
\begin{lemma}
    \label{lem:gluing:com_3}
    We have the following:
    $$\|\Ocomp^{\dagger}W_R^{\frm(\secp)}W_R^{\frm(\secp),\dagger}\Ocomp\Pi^{\good}\Pi^{\frl,3}\Pi_{\leq t} - \Pi^{\good}\Pi^{\frl,3}\Pi_{\leq t}\|_{\op}= O(t^{2}/2^{\secp})$$
\end{lemma}
\begin{lemma}
    \label{lem:gluing:com_4}
    We have the following:
    $$\|\Ocomp^{\dagger}W_R^{\frm(\secp)}W_R^{\frm(\secp),\dagger}\Ocomp\Pi^{\good}\Pi^{\frl,4}\Pi_{\leq t} - \Pi^{\good}\Pi^{\frl,4}\Pi_{\leq t}\|_{\op}= O(t^{2}/2^{\secp})$$
\end{lemma}

\begin{proof}[Proof of~\Cref{lem:gluing:fwd_1}]
We recall that we want to estimate: 
\begin{align*}
    \gamma =& \left\|\left(\Ocomp \Pi^{\good} W^{\glued} - W^{\frm(\secp)}\Ocomp\right)\Pi^{\good}\Pi^{\frl,1}\Pi_{\leq t}\right\|_{\op}\\
    \leq& \left\|\left(\Ocomp \Pi^{\good} W^{\glued} - W_L^{\frm(\secp)}\Ocomp\right)\Pi^{\good}\Pi^{\frl,1}\Pi_{\leq t}\right\|_{\op} \\
    &+\left\|W_R^{\frm(\secp),\dagger}\Ocomp\Pi^{\good}\Pi^{\frl,1}\Pi_{\leq t}\right\|_{\op} \\
    &+\left\|W_L^{\frm(\secp)}W_R^{\frm(\secp)}W_R^{\frm(\secp),\dagger}\Ocomp\Pi^{\good}\Pi^{\frl,1}\Pi_{\leq t}\right\|_{\op}\\
    \leq& \left\|\left(\Ocomp \Pi^{\good} V_L^{(3),\midd}V_L^{(2),\midd}V_L^{(1),\midd}\Pi^{\frl,1} - W_L^{\frm(\secp)}\Ocomp\right)\Pi^{\good}\Pi^{\frl,1}\Pi_{\leq t}\right\|_{\op} + O(t^2/2^{\secp})\\
    \leq& \left\|\left(\Ocomp \Pi^{\good} V_L^{(3),\midd}V_L^{(2),\midd}V_L^{(1),\midd} - W_L^{\frm(\secp)}\Ocomp\right)\Pi^{\good}\Pi^{\frl,1}\Pi_{\leq t}\right\|_{\op} + O(t^2/2^{\secp})\\
\end{align*}
Where the second line is by triangle inequality, and the third line is by~\Cref{lem:gluing:com_1}, and the fourth line is true because $\Pi^{\frl,1} = \left(I-V_R^{(1),\midd}V_R^{(1),\midd,\dagger}\right)$ and $\Pi^{\frl,1}\Pi^{\good} = \Pi^{\good}\Pi^{\frl,1}$. Next, we compute $$\left\|\left(\Ocomp \Pi^{\good} V_L^{(3),\midd}V_L^{(2),\midd}V_L^{(1),\midd} - W_L^{\frm(\secp)}\Ocomp\right)\Pi^{\good}\Pi^{\frl,1}\Pi_{\leq t}\right\|_{\op}.$$ 

\noindent We know that the subspace represented by $\Pi^{\good}$ is spanned by $\ket{x_0,w_1,x_1,\frG(\overline{S})}$ for $x_0,x_1\in\bit^{n}$, $w_1\in\bit^{\secp}$ and $\overline{S}$ is some good state parameter. Let $a = \coun(\overline{S})$ and $b = \leng(\overline{S})$.

\paragraph{\textbf{Computing $\Ocomp \Pi^{\good} V_L^{(3),\midd}V_L^{(2),\midd}V_L^{(1),\midd}\ket{x_0,w_1,x_1,\frG(\overline{S})}$:}} We start by computing: 
\begin{align*}
    \ket{\phi_1} =&  \Ocomp \Pi^{\good} V_L^{(3),\midd}V_L^{(2),\midd}V_L^{(1),\midd}\ket{x_0,w_1,x_1,\frG(\overline{S})}\\
    =&  \Ocomp \ket{\chi^{\frl,3}_{\frG(\overline{S}),(x_0,x_1)w_1}}\\
    =&  \Ocomp \frac{1}{2^n\sqrt{2^{\secp}-a}}\sum_{\substack{y_1,y_0\in\bit^{n}\\ w\in(\bit^{\secp}\setminus\Im(\overline{S}))}}\ket{y_0,w,y_1,\frG(\overline{S}\cup\set{(\cL\cL,(x_0,x_1),(y_0,y_1),w_1,w)})}\\
    =&  \frac{1}{2^n\sqrt{2^{\secp}-a}}\sum_{\substack{y_1,y_0\in\bit^{n}\\ w\in(\bit^{\secp}\setminus\Im(\overline{S}))}}\ket{y_0,w,y_1,\frF(\overline{S}\cup\set{(\cL\cL,(x_0,x_1),(y_0,y_1),w_1,w)})}\\
\end{align*}
Where the second line is by~\Cref{lem:LtoR:1}. 
\paragraph{\textbf{Computing $W_L^{\frm(\secp)}\Ocomp\ket{x_0,w_1,x_1,\frG(\overline{S})}$}} We start by computing: 
\begin{align*}
    \ket{\phi_2} =&  W_L^{\frm(\secp)}\Ocomp\ket{x_0,w_1,x_1,\frG(\overline{S})}\\
    =&  W_L^{\frm(\secp)}\ket{x_0,w_1,x_1,\frF(\overline{S})}\\
    =&  W_L^{\frm(\secp)}\frac{1}{\sqrt{2^{(b-a)n}\left((2^{\secp}-a)\ldots(2^{\secp}-b+1)\right)}} \sum_{\substack{\cU\in\bit^{(b-a)n}\\ \cV\in(\bit^{\secp}\setminus\Im(\overline{S}))^{b-a}_{\dist}}}\ket{x_0,w_1,x_1,\F(\overline{S},\cU,\cV)} \\ 
    =&  \frac{1}{\sqrt{2^{(b-a)n}\left((2^{\secp}-a)\ldots(2^{\secp}-b+1)\right)}} \sum_{\substack{\cU\in\bit^{(b-a)n}\\ \cV\in(\bit^{\secp}\setminus\Im(\overline{S}))^{b-a}_{\dist}}}\sum_{\substack{y_1,y_0\in\bit^{n}\\ w\in(\bit^{\secp}\setminus(\Im(\overline{S})\cup\cV))}}\\
    &\times\frac{1}{2^n\sqrt{2^{\secp}-b}}\ket{y_0,w,y_1,\F(\overline{S},\cU,\cV)\cup(\set{(x_0||w_1||x_1,y_0||w||y_1)},\empty)} \\ 
    =&  \frac{1}{2^n\sqrt{2^{(b-a)n}\left((2^{\secp}-a)\ldots(2^{\secp}-b)\right)}} \sum_{\substack{y_1,y_0\in\bit^{n}\\ w\in(\bit^{\secp}\setminus(\Im(\overline{S})))\\ \cU\in\bit^{(b-a)n}\\ \cV\in(\bit^{\secp}\setminus\Im(\overline{S})\cup\set{w})^{b-a}_{\dist}}}\\
    &\times\ket{y_0,w,y_1,\F(\overline{S}\cup\set{(\cL\cL,(x_0,x_1),(y_0,y_1),w_1,w)},\cU,\cV)} \\ 
    =&\frac{1}{2^n\sqrt{2^{\secp}-a}}\sum_{\substack{y_1,y_0\in\bit^{n}\\ w\in(\bit^{\secp}\setminus\Im(\overline{S}))}}\ket{y_0,w,y_1,\frF(\overline{S}\cup\set{(\cL\cL,(x_0,x_1),(y_0,y_1),w_1,w)})}
\end{align*}

\noindent Hence, we have $\ket{\phi_1}=\ket{\phi_2}$. Since the subspace represented by $\Pi^{\good}$ is spanned by $\ket{x_0,w_1,x_1,\frG(\overline{S})}$, hence $$\left\|\left(\Ocomp \Pi^{\good} V_L^{(3),\midd}V_L^{(2),\midd}V_L^{(1),\midd} - W_L^{\frm(\secp)}\Ocomp\right)\Pi^{\good}\Pi^{\frl,1}\Pi_{\leq t}\right\|_{\op} = 0.$$

\noindent Finally, we get 
$$\left\|\left(\Ocomp \Pi^{\good} W^{\gluedfwd} - W^{\frm(\secp)}\Ocomp\right)\Pi^{\good}\Pi^{\frl,1}\Pi_{\leq t}\right\|_{\op} = O(t^2/2^{\secp})$$
\end{proof}

\begin{proof}[Proof of~\Cref{lem:gluing:fwd_2}]
We recall that we want to estimate: 
\begin{align*}
    \gamma =& \left\|\left(\Ocomp \Pi^{\good} W^{\gluedfwd} - W^{\frm(\secp)}\Ocomp\right)\Pi^{\good}\Pi^{\frl,2}\Pi_{\leq t}\right\|_{\op}\\
    \leq& \left\|\left(\Ocomp \Pi^{\good} W^{\gluedfwd} - W_L^{\frm(\secp)}\Ocomp\right)\Pi^{\good}\Pi^{\frl,2}\Pi_{\leq t}\right\|_{\op} \\
    &+\left\|W_R^{\frm(\secp),\dagger}\Ocomp\Pi^{\good}\Pi^{\frl,2}\Pi_{\leq t}\right\|_{\op} \\
    &+\left\|W_L^{\frm(\secp)}W_R^{\frm(\secp)}W_R^{\frm(\secp),\dagger}\Ocomp\Pi^{\good}\Pi^{\frl,2}\Pi_{\leq t}\right\|_{\op}\\
    \leq& \left\|\left(\Ocomp \Pi^{\good} \left(\Pi^{\cL,32}-\Pi^{\cL,321}\right)V_L^{(3),\midd}V_L^{(2),\midd}V_R^{(1),\midd,\dagger} - W_L^{\frm(\secp)}\Ocomp\right)\Pi^{\good}\Pi^{\frl,2}\Pi_{\leq t}\right\|_{\op} \\
    &+ O(t^2/2^{\secp})
\end{align*}
Where the second line is by triangle inequality, and the third line is by~\Cref{lem:gluing:com_2}. Also notice that $\Pi^{\cL,321}V_L^{(3),\midd}V_L^{(2),\midd}V_R^{(1),\midd,\dagger}\Pi^{\good} = 0$.
Hence, we have \begin{align*}
    \gamma =& \left\|\left(\Ocomp \Pi^{\good} V_L^{(3),\midd}V_L^{(2),\midd}V_R^{(1),\midd,\dagger} - W_L^{\frm(\secp)}\Ocomp\right)\Pi^{\good}\Pi^{\frl,2}\Pi_{\leq t}\right\|_{\op} + O(t^2/2^{\secp})\\
\end{align*}

\noindent Next, we compute $$\left\|\left(\Ocomp \Pi^{\good} V_L^{(3),\midd}V_L^{(2),\midd}V_R^{(1),\midd,\dagger} - W_L^{\frm(\secp)}\Ocomp\right)\Pi^{\good}\Pi^{\frl,2}\Pi_{\leq t}\right\|_{\op}$$

\noindent We know by~\Cref{lem:proj:com1} that the subspace represented by $\Pi^{\good}\Pi^{\frl,2}$ is spanned by $\ket{\chi^{\frl,1}_{\overline{S},X,\vect{x},\vect{y},w_1}}_{\reg{AB\overline{ST}}}\ket{x'}_{\reg{C}}$s.

\paragraph{\textbf{Computing $\Ocomp \Pi^{\good} V_L^{(3),\midd}V_L^{(2),\midd}V_R^{(1),\midd,\dagger}\ket{\chi^{\frl,1}_{\overline{S},X,\vect{x},\vect{y},w_1}}_{\reg{AB\overline{ST}}}\ket{x'}_{\reg{C}}$:}} We start by computing: 
\begin{align*}
        \ket{\phi_1} =&  \Ocomp \Pi^{\good}V_L^{(3),\midd}V_L^{(2),\midd}V_R^{(1),\midd,\dagger} \left(\ket{\chi^{\frl,1}_{\overline{S},X,\vect{x},\vect{y},w_1}}_{\reg{AB\overline{ST}}}\ket{x'}_{\reg{C}}\right) \\
         =&  \Ocomp \Pi^{\good}\ket{\chi^{\frr,2}_{\overline{S},X,\vect{x}||x',\vect{y},w_1}}_{\reg{ABC\overline{ST}}} \\
         =&  \Ocomp \ket{\chi^{\frr,2}_{\overline{S},X,\vect{x}||x',\vect{y},w_1}}_{\reg{ABC\overline{ST}}} \\
         =&  \Ocomp \frac{1}{2^n\sqrt{2^{\secp}-a}}\sum_{\substack{y_0,y_1\in\bit^{n}\\ w\in(\bit^{\secp\setminus\Im(\overline{S})})}}\ket{y_0,w,y_1,\frG(\overline{S}\cup\set{(X\cR,\vect{x}||x',y_0||\vect{y}||y_1,w_1,w)})}_{\reg{ABC\overline{ST}}} \\
         =&  \frac{1}{2^n\sqrt{2^{\secp}-a}}\sum_{\substack{y_0,y_1\in\bit^{n}\\ w\in(\bit^{\secp\setminus\Im(\overline{S})})}}\ket{y_0,w,y_1,\frF(\overline{S}\cup\set{(X\cR,\vect{x}||x',y_0||\vect{y}||y_1,w_1,w)})}_{\reg{ABC\overline{ST}}} 
    \end{align*}
\noindent Where the above is by~\Cref{lem:LtoR:2}.

\paragraph{\textbf{Computing $W_L^{\frm(\secp)}\Ocomp\ket{\chi^{\frl,1}_{\overline{S},X,\vect{x},\vect{y},w_1}}_{\reg{AB\overline{ST}}}\ket{x'}_{\reg{C}}$}} We start by computing: 
\begin{align*}
    \ket{\phi_2} =&  W_L^{\frm(\secp)}\Ocomp\ket{\chi^{\frl,1}_{\overline{S},X,\vect{x},\vect{y},w_1}}_{\reg{AB\overline{ST}}}\ket{x'}_{\reg{C}}\\
     =&  W_L^{\frm(\secp)}\Ocomp\sum_{\substack{u\in\bit^{n}\\ v\in(\bit^{\secp}\setminus\Im(\overline{S}))}}\ket{u,v,x',\frG(\overline{S}\cup\set{(X\cR,\vect{x},\vect{y}||u,w_1,v)})}_{\reg{ABC\overline{ST}}}\\
     =&  W_L^{\frm(\secp)}\sum_{\substack{u\in\bit^{n}\\ v\in(\bit^{\secp}\setminus\Im(\overline{S}))}}\ket{u,v,x',\frF(\overline{S}\cup\set{(X\cR,\vect{x},\vect{y}||u,w_1,v)})}_{\reg{ABC\overline{ST}}}\\
    =&  \frac{1}{2^n\sqrt{2^{\secp}-a}}\sum_{\substack{y_0,y_1\in\bit^{n}\\ w\in(\bit^{\secp\setminus\Im(\overline{S})})}}\ket{y_0,w,y_1,\frF(\overline{S}\cup\set{(X\cR,\vect{x}||x',y_0||\vect{y}||y_1,w_1,w)})}_{\reg{ABC\overline{ST}}}
\end{align*}

\noindent Hence, we have $\ket{\phi_1}=\ket{\phi_2}$. Since the subspace represented by $\Pi^{\good}\Pi^{\frl,2}$ is spanned by $\ket{\chi^{\frl,1}_{\overline{S},X,\vect{x},\vect{y},w_1}}_{\reg{AB\overline{ST}}}\ket{x'}_{\reg{C}}$, hence $$\left\|\left(\Ocomp \Pi^{\good} V_L^{(3),\midd}V_L^{(2),\midd}V_R^{(1),\midd,\dagger} - W_L^{\frm(\secp)}\Ocomp\right)\Pi^{\good}\Pi^{\frl,2}\Pi_{\leq t}\right\|_{\op} = 0.$$

\noindent Finally, we get 
$$\left\|\left(\Ocomp \Pi^{\good} W^{\gluedfwd} - W^{\frm(\secp)}\Ocomp\right)\Pi^{\good}\Pi^{\frl,2}\Pi_{\leq t}\right\|_{\op} = O(t^2/2^{\secp})$$
\end{proof}

\begin{proof}[Proof of~\Cref{lem:gluing:fwd_3}]
We recall that we want to estimate: 
\begin{align*}
    \gamma =& \left\|\left(\Ocomp \Pi^{\good} W^{\gluedfwd} - W^{\frm(\secp)}\Ocomp\right)\Pi^{\good}\Pi^{\frl,3}\Pi_{\leq t}\right\|_{\op}\\
    \leq& \left\|\left(\Ocomp \Pi^{\good} W^{\gluedfwd} - (I-W_L^{\frm(\secp)}W_L^{\frm(\secp),\dagger})W_R^{\frm(\secp),\dagger}\Ocomp\right)\Pi^{\good}\Pi^{\frl,3}\Pi_{\leq t}\right\|_{\op} \\
    &+\left\|W_L^{\frm(\secp)}(I-W_R^{\frm(\secp)}W_R^{\frm(\secp),\dagger})\Ocomp\Pi^{\good}\Pi^{\frl,3}\Pi_{\leq t}\right\|_{\op}\\
    \leq& \left\|\left(\Ocomp \Pi^{\good} \left(\Pi^{\cL,3}-\Pi^{\cL,32}\right)V_L^{(3),\midd}V_R^{(2),\midd,\dagger}V_R^{(1),\midd,\dagger} - (I-W_L^{\frm(\secp)}W_L^{\frm(\secp),\dagger})W_R^{\frm(\secp),\dagger}\Ocomp\right)\Pi^{\good}\Pi^{\frl,3}\Pi_{\leq t}\right\|_{\op}\\
    &+O(t^2/2^{\secp})\\
    \leq& \underbrace{\left\|\left(\Ocomp \Pi^{\good} \left(\Pi^{\cL,3}\right)V_L^{(3),\midd}V_R^{(2),\midd,\dagger}V_R^{(1),\midd,\dagger} - W_R^{\frm(\secp),\dagger}\Ocomp\right)\Pi^{\good}\Pi^{\frl,3}\Pi_{\leq t}\right\|_{\op}}_{\gamma_1}\\
    &+ \underbrace{\left\|\left(\Ocomp \Pi^{\good} \left(\Pi^{\cL,32}\right)V_L^{(3),\midd}V_R^{(2),\midd,\dagger}V_R^{(1),\midd,\dagger} - (W_L^{\frm(\secp)}W_L^{\frm(\secp),\dagger})W_R^{\frm(\secp),\dagger}\Ocomp\right)\Pi^{\good}\Pi^{\frl,3}\Pi_{\leq t}\right\|_{\op}}_{\gamma_2}\\
    &+O(t^2/2^{\secp})
\end{align*}
Where the second line is by triangle inequality, and the third line is by~\Cref{lem:gluing:com_3}, and the last line is by triangle inequality. Next, we compute $\gamma_1$ and $\gamma_2$. Before computing this, we know by~\Cref{lem:proj:com2,lem:proj:com3} that the subspace represented by $\Pi^{\good}\Pi^{\frl,2}$ is spanned by $\ket{\chi^{\frl,2}_{\overline{S},X,\vect{x},\vect{y},w_1}}_{\reg{ABC\overline{ST}}}$s where $\leng(\vect{x})> 2$.

\paragraph{\textbf{Computing $\gamma_1$:}} We start by looking at some fixed $\overline{S},X,\vect{x}||x',\vect{y},w_1$ with $a = \coun{S}$ and $b = \leng(\overline{S})\cup\leng(\vect{x})$. 
Then we will show $$\Big(\Ocomp \Pi^{\good} V_L^{(3),\midd}V_R^{(2),\midd,\dagger}V_R^{(1),\midd,\dagger} - W_R^{\frm(\secp),\dagger}\Ocomp\Big)\ket{\chi^{\frl,2}_{\overline{S},X,\vect{x}||x',\vect{y},w_1}}_{\reg{ABC\overline{ST}}} = 0$$
We start by computing the first term (call it $\ket{\phi_1}$):
\begin{align*}
    \ket{\phi_1} =& \Ocomp \Pi^{\good} V_L^{(3),\midd}V_R^{(2),\midd,\dagger}V_R^{(1),\midd,\dagger}\ket{\chi^{\frl,2}_{\overline{S},X,\vect{x}||x',\vect{y},w_1}}_{\reg{ABC\overline{ST}}}\\ 
    =& \Ocomp \Pi^{\good}\ket{\chi^{\frr,1}_{\overline{S},X,\vect{x},\vect{y},w_1}}_{\reg{AB\overline{ST}}}\ket{x'}_{\reg{C}} \\
    =& \Ocomp \frac{1}{\sqrt{2^{n}(2^{\secp}-a)}}\sum_{\substack{y\in\bit^{n}\\ w\in(\bit^{\secp}\setminus\Im(\overline{S}))}}\ket{y,w,x',\frG(\overline{S}\cup\set{(X\cL,\vect{x},\vect{y},w_1,w)})}_{\reg{ABC\overline{ST}}}\\
    =& \frac{1}{\sqrt{2^{n}(2^{\secp}-a)}}\sum_{\substack{y\in\bit^{n}\\ w\in(\bit^{\secp}\setminus\Im(\overline{S}))}}\ket{y,w,x',\frF(\overline{S}\cup\set{(X\cL,\vect{x},\vect{y},w_1,w)})}_{\reg{ABC\overline{ST}}}
\end{align*}
\noindent Where the above is by~\Cref{lem:LtoR:2}. Next, we compute the second term (call it $\ket{\phi_2}$):
\begin{align*}
    \ket{\phi_2} =& W_R^{\frm(\secp),\dagger}\Ocomp\ket{\chi^{\frl,2}_{\overline{S},X,\vect{x}||x',\vect{y},w_1}}_{\reg{ABC\overline{ST}}}\\ 
    =& W_R^{\frm(\secp),\dagger}\Ocomp \frac{1}{2^n\sqrt{2^{\secp}-a}}\sum_{\substack{y_0,y_1\in\bit^{n}\\ w\in(\bit{\secp}\setminus\Im(\overline{S}))}} \ket{y_0,w,y_1,\frG(\overline{S}\cup\set{(X\cR,\vect{x}||x',\vect{y},w_1,w)})}\\
    =& W_R^{\frm(\secp),\dagger}\frac{1}{2^n\sqrt{2^{\secp}-a}}\sum_{\substack{y_0,y_1\in\bit^{n}\\ w\in(\bit{\secp}\setminus\Im(\overline{S}))}} \ket{y_0,w,y_1,\frF(\overline{S}\cup\set{(X\cR,\vect{x}||x',\vect{y},w_1,w)})}\\
    =& \frac{1}{\sqrt{2^{n}(2^{\secp}-a)}}\sum_{\substack{y\in\bit^{n}\\ w\in(\bit^{\secp}\setminus\Im(\overline{S}))}}\ket{y,w,x',\frF(\overline{S}\cup\set{(X\cL,\vect{x},\vect{y},w_1,w)})}_{\reg{ABC\overline{ST}}}
\end{align*}
\noindent Hence $\ket{\phi_1}=\ket{\phi_2}$. Hence, we get $\gamma_1=0$.

\noindent Similarly, we can show that $\gamma_2=O(t^2/2^{\secp})$.
\noindent Hence combining, we get 
$$\left\|\left(\Ocomp \Pi^{\good} W^{\gluedfwd} - W^{\frm(\secp)}\Ocomp\right)\Pi^{\good}\Pi^{\frl,3}\Pi_{\leq t}\right\|_{\op} = O(t^{2}/2^{\secp}).$$
\end{proof}

\begin{proof}[Proof of~\Cref{lem:gluing:fwd_4}]
We recall that we want to estimate: 
\begin{align*}
    \gamma =& \left\|\left(\Ocomp \Pi^{\good} W^{\gluedfwd} - W^{\frm(\secp)}\Ocomp\right)\Pi^{\good}\Pi^{\frl,4}\Pi_{\leq t}\right\|_{\op}\\
    \leq& \left\|\left(\Ocomp \Pi^{\good} W^{\gluedfwd} - (I-W_L^{\frm(\secp)}W_L^{\frm(\secp),\dagger})W_R^{\frm(\secp),\dagger}\Ocomp\right)\Pi^{\good}\Pi^{\frl,4}\Pi_{\leq t}\right\|_{\op} \\
    &+\left\|W_L^{\frm(\secp)}(I-W_R^{\frm(\secp)}W_R^{\frm(\secp),\dagger})\Ocomp\Pi^{\good}\Pi^{\frl,4}\Pi_{\leq t}\right\|_{\op}\\
    \leq& \left\|\left(\Ocomp \Pi^{\good} \left(I-\Pi^{\cL,3}\right)V_R^{(3),\midd,\dagger}V_R^{(2),\midd,\dagger}V_R^{(1),\midd,\dagger} - (I-W_L^{\frm(\secp)}W_L^{\frm(\secp),\dagger})W_R^{\frm(\secp),\dagger}\Ocomp\right)\Pi^{\good}\Pi^{\frl,4}\Pi_{\leq t}\right\|_{\op}\\
    &+O(t^2/2^{\secp})\\
    \leq& \underbrace{\left\|\left(\Ocomp \Pi^{\good} V_R^{(3),\midd,\dagger}V_R^{(2),\midd,\dagger}V_R^{(1),\midd,\dagger} - W_R^{\frm(\secp),\dagger}\Ocomp\right)\Pi^{\good}\Pi^{\frl,4}\Pi_{\leq t}\right\|_{\op}}_{\gamma_1}\\
    &+ \underbrace{\left\|\left(\Ocomp \Pi^{\good} \left(\Pi^{\cL,3}\right)V_R^{(3),\midd,\dagger}V_R^{(2),\midd,\dagger}V_R^{(1),\midd,\dagger} - (W_L^{\frm(\secp)}W_L^{\frm(\secp),\dagger})W_R^{\frm(\secp),\dagger}\Ocomp\right)\Pi^{\good}\Pi^{\frl,4}\Pi_{\leq t}\right\|_{\op}}_{\gamma_2}\\
    &+O(t^2/2^{\secp})
\end{align*}
Where the second line is by triangle inequality, and the third line is by~\Cref{lem:gluing:com_4}, and the last line is by triangle inequality. Next, we compute $\gamma_1$ and $\gamma_2$. Before computing this, we know by~\Cref{lem:proj:com3} that the subspace represented by $\Pi^{\good}\Pi^{\frl,4}$ is spanned by $\ket{\chi^{\frl,3}_{\overline{S},(x_0,x_1),w_1}}_{\reg{ABC\overline{ST}}}$s. 

\paragraph{\textbf{Computing $\gamma_1$:}} We start by looking at some fixed $\overline{S},(x_0,x_1),w_1$ with $a = \coun{S}$ and $b = \leng(\overline{S})$. 
Then we will show $$\Big(\Ocomp \Pi^{\good} V_R^{(3),\midd,\dagger}V_R^{(2),\midd,\dagger}V_R^{(1),\midd,\dagger} -W_R^{\frm(\secp),\dagger}\Ocomp\Big)\ket{\chi^{\frl,3}_{\overline{S},(x_0,x_1),w_1}}_{\reg{ABC\overline{ST}}} = 0$$
We start by computing the first term (call it $\ket{\phi_1}$):
\begin{align*}
    \ket{\phi_1} =& \Ocomp \Pi^{\good} V_R^{(3),\midd,\dagger}V_R^{(2),\midd,\dagger}V_R^{(1),\midd,\dagger}\ket{\chi^{\frl,3}_{\overline{S},(x_0,x_1),w_1}}_{\reg{ABC\overline{ST}}}\\ 
    =& \Ocomp \Pi^{\good}\ket{x_0,w_1,x_1,\frG(\overline{S})}\\
    =& \ket{x_0,w_1,x_1,\frF(\overline{S})}
\end{align*}
Where the above is by~\Cref{lem:LtoR:1}. Next, we compute the second term (call it $\ket{\phi_2}$):
\begin{align*}
    \ket{\phi_2} =& W_R^{\frm(\secp),\dagger}\Ocomp\ket{\chi^{\frl,3}_{\overline{S},(x_0,x_1),w_1}}_{\reg{ABC\overline{ST}}}\\ 
    =& W_R^{\frm(\secp),\dagger}\Ocomp \frac{1}{2^n\sqrt{2^{\secp}-a}}\sum_{\substack{y_0,y_1\in\bit^{n}\\ w\in(\bit{\secp}\setminus\Im(\overline{S}))}} \ket{y_0,w,y_1,\frG(\overline{S}\cup\set{(\cL\cL,(x_0,x_1),(y_0,y_1),w_1,w)})}\\
    =& W_R^{\frm(\secp),\dagger}\frac{1}{2^n\sqrt{2^{\secp}-a}}\sum_{\substack{y_0,y_1\in\bit^{n}\\ w\in(\bit{\secp}\setminus\Im(\overline{S}))}} \ket{y_0,w,y_1,\frF(\overline{S}\cup\set{(\cL\cL,(x_0,x_1),(y_0,y_1),w_1,w)})}\\
    =& \ket{x_0,w_1,x_1,\frF(\overline{S})}
\end{align*}
\noindent Hence $\ket{\phi_1}=\ket{\phi_2}$. Hence, we get $\gamma_1=0$.

\noindent Similarly, we can show that $\gamma_2=O(t^2/2^{\secp})$.

Hence combining, we get 
$$\left\|\left(\Ocomp \Pi^{\good} W^{\gluedfwd} - W^{\frm(\secp)}\Ocomp\right)\Pi^{\good}\Pi^{\frl,4}\Pi_{\leq t}\right\|_{\op} = O(t^{2}/2^{\secp}).$$
\end{proof}

\end{document}